\documentclass{amsart}
\usepackage{amsaddr}
\usepackage{amsmath}
\usepackage{amssymb}
\usepackage{textcomp}
\usepackage{gensymb}
\usepackage{dsfont}
\usepackage{graphicx}
\usepackage{geometry}
\usepackage{tabularray}
\usepackage{cite}
\usepackage{textgreek}
\usepackage[math]{cellspace}
\usepackage{comment}
\title{Quantifying T cell morphodynamics and migration in 3D collagen matrices}
\author{Yeeren I. Low}
\address{Department of Physics, McGill University, Montr\'{e}al, Qu\'{e}bec, Canada}
\curraddr{Department of Physics, University of Vermont, Burlington, Vermont, USA}
\email{yeeren.low@mail.mcgill.ca}

\begin{document}
	\begin{abstract}
		T cells undergo large shape changes (morphodynamics) when migrating. While progress has been made elucidating the molecular basis of cell migration, statistical characterization of morphodynamics and migration has been limited, particularly in physiologically realistic 3D environments. A previous study (H.\ Cavanagh \textit{et al.}, \textit{J.\ R.\ Soc.\ Interface} \textbf{19}: 20220081) found discrete states of dynamics as well as periodic oscillations of shape. However, we show that these results are due to artifacts of the analysis methods. Here, we present a revised analysis of the data, applying a method based on an underdamped Langevin equation. We find that different shape modes have different correlation times. We also find novel non-Gaussian effects. This study provides a framework in which quantitative comparisons of cell morphodynamics and migration can be made, e.g.\ between different biological conditions or mechanistic models.
	\end{abstract}
	\maketitle
	
	\section{Introduction}
	Efficient migration of leukocytes (immune cells) is crucial to mounting an effective immune response. Leukocytes must travel through various microenvironments in order to scan tissues for targets such as pathogens, antigen, or other cells \cite{MolBiolCell}.
	
	Cell migration depends fundamentally on remodeling of the actin cytoskeleton. Leukocyte migration can occur spontaneously or as a result of extracellular cues such as chemokine, and is enabled by protrusion of pseudopods containing a branched actin network at the front and contraction of actomyosin bundles at the back \cite{MechanicalModes2009}, occurring in some contexts as a cyclical process \cite{3DLeukocyteMotility1992}. It has been proposed that actin protrusions play an exploratory role but are dispensable for migration \cite{Leithner2016, IntegratingPhysicalMolecularInsights}. Meanwhile, an actin cortex maintains the integrity of the cell shape. The cells have a polarized shape, with a rounded or wide cell front while the cell rear contains a thin cylindrical structure called the uropod \cite{Dupre2015}. Membrane material for large shape changes is provided by microvilli, which cover the cells \cite{IfYouPleaseDrawMeACell2020}.
	
	Leukocytes utilize a fast mode migration termed ``amoeboid'', so named due to their large shape changes and similarity to the social amoeba \textit{Dictyostelium discoideum} \cite{AmoeboidLeukocyteCrawling2001}. This migration mode is further characterized by weak adhesion and typically does not involve proteolysis of the extracellular matrix \cite{InterstitialLeukocytesMigrationImmuneFunction2008, FocalAdhesionIndependentCellMigration2016, PrinciplesLeukocyteMigrationStrategies2020}. While many studies have analyzed cell migration in 2D, in the context of immune response it is more physiologically realistic to consider 3D migration \cite{DimensionsCellMigration2013}. On 2D surfaces, leukocytes use adhesion receptors (such as integrins) to anchor the cell during protrusion \cite{Mechanisms3DCellMigration2019}; however, in 3D environments they are able to use an adhesion-independent mode of migration, using a ``flowing and squeezing'' mechanism instead \cite{RapidLeukocyteMigration2008}, and they are also able to use topography to generate motion \cite{CellularLocomotionTopography2020}. However, determination of the actual mechanism underlying migration is complicated by adaptability \cite{FocalAdhesionIndependentCellMigration2016}.
	
	Leukocyte migration can be characterized as a search problem, and different types of random walks have been proposed to describe it \cite{TCellMigrationSearchStrategiesMechanisms2016}. Migration is modulated by both environmental and cell-intrinsic factors \cite{PMrassReview2010, WeningerLeukocyteMigration2014, MunozTCellMigrationLymphNodes2014}, and molecular perturbations are seen to modify turning behavior \cite{Myo1g2014, PMrassROCK2017}.
	
	Turning to the underpinnings of cell locomotion, actin waves have been observed in a variety of cell types and are thought to underlie cell migration. They are hypothesized to arise from an excitable system \cite{AllardMogilner2013, InagakiKatsuno2017}. Membrane tension is proposed to act as a global negative feedback to restrict protrusive activity around the cell surface \cite{SahaJoiningForces2018}. In 2D migration of the social amoeba \textit{Dictyostelium discoideum}, the molecular workings of the excitable system have been recently elucidated \cite{KuhnDictyostelium2021, ShiIglesias2013, LiExcitableNetworks2020, PalExcitableSignalTransductionNetworks2019, ChengRolesSignaling2020}. For cell migration more generally, there have been significant modeling efforts, particularly in the case of 2D migration \cite{ButtenschonReviewModels2020, SunZamanModeling2017, BanerjeeActinCytoskeleton2020, HolmesComparisonComputationalModels2012, CallanJonesAmoeboidMotility2022, DanuserMathematicalModeling2013}. Modeling biochemical reactions along with cellular shape change, however, poses particular challenges due to the so-called ``moving boundary problem'' \cite{StinnerBiochemistryChangingGeometries2020, DiNapoliComputationalAnalysis2021}. Swimming in 3D of \textit{Dictyostelium} has been modeled by a reaction-diffusion system on the membrane \cite{CampbellComputationalAmoeboidSwimming2017}. Both 2D crawling and 3D swimming of \textit{Dictyostelium} occur by protrusions forming at the front of the cell, which bifurcate and translocate toward the rear of the cell \cite{BarryBretscherSwimming2010, BaeBodenschatzSwimming2010, VanHaastertWalkingGlidingSwimming2011, DriscollCellShapeDynamics2012}. These shape changes are time-irreversible in accordance with Purcell's scallop theorem \cite{PurcellLowReynoldsNumber1976}, although shape changes are not strictly required for swimming \cite{FarutinCrawlingFluid2019, OthmerEukaryoticCellDynamics2018}. Time-irreversible force dynamics has also been observed for 3D mesenchymal migration \cite{GodeauCellMigrationCorrelation2022}. Meanwhile, links between cell shape and migration have been studied \cite{BodorCellShapes2020}. The connection between biology and mechanics is known as mechanobiology, which is a growing field \cite{CellMechanicsMechanobiology2023}.
	
	To model cellular shape change and migration, rather than taking a ``bottom-up'' or mechanistic approach to modeling, where biochemical reactions are posited and force balance equations are written, we opt for a ``top-down'' or data-driven approach, where experimental data is characterized by measured statistical coefficients. This approach allows for identification of important quantities governing the dynamics, which could be compared between different biological conditions or used as benchmarks for mechanistic models. In the words of \cite{LiDictyDynamics2011}, such characterization ``is important because it defines the motion that the bottom-up approach attempts to explain''. Data-driven modeling of cell migration has been done mostly in 2D \cite{SelmecziCellMotility2008, LiPersistentCellMotion2008, LiDictyDynamics2011, BodekerAmoeboidMotion2010, AmselemDictyosteliumChemotaxis2012, BosgraafPseudopodia2009, VanHaastertCorrelatedRandomWalk2010, VanHaastertStochasticModelChemotaxis2010, VanHaastertAmoeboidPseudopodExtension2020, VanHaastertMemoryAmoeboid2021, LiuTLymphocyteMigration2015, MetznerSuperstatisticalRandomWalks2015, CherstvyNonGaussianityAmoeboid2018, MitterwallnerNonMarkovianMotility2020}. Similar methods have been used to characterize dynamics of the nematode worm \textit{Caenorhabditis elegans} shape, motion, and neural activity \cite{StephensDimensionalityDynamicsCElegans2008, StephensModesMovementCElegans2010, StephensLongTimescalesCElegans2011, BrownBehavioralMotifsCElegans2012, YeminiCElegansPhenotypes2013, GomezMarinHierarchicalCompressionCElegans2016, JaverBehaviouralFeaturesCElegans2018, DanielsCElegansEscape2019, LindermanHierarchical2019, CostaLocallyLinear2019, AhamedContinuousComplexityCElegans2021, CostaMaximallyPredictiveStates2023, costa2023markovian, CostaFluctuating2024} and cell migration confined to adhesive micropatterns \cite{BrucknerStochasticDynamics2019, BrucknerBehaviouralVariability2020, FinkAreaGeometryDependence2020, BrucknerCellCellInteractions2021, BrucknerGeometryAdaptation2022}. A review of top-down modeling efforts applied to cell migration is given in \cite{BrucknerLearning2024}.
	
	T cell morphodynamics and migration in 3D collagen matrices have been previously analyzed in \cite{Cavanagh2022}. However, that study contains several major flaws and shortcomings which will be discussed in the main text. This paper presents a reanalysis of the data in \cite{Cavanagh2022}.
	
	\section{Dimensionality reduction of the shape of motile cells}
	First, motile cells were distinguished from sessile cells based on trajectories of the cell centroid. Due to the small number of cells measured, this could be done manually, with the trajectories of motile cells spanning a distance (the maximum distance between two points on a trajectory) of at least $\approx$ 20 \textmu m. A total of 10 motile cells were detected, all recorded with a frame interval of 4.17 s.
	
	Next, we analyzed the shapes of the motile cells. Sessile cells are spherical, whereas motile cells assume a polarized shape \cite{FriedlTLymphocytes1998}. The study \cite{Cavanagh2022} uses descriptors of the cell surface based on spherical harmonics $Y_l^m$, and taking for each value of $l$ the squared magnitude summed over $m$. However, this loses information. The study \cite{Cavanagh2022} partially remedies this by tracking the uropod and recording the distance between the uropod and centroid. We can do better by taking moments of harmonic polynomials relative to the uropod\textendash centroid axis \cite{BodorCellShapes2020}; we use this axis as a proxy for the polarization axis. We use a convention for $Y_l^m$ without Condon\textendash Shortley phase, normalized according to:
	\begin{equation}
		\frac{1}{4 \pi} \int \mathrm d \Omega \, {Y_l^m}^* Y_{l'}^{m'} = \delta_{ll'} \delta_{mm'}
	\end{equation}
	(asterisk denoting complex conjugate). We define central moments:
	\begin{equation}
		M_l^m := \int \mathrm d \mathcal A \, r^l Y_l^m(\theta, \phi),
	\end{equation}
	where $(r,\theta,\phi)$ are spherical coordinates relative to the centroid of the cell surface and the uropod\textendash centroid axis, and normalized moments:
	\begin{equation}
		m_l^m := \frac{M_l^m}{\mathcal A^{(l+2)/2}},
	\end{equation}
	where $\mathcal A$ is the cell surface area. Due to translational symmetry, there are no $l = 1$ terms. Rotational invariance is broken by the choice of polarization axis\footnote{This leads to the possibility of describing shape dynamics using linear equations (section ``Linear Gaussian model''), which is much simpler than a rotationally invariant description which necessitates complicated nonlinearities \cite{Ohta2017}. Such a simplification is biologically meaningful as cell polarity is maintained by spatial distributions of molecular components.}. The next question is how to scale the quantities with different values of $l$. We consider a small deviation from a spherical surface, represented by $r = R(1 + \epsilon_l^m Y_l^m)$ for some constant $R$ (using real spherical harmonics\footnote{For $m>0$, the use of real spherical harmonics leads to a factor of $\sqrt{2}$ in the definition of $\epsilon_l^m$ relative to complex spherical harmonics. If we refer to the coefficients of real spherical harmonics as $\Re \epsilon_l^m$ and $\Im \epsilon_l^m$ ($\Re$ and $\Im$ denoting real and imaginary parts, respectively), as we do here, then the squared magnitude of the deformation mode is $|\epsilon_l^m|^2$ rather than $2 |\epsilon_l^m|^2$.}). For $l \ge 2$, to order $\mathcal O(\epsilon_l^m)$, we have:
	\begin{equation}
		m_l^m = \frac{l+2}{(4 \pi)^{l/2}} \epsilon_l^m.
	\end{equation}
	Thus, we define our shape variables $s_l^m$ as:
	\begin{equation}
		s_l^m := \frac{(4\pi)^{l/2}}{l+2} m_l^m,
	\end{equation}
	for $l \ge 2$. For $l = 0$, we define $s_0^0 := \log (\mathcal A/4 \pi)/2$ so that a change $r \to (1+\epsilon) r$ results in a corresponding change $s_0^0 \to s_0^0 + \epsilon$. In addition, if the uropod\textendash centroid axis is called the $z$-axis, we define components of the velocity $v^0 := v^z$ and $v^1 := v^x + i v^y$, as well as orientational changes $\Delta \theta e^{i \phi}$ where $\Delta \theta$ is the angle between the uropod\textendash centroid axes at the initial and final time-steps, and $\phi$ is the azimuth of the uropod\textendash centroid axis at the final time-step in the coordinate system of the initial time-step.
	
	Lastly, we examine the variances of the shape variables, but before doing so, we subtract the cell-wise mean of the shape parameters. Without doing so, we may obtain a situation where we capture intercellular rather than intracellular variability \cite{MechanismOfShapeDetermination2008}. We retain moments up to order $l = 3$, as this is the minimal possible value of $l$ for which amoeboid swimming by means of shape change is possible \cite{FarutinAmoeboidSwimming2013}. The $l = 2$, $m = 0$ mode is simply elongation of the cell along its axis (the $z$-axis). The $l = 2$, $m = 1$ mode describes tilting of the cell and is positive when the cell front ($z>0$) deviates in the $+x$-direction while the cell back ($z<0$) deviates in the $-x$-direction. The $l = 2$, $m = 2$ mode describes lateral elongation of the cell and is positive when the elongation is along the $x$-axis. The $l = 3$, $m = 0$ mode describes broken symmetry along the polarization axis and is positive when the cell front is narrower than the cell back. Thus the ``stereotypical'' shape of a polarized cell, which is wide at the front, has $s^0_3 < 0$. The $l = 3$, $m = 1$ mode describes a bent shape, and is positive when the front and back of the cell deviate in the $+x$-direction while in between it deviates in the $-x$-direction. The $l = 3$, $m = 2$ mode describes differential lateral elongation, and positive when the front of the cell is elongated along the $x$-axis while the back of the cell is elongated along the $y$-axis. The $l = 3$, $m = 3$ mode describes a triangular deformation in the $xy$-plane, and is positive when the shape is elongated along the $(1,0)$ and $(-1/2,\pm \sqrt 3/2)$ directions. In Table \ref{tab:corning-vars}, 95\% confidence intervals calculated using a $t$-distribution are listed. The only non-zero quantities are invariant under a change of definition of azimuth $\phi \to \phi + \Delta \phi$, due to symmetry in the description. We see that the variances of $s_0^0$ and $s_3^3$ are small compared to the other shape variables, so we ignore them. We also see that, up to this point in the analysis, there is no chirality; chirality would be manifested as quantities not being invariant under the transformation $\phi \to -\phi$.
	
	\begin{table}
		\begin{center}
			\caption{Quantities with 95\% confidence intervals.}
			\label{tab:corning-vars}
			\begin{tblr}{Q[c,m]|Q[c,m]|}
				\hline
				$\exp(\langle s_0^0\rangle)$ & $9.2 \pm 0.8$ \textmu m \\ \hline
				$\langle s_2^0 \rangle$ & $0.16 \pm 0.07$ \\ \hline
				$\langle s_3^0 \rangle$ & $-0.12 \pm 0.03$ \\ \hline
				$\langle v^0 \rangle$ & $0.28 \pm 0.08$ \textmu m/fr \\ \hline \hline
				$\operatorname{Var}(s_0^0)$ & $(1.1 \pm 0.8) \times 10^{-3}$ \\ \hline
				$\operatorname{Var}(s_2^0)$ & $(5.6 \pm 2.2) \times 10^{-3}$ \\ \hline
				$\langle s_2^1 {s_2^1}^* \rangle$ & $(8.7 \pm 2.8) \times 10^{-3}$ \\ \hline
				$\langle s_2^2 {s_2^2}^* \rangle$ & $(6.7 \pm 2.1) \times 10^{-3}$ \\ \hline
				$\operatorname{Var}(s_3^0)$ & $(2.8 \pm 1.1) \times 10^{-3}$ \\ \hline
				$\langle s_3^1 {s_3^1}^* \rangle$ & $(3.2 \pm 1.4) \times 10^{-3}$ \\ \hline
				$\langle s_3^2 {s_3^2}^* \rangle$ & $(2.2 \pm 1.0) \times 10^{-3}$ \\ \hline
				$\langle s_3^3 {s_3^3}^* \rangle$ & $(0.8 \pm 0.5) \times 10^{-3}$ \\ \hline \hline
				$\operatorname{Cov}(s_2^0,s_3^0)$ & $(-1.0 \pm 1.1) \times 10^{-3}$ \\ \hline
				$\langle s_2^1 {s_3^1}^* \rangle$ & $(1.1 \pm 0.7) \times 10^{-3} + (0.2 \pm 0.8) \times 10^{-3} i$ \\ \hline
				$\langle s_2^2 {s_3^2}^* \rangle$ & $(2.6 \pm 1.0) \times 10^{-3} + (0.1 \pm 0.3) \times 10^{-3} i$ \\ \hline
			\end{tblr}
		\end{center}
	\end{table}
	
	Next, we investigate whether dynamics is stationary. We will see later that a description based on overdamped Langevin equations \cite{Frishman2020} is not sufficient on the measured time-scales, i.e., an underdamped description \cite{Bruckner2020} is necessary.\footnote{Non-Markovianity of cell shapes in 2D for measurements with a 3 s frame interval was suggested in \cite{TweedyScreening2019}. However, quantities were linearly interpolated between time-points, which is expected to introduce artifactual non-Markovianity.} We use the characterization based on second-order time-symmetric and -antisymmetric quantities \cite{low2023second} and perform linear regression of these quantities with respect to time. Before statistical analysis, we also scale each cell's quantities by a suitable factor related to the cell-wise variances of the variables. Using a $t$-test\footnote{Due to a population of only 10 cells, we opted for a $t$-test instead of a bootstrap \cite{IntroductionBootstrap1994}. The reason is that if 10 quantities are sampled i.i.d.\ from a distribution with median 0, then with probability $2^{-9}$, all values will have the same sign. Thus, we cannot estimate $p$-values below $2^{-9}$ using the bootstrap. However, the Holm\textendash Bonferroni correction demands estimation of such $p$-values, as will be seen. As for Gaussianity, while instantaneous quantities may not be normally distributed, we would expect that the time-average approximately obeys a Gaussian distribution. Appendix D contains a discussion of the validity of this approximation.} with Holm\textendash Bonferroni correction \cite{Holm1979}\footnote{An alternative is the Hochberg procedure \cite{Hochberg1988}, which is more powerful than the Holm\textendash Bonferroni procedure. It is based on the Simes test \cite{Simes1986}, which is conservative for tests that are positively dependent in a certain sense \cite{SarkarChang1997, Sarkar1998}. It may be expected to be valid in the case of a jointly Gaussian distribution, but to our knowledge, at this point in time there is no proof of this (or proof to the contrary). In all the cases investigated in this study, identical results were obtained with the Hochberg procedure as with the Holm\textendash Bonferroni procedure.}, no statistically significant deviation from stationarity was detected.
	
	\section{Linear Gaussian model}
	Here, we fit a linear Gaussian model to the dynamics. In such a model, only quantities with the same (absolute value of) $m$ can couple to each other. For $m=0$, the model is:
	\begin{equation}
		\begin{pmatrix}
			\ddot s_2^0 \\
			\ddot s_3^0 \\
			\dot v^0
		\end{pmatrix} = \mathbf A_{\mathbf x}^0 \Delta \mathbf s_{2:3}^0 + \mathbf A_{\mathbf v}^0 \begin{pmatrix}
		\dot s_2^0 \\
		\dot s_3^0 \\
		\Delta v^0
		\end{pmatrix} + \boldsymbol \xi^0, \quad \langle \boldsymbol \xi^0(t) \boldsymbol \xi^0(t')^{\mathsf T} \rangle = 2 \mathbf D^0 \delta(t - t'),
	\end{equation}
	where $\Delta$ denotes deviation from the cell-wise mean, $\mathbf s_{2:3}^0 := (s_2^0, s_3^0)^{\mathsf T}$, $\boldsymbol \xi^0$ is zero-mean Gaussian white noise, and $\mathbf A_{\mathbf x}^0$, $\mathbf A_{\mathbf v}^0$, and $\mathbf D^0$ are constant matrices. We used the procedure described in \cite{Bruckner2020} to infer dynamics; however, we found it to be necessary to fit a continuous-time model exactly (see Appendix A).
	
	We see that on time-scales $\gtrsim$ 30 s, the shape autocovariances decay exponentially. We computed a decay rate of $0.019 \pm 0.012$ $\mathrm s^{-1}$ (95\% confidence interval, $t$-distribution, excluding one cell whose parameters could not be successfully fitted) corresponding to a correlation time of 51 s. Next, we used the characterization based on time-symmetric and -antisymmetric quantities \cite{low2023second}, and normalized each cell's quantities by a suitable factor related to the variances, e.g.\ $\langle \Delta s_2^0 \Delta s_3^0 \rangle$ is divided by $\sqrt{\langle (\Delta s_2^0)^2 \rangle \langle (\Delta s_3^0)^2 \rangle}$. For stationary demeaned variables $x^i$ and $x^j$, we used the antisymmetric estimator $L(x^i, x^j) = \langle x^i \dot x^j \rangle - \langle \dot x^i x^j \rangle$ where $L(\cdot,\cdot)$ is the angular momentum \cite{low2023second}. As before, we used a $t$-test with Holm\textendash Bonferroni correction to evaluate statistical significance. We evaluated quantitative significance of effects using the procedure described in \cite{low2023second} for the case of almost-Markovian dynamics. For quantities involving fluctuations (diffusivities), we do the comparison using the discrete-time estimators described in \cite{Bruckner2020} (this is done throughout). The quantities with both statistical and quantitative significance are $L(\Delta s_2^0, \Delta s_3^0) < 0$ (adjusted $p = 0.008$) and $L(\dot s_2^0, \dot s_3^0) < 0$ (adjusted $p = 1.4 \times 10^{-4}$). These are still statistically significant when all values of $m$ are considered in the multiple hypothesis test. It is worth noting that the signs of these quantities were consistent across all cells. This is in accordance with previous observations of time-irreversibility \cite{BarryBretscherSwimming2010, BaeBodenschatzSwimming2010, VanHaastertWalkingGlidingSwimming2011, DriscollCellShapeDynamics2012}.
	
\begin{figure}
		\begin{center}
			\includegraphics[scale=0.5]{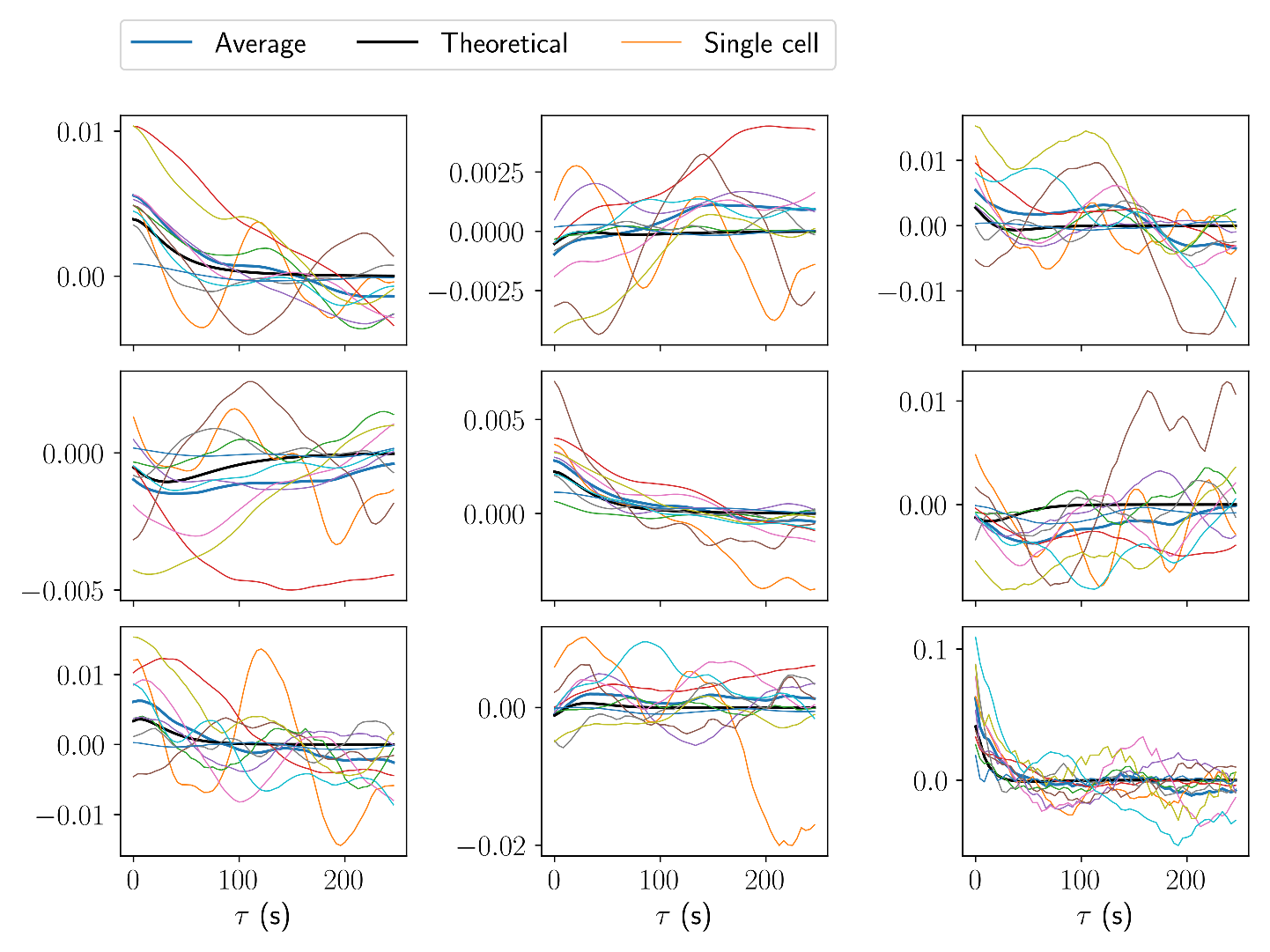}
		\end{center}
		\caption{Covariance functions with $m=0$. Velocities are in units of \textmu m/fr.}
		\label{fig:cov-func-0}
	\end{figure}

	We compared the measured and theoretical covariance functions, with mostly good agreement (Fig.\ \ref{fig:cov-func-0}). There seem to be some possible discrepancies between theory and experiment at zero time-lag; however, these are not statistically significant. Also, $\langle \Delta s^0_3(t+\tau) \Delta v^0(t) \rangle$ ($\tau > 0$) seems to possibly differ between theory and experiment. We tested the integral of the covariance function (Eq.\ \eqref{eq:exp-int} with the integral restricted to $t>t'$ and $S = 100$ s) and obtained an unadjusted $p$-value of $1.4 \times 10^{-3}$. Applying Holm\textendash Bonferroni correction to the 29 covariance functions, this gives an adjusted $p$-value of 0.04, barely qualifying as statistically significant. When combined with the other tests being performed, we may consider this as a non-significant result.
	
\begin{figure}
		\begin{center}
			\includegraphics[scale=0.5]{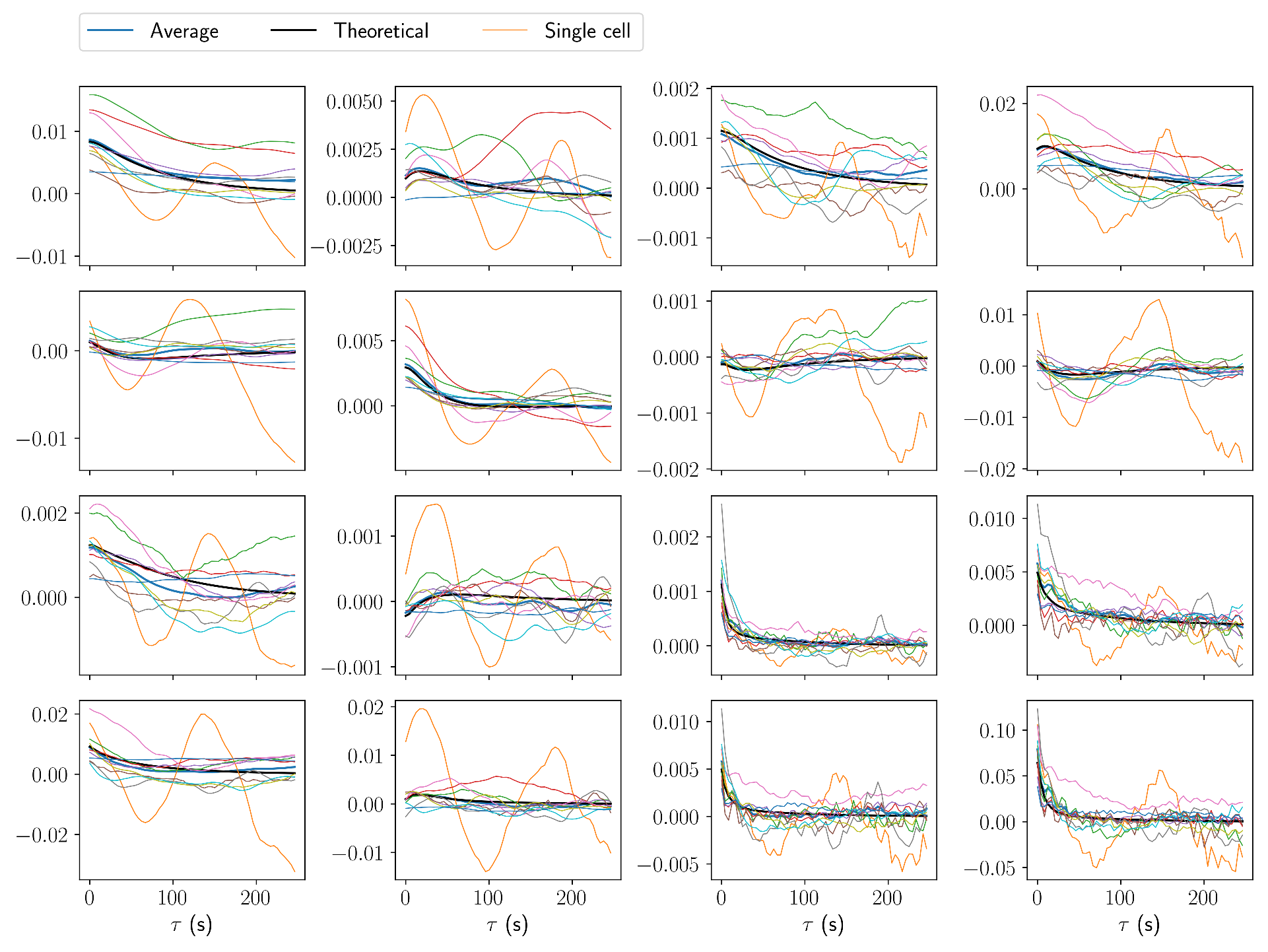}
		\end{center}
		\caption{Covariance functions with $m=1$. Velocities are in units of \textmu m/fr.}
		\label{fig:cov-func-1}
	\end{figure}

	We now turn to the $m=1$ dynamics, described by:
	\begin{equation}
		\begin{pmatrix}
			\ddot s_2^1 \\
			\ddot s_3^1 \\
			\mathrm d (\dot \theta e^{i \phi})/\mathrm d t \\
			\dot v^1
		\end{pmatrix} = \mathbf A_{\mathbf x}^1 \mathbf s_{2:3}^1 + \mathbf A_{\mathbf v}^1 \begin{pmatrix}
		\dot s_2^1 \\
		\dot s_3^1 \\
		\dot \theta e^{i \phi} \\
		v^1
		\end{pmatrix} + \boldsymbol \xi^1, \quad \langle \boldsymbol \xi^1(t) \boldsymbol \xi^1(t')^{\mathsf H} \rangle = 2 \mathbf D^1 \delta(t - t'),
	\end{equation}
	where $\mathbf s_{2:3}^1 := (s_2^1, s_3^1)^{\mathsf T}$, $\boldsymbol \xi^1$ is zero-mean complex Gaussian white noise statistically invariant under a shift of azimuth $\boldsymbol \xi^1 \to e^{i \Delta \phi} \boldsymbol \xi^1$ ($\mathsf H$ denotes Hermitian conjugate), and $\mathbf A_{\mathbf x}^1$, $\mathbf A_{\mathbf v}^1$, and $\mathbf D^1$ are constant matrices. Now, we see a difference in the correlation times of $s_2^1$ and $s_3^1$. We calculated correlation times of 85 s and 27 s\footnote{Fitting continuous-time parameters of the $m=1$ component of the full underdamped Langevin equation for individual cells was unsuccessful for half of the cells. Therefore, we estimated decay rates of individual cells for the $l=2$ and $l=3$ components by fitting exponential decays to the autocovariance functions at time-lags $\tau=0$ and $\tau=33.4$ s. The results are $0.010 \pm 0.006$ $\mathrm s^{-1}$ and $0.026 \pm 0.008$ $\mathrm s^{-1}$, respectively.}. The smaller decay rate was always associated with an eigenvector with larger $s_2^1$ component. The statistically and quantitatively significant quantities are listed in Table \ref{tab:quants-1}. Again, we see time-irreversibility in the same sense as before. In addition, we have the time-antisymmetric quantities $\Re \langle s_2^1 \dot \theta e^{-i \phi} \rangle, \Re \langle s_2^1 {v^1}^* \rangle > 0$ ($\Re$ denoting real part). We also have correlated fluctuations $\Re D_{2 \theta}^1 < 0$ (putatively) and $\Re D_{3 \theta}^1, \Re D_{\theta v}^1 > 0$. The signs of the quantities in Table \ref{tab:quants-1} were consistent across all cells, except a single value of $\Re D_{2 \theta}^1$ with normalized magnitude a factor of 0.25 times the mean, and a single value of $\Re D_{3 \theta}^1$ with normalized magnitude a factor of $10^{-3}$ times the mean. As with the $m=0$ quantities, on time-scales $\gtrsim$ 30 s the shape autocovariances decay exponentially. Theoretical and experimental covariance functions agreed well (Fig.\ \ref{fig:cov-func-1}). In contrast with $m=0$, we now have the possibility of chirality, which would manifest as non-zero imaginary parts. However, no chirality was detected. The imaginary part of the $m=1$ covariance function did not have any discernible deviation from 0 (not shown).
	
	\begin{table}
		\begin{center}
			\caption{Statistically and quantitatively significant quantities for $m=1$. Except for $\Re D_{2 \theta}^1<0$, these are still statistically significant when all values of $m$ are considered in the multiple hypothesis test.}
			\label{tab:quants-1}
			\begin{tblr}{|Q[c,m]|Q[c,m]|}
				Quantity and sign & Adjusted $p$-value \\ \hline
				$\Re L(s_2^1, {s_3^1}^*) < 0$ & $0.020$ \\ \hline
				$\Re \langle s_2^1 \dot \theta e^{-i \phi} \rangle > 0$ & $4.2 \times 10^{-5}$ \\ \hline
				$\Re \langle s_2^1 {v^1}^* \rangle > 0$ & $1.4 \times 10^{-3}$ \\ \hline
				$\Re \langle \dot \theta e^{i \phi} {v^1}^* \rangle > 0$ & $1.6 \times 10^{-7}$ \\ \hline
				$\Re L(\dot s_2^1, {{}\dot s_3^1}^*) < 0$ & $9.6 \times 10^{-3}$ \\ \hline
				$\Re D_{2 \theta}^1 < 0$ & $0.043$ \\ \hline
				$\Re D_{3 \theta}^1 > 0$ & $1.0 \times 10^{-3}$ \\ \hline
				$\Re D_{\theta v}^1 > 0$ & $5.0 \times 10^{-5}$ \\ \hline
			\end{tblr}
		\end{center}
	\end{table}
	
\begin{figure}
		\begin{center}
			\includegraphics[scale=0.5]{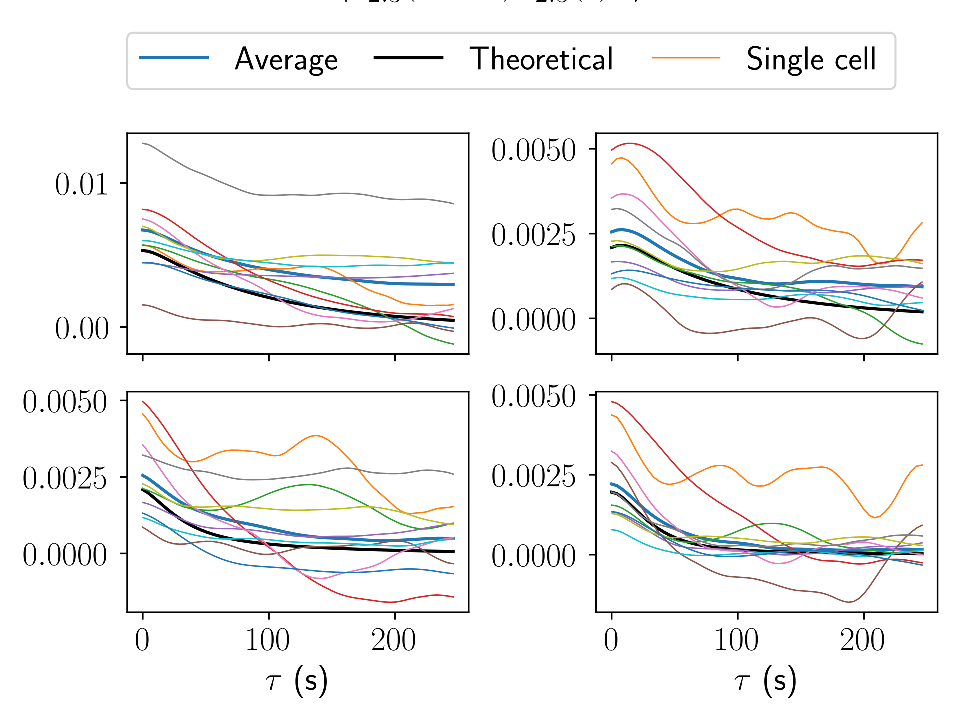}
		\end{center}
		\caption{Covariance functions with $m=2$.}
		\label{fig:cov-func-2}
	\end{figure}

	Finally, we have the $m=2$ dynamics, which is described by:
	\begin{equation}
		\ddot {\mathbf s}_{2:3}^2 = \mathbf A_{\mathbf x}^2 \mathbf s_{2:3}^2 + \mathbf A_{\mathbf v}^2 \dot {\mathbf s}_{2:3}^2 + \boldsymbol \xi^2, \quad \langle \boldsymbol \xi^2(t) \boldsymbol \xi^2(t')^{\mathsf H} \rangle = 2 \mathbf D^2 \delta(t - t'),
	\end{equation}
	where $\mathbf s_{2:3}^2 := (s_2^2, s_3^2)^{\mathsf T}$, $\boldsymbol \xi^2$ is zero-mean complex Gaussian white noise statistically invariant under a shift of azimuth $\boldsymbol \xi^2 \to e^{i \Delta \phi} \boldsymbol \xi^2$, and $\mathbf A_{\mathbf x}^2$, $\mathbf A_{\mathbf v}^2$, and $\mathbf D^2$ are constant matrices. We again have a difference in the correlation times of $s_2^2$ and $s_3^2$. We computed decay rates of $(8.8 \pm 6.0) \times 10^{-3}$ $\mathrm s^{-1}$ and $0.040 \pm 0.017$ $\mathrm s^{-1}$, corresponding to correlation times of 99 s and 20 s. The statistically and quantitatively significant quantities are listed in Table \ref{tab:quants-2}, again with time-irreversibility in the same sense as before, and no chirality detected. The signs of these quantities were consistent across all cells. As before, on time-scales $\gtrsim$ 30 s the shape autocovariances decay exponentially. Theoretical and experimental covariance functions agreed well (Fig.\ \ref{fig:cov-func-2}), except for a possible discrepancy of $\langle s^2_2(t+\tau) s^2_2(t)^* \rangle$. However, testing the integral of the covariance function (Eq.\ \eqref{eq:exp-int} with $S = 200$ s) gives a $p$-value of 0.026 which is not statistically significant when multiple hypotheses are taken into account. Again, no chirality was detected and the imaginary part of the $m=2$ covariance function did not have any discernible deviation from 0 (not shown).
	
	\begin{table}
		\begin{center}
			\caption{Statistically and quantitatively significant quantities for $m=2$. These are still statistically significant when all values of $m$ are considered in the multiple hypothesis test.}
			\label{tab:quants-2}
			\begin{tblr}{|Q[c,m]|Q[c,m]|}
				Quantity and sign & Adjusted $p$-value \\ \hline
				$\Re \langle s_2^2 {s_3^2}^* \rangle > 0$ & $1.7 \times 10^{-7}$ \\ \hline
				$\Re L(s_2^2, {s_3^2}^*) < 0$ & $4.4 \times 10^{-4}$ \\ \hline
				$\Re \langle \dot s_2^2 {{}\dot s_3^2}^* \rangle > 0$ & $4.2 \times 10^{-7}$ \\ \hline
				$\Re L(\dot s_2^2, {{}\dot s_3^2}^*) < 0$ & $2.6 \times 10^{-4}$ \\ \hline
				$\Re D_{23}^2 > 0$ & $3.5 \times 10^{-6}$ \\ \hline
			\end{tblr}
		\end{center}
	\end{table}
	
	We may compare our results to diffusion on a sphere. The spherical Laplacian has eigenfunctions $Y_l^m(\theta,\phi)$ with eigenvalues $-l(l+1)$. We observe a difference in correlation times for $l=2$ and $l=3$ for $m>0$, but not for $m=0$.
	
	\section{Previous analysis}
	The previous study \cite{Cavanagh2022} claimed oscillatory autocovariance functions (ACFs) together with much longer correlation times ($\ge$ 150 s). However, the power spectra (Supplementary Figure 5c in \cite{Cavanagh2022}) do not contain a peak at non-zero frequency. This suggests that the supposed oscillations in the ACFs are statistical noise. Thus, the fitting method which uses the peaks of the measured ACFs grossly overestimates the correlation times. A rough comparison of the power spectra of cells in the ``run'' mode to a Lorentzian function (the power spectrum of an Ornstein\textendash Uhlenbeck process) suggests correlation times of approximately 60 s for all three principal components, in accordance with our results for the $m=0$ modes. The results for the cells in the ``stop'' mode are suspect, as it is known that (and has been mentioned earlier that) sessile cells are spherical \cite{FriedlTLymphocytes1998}. Indeed, manual inspection of the trajectories of the cell centroid reveals that two out of the four cells classified as ``stop'' in \cite{Cavanagh2022} are in fact motile.
	
\begin{figure}
		\begin{center}
			\includegraphics[scale=0.5]{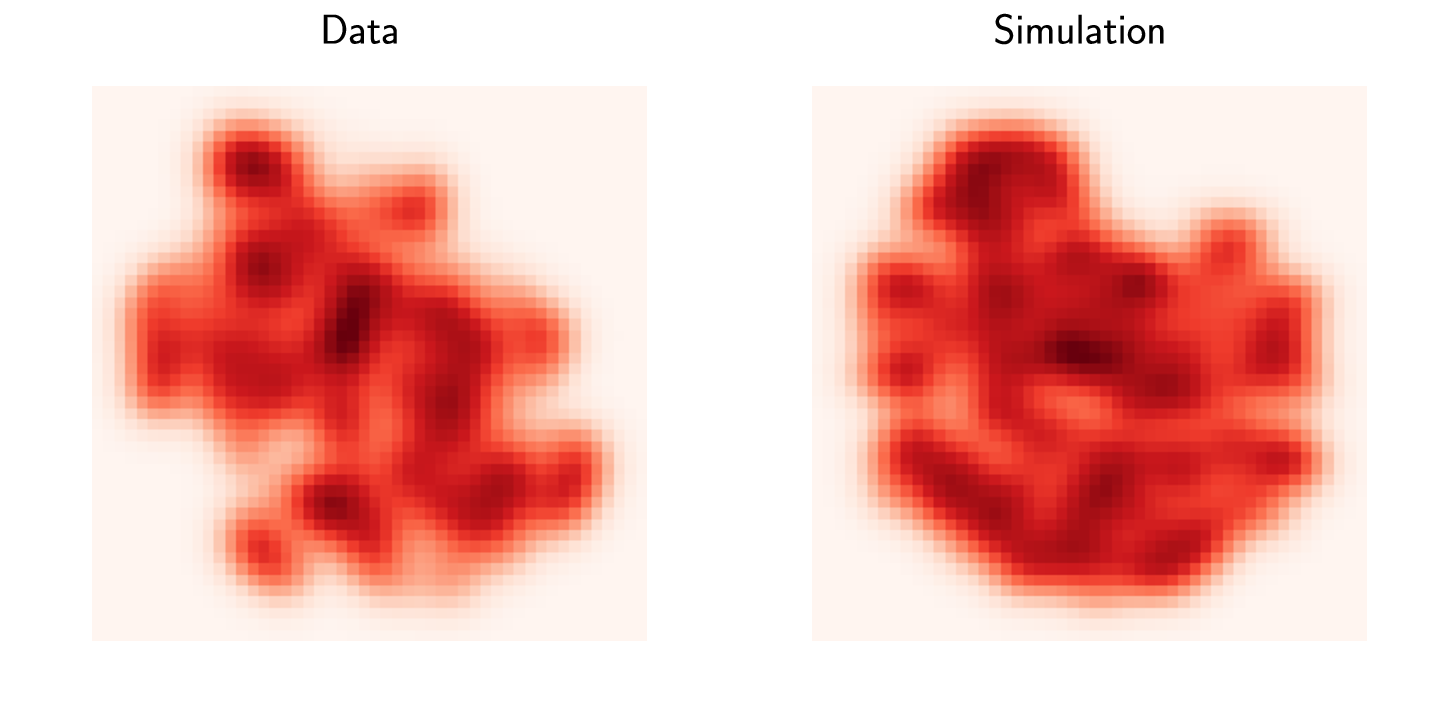}
		\end{center}
		\caption{Probability distributions obtained from applying t-SNE to real data and simulated data from a linear Gaussian model.}
		\label{fig:tsne}
	\end{figure}

	Next, it was claimed that morphodynamics can be described as inhabiting a set of discrete states. This conclusion was reached by using wavelet analysis along with t-SNE. However, applying the t-SNE algorithm to finite data can result in artifactual non-uniform probability distributions. To test for this possibility, the analysis procedure was applied to the $m=0$ shape variables and compared with the linear Gaussian model from the previous section simulated for 1764 frames (the same number of frames as in the dataset of motile cells), with very similar results (Fig.\ \ref{fig:tsne}; also compare Supplementary Figure 9a in \cite{Cavanagh2022}). Thus, the appearance of multiple peaks in the probability distribution seems to be an artifact of t-SNE.
	
	Next, in \cite{Cavanagh2022} it was claimed that the dynamics of the second principal component contain oscillations with period 100 s. However, as was mentioned earlier, there is no peak in the power spectra at non-zero frequency. This suggests a different explanation for the observed oscillations. The wavelet transform is obtained by convoluting the time-series with the wavelet, which means that the Fourier transforms of the time-series and the wavelet are multiplied. The maximum power of the first and second derivatives of a Gaussian with standard deviation $\sigma$ occurs at angular frequencies of $\sigma^{-1}$ and $\sqrt{2} \sigma^{-1}$, respectively. The so-called ``width of influence'' of the wavelet is taken to be approximately $6 \sigma$, according to the code provided in \cite{Cavanagh2022}. It was claimed that the entropy\footnote{Not to be confused with physical entropy production, which is related to the probability ratio between forward and backward trajectories \cite{SeifertStochasticThermodynamics2019}.} is minimized when the ratio of the width of the Mexican hat to the width of the derivative of a Gaussian is 1.5, while the oscillation frequency is located at the maximum of the power spectra of the wavelets. Together, these strongly support the explanation of oscillations as an artifact of the wavelet transform.
	
	Lastly, because wavelets have zero integral, the analysis of dynamics in \cite{Cavanagh2022} removes information about location in shape space. In contrast, we analyze dynamics as a function of shape.
	
	\section{Non-Gaussian effects}
	
	Following \cite{low2023second}, we characterize the third-order dynamics by the quantities $\langle x^i x^j x^k \rangle$, $\langle x^i x^j \dot x^k \rangle$ where:
	\begin{equation}
		\langle x^i x^j \dot x^k \rangle + \langle x^i \dot x^j x^k \rangle + \langle \dot x^i x^j x^k \rangle = 0,
	\end{equation}
	$\langle x^i \dot x^j \dot x^k \rangle$,
	\begin{equation}
		\widetilde{L}(x^i, \dot x^j, \dot x^k) := \frac 1 2 \left[{L(x^i \dot x^j, \dot x^k) - L(x^i \dot x^k, \dot x^j)}\right],
	\end{equation}
	$\langle \dot x^i \dot x^j \dot x^k \rangle$, and $L(\dot x^i \dot x^j, \dot x^k)$, where:
	\begin{equation}
		L(\dot x^i \dot x^j, \dot x^k) + L(\dot x^j \dot x^k, \dot x^i) + L(\dot x^k \dot x^i, \dot x^j) = 0.
	\end{equation}
	\footnote{It is shown in \cite{Bruckner2020} how to extract terms in the Langevin equation from trajectories. However, the time-step is too large relative to dynamics to justify a discrete-time approach, and it would be extremely troublesome to relate discrete-time measurements to continuous-time parameters. Thus, we opted not to do this at all.}In the case of almost-Markovian dynamics, the first four sets of quantities form one group, while the last two sets of quantities form another group \cite{low2023second}. For our system, in the first group, there are 389 independent quantities, of which 75 are purely real and 12 are purely imaginary. In the second group, there are 338 independent quantities, of which 72 are purely real and none are purely imaginary. We apply Holm\textendash Bonferroni correction independently to each group and evaluate quantitative significance according to \cite{low2023second}. As before, prior to statistical testing, we divide quantities for each cell by a factor related to the variances, e.g.\ $\langle \Delta s^0_i \Delta s^0_j \Delta s^0_k \rangle$ is divided by $\sqrt{\langle (\Delta s^0_i)^2 \rangle \langle (\Delta s^0_j)^2 \rangle \langle (\Delta s^0_k)^2 \rangle}$. As with the second-order quantities, we tested these for stationarity by using linear regression with time but found no statistically significant trends. We estimate the contribution of measurement error using the procedure described in \cite{Bruckner2020,low2023second} (see Appendix B). From the linear Gaussian model, it appears that for the time-step used here, the measurement error is overestimated. Besides, estimated measurement error for individual cells may not be reliable. Thus, we take an informal approach and compare the population-averaged estimate of the contribution of measurement error to the population-averaged estimate of the quantity in question.
	
	The statistically and quantitatively significant non-Gaussian effects are tabulated in Table \ref{tab:non-gaussian}. No chirality was detected. First, we note that the existence of non-Gaussian effects depends on the choice of variables. This is particularly relevant in our situation, where although we defined our shape variables based on linearization about a spherical shape, in the regime of actual shapes obtained the mapping from the deformation coefficients defining the distance function $r(\theta, \phi)$ to our shape variables is highly nonlinear (see Appendix C). We have attempted to roughly estimate the contribution of this nonlinearity to the non-Gaussian coefficients, assuming linear Gaussian dynamics for the deformation coefficients, orientational dynamics, and centroid motion. These are listed in the ``Est.''\ columns in Table \ref{tab:non-gaussian}, expressed as a fraction of the measured value. We see that for most quantities involving shape alone, there is a significant possibility of non-Gaussian effects being explained by the nonlinear shape mapping. However, for the quantities involving orientational dynamics or centroid motion, the estimated contribution is small and thus we think it is likely not due to the nonlinearity of shape coordinates. Thus, we focus our attention on these quantities.
	
	\begin{table}
		\begin{center}
			\caption{Statistically and quantitatively significant non-Gaussian effects. Parentheses indicate effects that are not statistically significant when the entire set of quantities is considered.}
			\label{tab:non-gaussian}
			\begin{tblr}{|Q[c,m]|Q[c,m]|Q[c,m]||Q[c,m]|Q[c,m]|Q[c,m]|}
				Quantity and sign & Adj.\ $p$-value & Est. & Quantity and sign & Adj.\ $p$-value & Est. \\ \hline
				$\Re \langle s^1_2 s^1_2 {s^2_2}^* \rangle > 0$ & $0.019$ & $0.82$ & $\Re \langle s^1_2 \dot \theta e^{i \phi} {{}\dot s^2_3}^* \rangle < 0$ & $0.024$ & $-0.25$ \\ \hline
				$\Re \langle s^1_2 s^1_3 {s^2_3}^*\rangle > 0$ & $7.5 \times 10^{-4}$ & $0.45$ & $\Re \langle s^1_2 v^1 {{}\dot s^2_2}^* \rangle < 0$ & $2.9 \times 10^{-3}$ & $0.10$ \\ \hline
				$\Re \langle s^1_3 s^1_3 {s^2_2}^* \rangle > 0$ & $7.5 \times 10^{-4}$ & $0.81$ & $\Re \langle s^1_3 \dot \theta e^{i \phi} {{}\dot s^2_3}^* \rangle < 0$ & $4.0 \times 10^{-4}$ & $0.09$ \\ \hline
				$\Re \langle \Delta s^0_2 \dot s^1_2 \dot \theta e^{-i \phi} \rangle < 0$ & $(0.029)$ & $0.02$ & $\Re \widetilde{L}(s^1_2, \dot s^1_2, {{}\dot s^2_3}^*) < 0$ & $0.022$ & $0.31$ \\ \hline
				$\Re \langle \Delta s^0_3 \dot s^1_3 \dot \theta e^{-i \phi} \rangle < 0$ & $(0.031)$ & $0.05$ & $\Re \langle \Delta s^0_2 \, \mathrm d [\dot s^1_2, {{}\dot s^1_3}^*]/\mathrm d t \rangle < 0$ & $(0.026)$ & $0.05$ \\ \hline
				$\Re \langle s^1_2 \dot s^0_2 \dot \theta e^{-i \phi} \rangle > 0$ & $2.8 \times 10^{-4}$ & $-0.07$ & $\Re \langle \Delta s^0_2 \, \mathrm d [\dot s^1_2, \dot \theta e^{-i \phi}]/\mathrm d t \rangle < 0$ & $2.1 \times 10^{-3}$ & $0.01$ \\ \hline
				$\Re \langle s^1_2 \dot s^0_2 {v^1}^* \rangle > 0$ & $5.0 \times 10^{-3}$ & $-0.02$ & $\Re \langle s^1_2 \, \mathrm d [\dot s^0_2, \dot \theta e^{-i \phi}]/\mathrm d t \rangle > 0$ & $8.8 \times 10^{-4}$ & $-0.04$ \\ \hline
				$\Re \langle s^1_3 \dot s^0_2 \dot \theta e^{-i \phi} \rangle > 0$ & $0.016$ & $0.16$ & $\Re \langle s^1_2 \, \mathrm d [\dot s^0_2, {v^1}^*]/\mathrm d t \rangle > 0$ & $6.4 \times 10^{-3}$ & $0.01$ \\ \hline
				$\Re \langle s^1_3 \dot s^0_2 {v^1}^* \rangle > 0$ & $0.017$ & $0.08$ & $\Re \langle s^2_2 \, \mathrm d [\dot s^0_2, {{}\dot s^2_2}^*]/\mathrm d t \rangle < 0$ & $(0.047)$ & $0.87$ \\ \hline
				$\Re \langle s^2_2 \dot s^0_2 {{}\dot s^2_2}^* \rangle < 0$ & $0.025$ & $1.07$ & $\Re \langle s^2_2 \, \mathrm d [\dot s^0_3, {{}\dot s^2_3}^*]/\mathrm d t \rangle < 0$ & $1.1 \times 10^{-3}$ & $1.07$ \\ \hline
				$\Re \langle s^2_2 \dot s^0_3 {{}\dot s^2_3}^* \rangle < 0$ & $2.9 \times 10^{-3}$ & $1.49$ & $\Re \langle s^2_3 \, \mathrm d [\dot s^0_2, {{}\dot s^2_2}^*]/\mathrm d t \rangle < 0$ & $0.022$ & $1.08$ \\ \hline
				$\Re \langle {s^2_2}^* \dot s^1_2 v^1 \rangle > 0$ & $4.0 \times 10^{-4}$ & $-0.08$ & $\Re \langle s^2_3 \, \mathrm d [\dot s^0_3, {{}\dot s^2_2}^*]/\mathrm d t \rangle < 0$ & $0.024$ & $0.43$ \\ \hline
				$\Re \langle {s^2_2}^* \dot s^1_3 \dot s^1_3 \rangle > 0$ & $3.0 \times 10^{-3}$ & $1.19$ & $\Re \langle {s^2_2}^* \, \mathrm d [\dot s^1_3, \dot s^1_3]/\mathrm d t \rangle > 0$ & $4.0 \times 10^{-3}$ & $0.79$ \\ \hline
				$\Re \langle {s^2_3}^* \dot s^1_2 \dot s^1_3 \rangle > 0$ & $0.012$ & $-0.18$ & $\Re \langle s^1_2 \, \mathrm d [\dot s^1_2, {{}\dot s^2_2}^*]/\mathrm d t \rangle > 0$ & $0.017$ & $1.05$ \\ \hline
				$\Re \langle {s^2_3}^* \dot s^1_2 \dot \theta e^{i \phi} \rangle > 0$ & $2.0 \times 10^{-3}$ & $-0.14$ & $\Re \langle s^1_2 \, \mathrm d [\dot \theta e^{i \phi}, {{}\dot s^2_2}^*]/\mathrm d t \rangle < 0$ & $9.3 \times 10^{-3}$ & $0.13$ \\ \hline
				$\Re \langle {s^2_3}^* \dot s^1_2 v^1 \rangle > 0$ & $2.7 \times 10^{-5}$ & $-0.00$ & $\Re \langle s^1_3 \, \mathrm d [\dot s^1_2, {{}\dot s^2_3}^*]/\mathrm d t \rangle > 0$ & $2.0 \times 10^{-4}$ & $0.37$ \\ \hline
				$\Re \langle s^1_2 \dot s^1_2 {{}\dot s^2_2}^* \rangle > 0$ & $(0.028)$ & $1.22$ & $\Re \langle (\dot \theta e^{i \phi})^2 {{}\dot s^2_2}^* \rangle < 0$ & $0.011$ & $0.24$ \\ \hline
				$\Re \langle s^1_2 \dot s^1_3 {{}\dot s^2_3}^* \rangle > 0$ & $6.6 \times 10^{-4}$ & $0.77$ & $\Re L(v^1 {{}\dot s^2_2}^*, \dot s^1_3) > 0$ & $(0.038)$ & $-0.07$ \\ \hline
				$\Re \langle s^1_2 \dot \theta e^{i \phi} {{}\dot s^2_2}^* \rangle < 0$ & $7.8 \times 10^{-5}$ & $0.19$ & & & \\ \hline
			\end{tblr}
		\end{center}
	\end{table}
	
	First, we have $\Re \langle \Delta s^0_2 \dot s^1_2 \dot \theta e^{-i \phi} \rangle, \Re \langle \Delta s^0_2 \, \mathrm d [\dot s^1_2, \dot \theta e^{-i \phi}]/\mathrm d t \rangle < 0$, where $[\cdot,\cdot]$ is the covariation defined by:
\begin{equation}
	[x,y](t=0)=0, \quad \frac{\mathrm d[x,y](t)}{\mathrm d t} = \lim_{\tau \to 0^+} \frac{(x(t+\tau)-x(t))(y(t+\tau)-y(t))}{\tau}.
\end{equation}
	This expresses a negative correlation between elongation and joint fluctuations of $\dot s^1_2$ and $\dot \theta e^{i \phi}$. This makes sense as a more elongated shape will be less affected by fluctuations of direction at the front. We also have (putatively) $\Re \langle \Delta s^0_3 \dot s^1_2 \dot \theta e^{-i \phi} \rangle < 0$, which expresses a positive correlation between widening and joint fluctuations of $\dot s^1_2$ and $\dot \theta e^{i \phi}$, which also makes sense. Next, we have quantities of the opposite sign: $\Re \langle s^1_2 \dot s^0_2 \dot \theta e^{-i \phi} \rangle,\Re \langle s^1_2 \, \mathrm d [\dot s^0_2, \dot \theta e^{-i \phi}]/\mathrm d t \rangle, \Re \langle s^1_2 \dot s^0_2 {v^1}^* \rangle, \Re \langle s^1_2 \, \mathrm d [\dot s^0_2, {v^1}^*]/\mathrm d t \rangle, \Re \langle s^1_3 \dot s^0_2 \dot \theta e^{-i \phi} \rangle, \Re \langle s^1_3 \dot s^0_2 {v^1}^* \rangle > 0$. These also make sense as a non-straight cell will change direction more if it elongates faster. It is notable that in the case of a martingale in Markovian dynamics, opposite signs are required for vanishing of the third-order angular momentum \cite{low2023second}. This can be intuitively understood as follows. Suppose we start with a low $s^0_2$ with $s^1_2 = 0$; we then have strongly correlated fluctuations of $\dot s^1_2$ and $\dot \theta e^{i \phi}$. Now, we have a value of $s^1_2$ positively correlated with the fluctuation of $\dot \theta e^{i \phi}$. If $\dot \theta e^{i \phi}$ were to continue in the same direction, this would imply an increase in $s^0_2$, meaning an effect in the opposite direction as the low $s^0_2$ we started with. However, it is possible that there are third-order time-antisymmetric quantities which are quantitatively significant but that we do not have the necessary statistics to resolve them.
	
	Next, we have another pair of sets of quantities with opposite sign: $\Re \langle {s^2_2}^* \dot s^1_2 v^1 \rangle, \Re \langle {s^2_3}^* \dot s^1_2 \dot \theta e^{i \phi} \rangle, \Re \langle {s^2_3}^* \dot s^1_2 v^1 \rangle > 0$, whereas $\Re \langle s^1_2 \dot \theta e^{i \phi} {{}\dot s^2_2}^* \rangle, \Re \langle s^1_2 \dot \theta e^{i \phi} {{}\dot s^2_3}^* \rangle, \Re \langle s^1_2 v^1 {{}\dot s^2_2}^* \rangle, \Re \langle s^1_3 \dot \theta e^{i \phi} {{}\dot s^2_3}^* \rangle, \Re \langle s^1_2 \, \mathrm d [\dot \theta e^{i \phi}, {{}\dot s^2_2}^*]/\mathrm d t \rangle < 0$. The first set of quantities says that joint fluctuations of shape and direction tend to occur along lateral elongation of the front of the cell. The second set of quantities is somewhat less intuitive, but can be understood as follows. If an $m=1$ cell deformation mode and the cell's direction of motion or change in orientation are aligned, then the front widens laterally perpendicular to the deformation mode. If they are anti-aligned, then the front widens laterally parallel to the deformation mode. Again, as in the previous case, the quantities of opposite signs make opposite contributions to the third-order angular momentum.
	
	Lastly, we have the two quantities from the second group of the partition mentioned at the beginning of this section. The first quantity, $\Re \langle (\dot \theta e^{i \phi})^2 {{}\dot s^2_2}^* \rangle < 0$, is time-antisymmetric and can be interpreted as a negative correlation between $\dot s^2_2$ and orientational changes. The second quantity, $\Re L(v^1 {{}\dot s^2_2}^*, \dot s^1_3) > 0$, is time-symmetric and can be interpreted as a positive correlation between the acceleration $\ddot s^1_3$ and $v^1 {{}\dot s^2_2}^*$, which is admittedly difficult to intuit.
	
	The signs of all the quantities listed in Table \ref{tab:non-gaussian} are consistent across all cells, except $\Re \langle s^1_2 s^1_2 {s^2_2}^* \rangle$ for one cell of normalized magnitude 0.11 times that of the mean, and $\Re \langle {s_2^2}^* \dot s^1_3 \dot s^1_3 \rangle$ for one cell of normalized magnitude 0.19 times that of the mean.
	
\begin{figure}
		\begin{center}
			\includegraphics[scale=0.5]{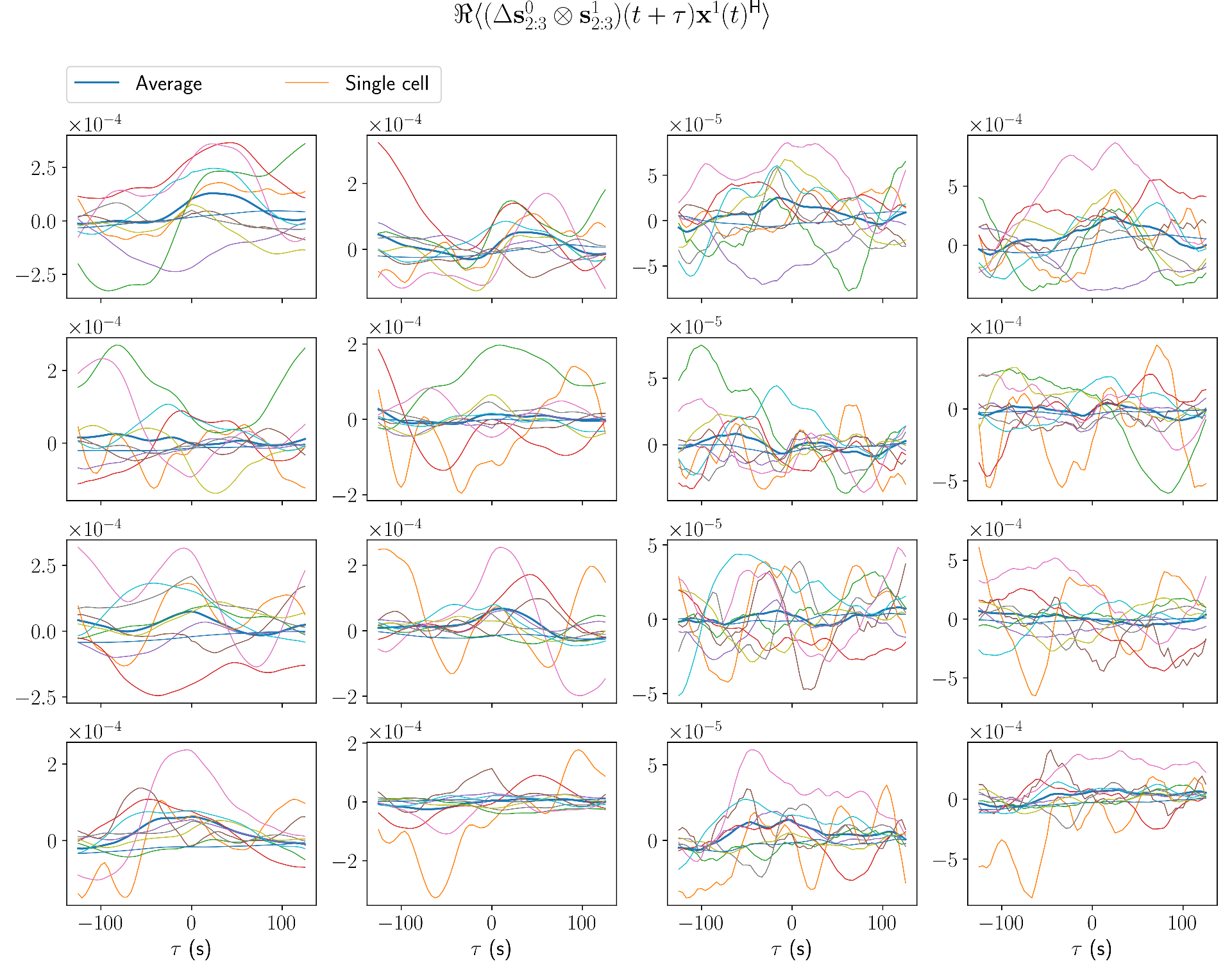}
		\end{center}
		\caption{Third-order covariance functions. Velocities are in units of \textmu m/fr.}
		\label{fig:plot-011}
	\end{figure}

\begin{figure}
		\begin{center}
			\includegraphics[scale=0.5]{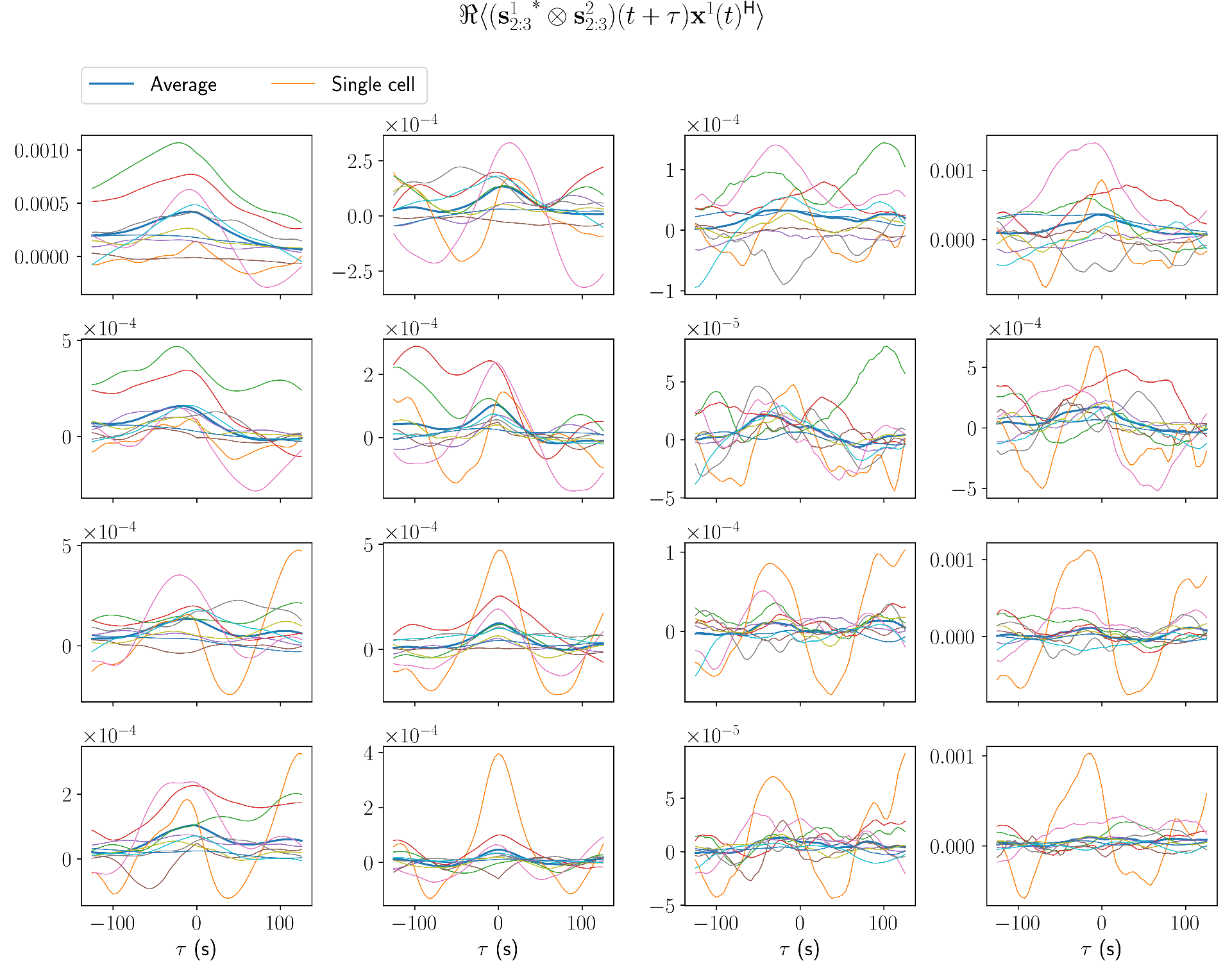}
		\end{center}
		\caption{Third-order covariance functions. Velocities are in units of \textmu m/fr.}
		\label{fig:plot-121}
	\end{figure}

	Ideally, the next step would be to investigate the third-order covariance functions and evaluate if they are consistent with a description based on an underdamped Langevin equation. However, given that there are hundreds of coefficients, such an evaluation would be difficult mathematically as well as visually with a sample size as small as ours. Nevertheless, we have included some third-order covariance functions for illustration (Figs.\ \ref{fig:plot-011} and \ref{fig:plot-121}).
	
	\section{Population variability}
	
	So far, we have looked at population averages. To investigate population variability, we need an estimate of the variability of time-averages for a stochastic process. We consider population variability of first- and second-order quantities that are invariant under a shift of azimuth. However, before we do so, we need to address the issue of unbiased estimation of the covariance matrix. We consider the Langevin equation Eq.\ \eqref{eq:linear-gaussian-langevin}, with state vector $\mathbf y := (\mathbf x^{\mathsf T}, \dot{\mathbf x}^{\mathsf T})^{\mathsf T}$. Consider a trajectory of length $T$, and let time-averages be denoted by an overline. We have:
	\begin{equation} \label{eq:adjust}
		\langle (\overline{\mathbf y}) (\overline{\mathbf y}^{\mathsf T}) \rangle = - \left[{\frac{\boldsymbol{\Gamma}^{-1}}{T} + \frac{\boldsymbol{\Gamma}^{-2}}{T^2}(\mathds{1} - e^{\boldsymbol{\Gamma}T})}\right] \mathbf C - \mathbf C \left[{\frac{\boldsymbol{\Gamma}^{-1}}{T} + \frac{\boldsymbol{\Gamma}^{-2}}{T^2}(\mathds{1} - e^{\boldsymbol{\Gamma}T})}\right]^{\mathsf T},
	\end{equation}
	where $\mathds 1$ is the identity matrix. For variables $y^i$ and $y^j$, at least one of whose mean is unknown (i.e., $s^0_2$, $s^0_3$, or $v^0$), the naively measured value of $C^{ij}$ is $\overline{y^i y^j} - \overline{y^i} \cdot \overline{y^j}$, whose expectation is (the true) $C^{ij}$ minus the $(i,j)$ component of Eq.\ \eqref{eq:adjust}. We use this and similar equalities to estimate dynamics accounting for the finite length of trajectories.
	
	Now, for any two stationary quantities $w$ and $z$, we have:
	\begin{equation} \label{eq:cov-pop}
		\begin{aligned}
			\operatorname{Cov}(\overline w, \overline z) &= \frac{1}{T^2} \int_0^T \mathrm d t \int_0^T \mathrm d t' \, \langle (w(t) - \langle w \rangle) (z(t) - \langle z \rangle) \rangle \\
			&\approx \frac{1}{T} \int_{-\infty}^{+\infty} \mathrm d \tau \, \langle (w(t) - \langle w \rangle) (z(t+\tau) - \langle z \rangle) \rangle.
		\end{aligned}
	\end{equation}
	For an unbiased estimate of population covariance, we need to subtract this quantity from the observed population covariance. However, if we want to estimate this quantity from data, we must truncate the integral \cite{low2023second}. We are thus led to consider experimental quantities of the form:
	\begin{equation} \label{eq:exp-int}
		\frac{1}{T} \int_{\substack{0 \le t < T \\ 0 \le t' < T \\ |t'-t|<S}} \mathrm d t \, \mathrm d t' \, (w(t) - \overline w) (z(t') - \overline z) = - \frac{1}{T} \int_{\substack{0 \le t < T \\ 0 \le t' < T \\ |t'-t| \ge S}} \mathrm d t \, \mathrm d t' \, (w(t) - \overline w) (z(t') - \overline z)
	\end{equation}
	where $0<S<T$. (If $w$ and $z$ are known to have zero mean, then $\overline w$ and $\overline z$ are not included in the integrands and instead $T \overline w \cdot \overline z$ is added to the r.h.s.) We chose $S = 100 \textrm{ s}$. Our procedure is to calculate the theoretical (Gaussian) and experimental values for Eq.\ \eqref{eq:exp-int} and use the correction to adjust the theoretically calculated value for Eq.\ \eqref{eq:cov-pop}. In our calculations, we neglected terms of the form $e^{\boldsymbol{\Gamma}(T-S)}$ in quantities not involving $s^0_0$. The following relation is useful in combining terms:
	\begin{equation}
		\langle \overline{y^i} y^j(t) \rangle = \langle y^i(T-t) \overline{y^j} \rangle.
	\end{equation}
	In estimating the corrections, similarly to the previous sections, we normalized values according to $\mathbf C$ and $\mathbf D$ before averaging over cells. The deviations from theoretical values also provides a test for temporal heterogeneity \cite{low2023second}; nothing statistically significant was found. For any pair of quantities, the estimated covariance matrix due to within-trajectory stochasticity (Eq.\ \eqref{eq:cov-pop}) is positive definite. To estimate population variances, if necessary we first adjusted each cell's value by the expected mean, and then calculated a $\chi^2$ value according to Eq.\ \eqref{eq:chi-sq}. We use this value to calculate statistical significance as well as to obtain an unbiased estimate of population variance. We evaluate a ``reference'' variance value prescribed in the same way as deviations between theory and experiment \cite{low2023second}, and we consider population variance to be quantitatively significant if the unbiased estimate is more than 0.09 (the square of 0.3, the ratio proposed in \cite{low2023second}) times this value. The results for first-order quantities are given in Table \ref{tab:pop-var-1} and the results for second-order quantities are given in Table \ref{tab:pop-var-2}. In the latter case, we estimated an adjustment for statistical significance due to non-Gaussianity of the quantities (see Appendix D). We found significant population variability for many quantities; however, as with population averages, no population variability of chirality was detected.

	\begin{table}
		\begin{center}
			\caption{Population variability of first-order quantities. The ratios of estimated population variance to reference values are indicated.}
			\label{tab:pop-var-1}
			\begin{tblr}{|Q[c,m]|Q[c,m]|Q[c,m]|}
				Quantity & Ratio & $p$-value \\ \hline
				$\langle s^0_0 \rangle$ & 3.62 & $< 10^{-8}$ \\ \hline
				$\langle s^0_2 \rangle$ & 1.85 & $< 10^{-8}$ \\ \hline
				$\langle s^0_3 \rangle$ & 0.61 & $1.7 \times 10^{-6}$ \\ \hline
				$\langle v^0 \rangle$ & 0.25 & $< 10^{-8}$ \\ \hline
			\end{tblr}
		\end{center}
	\end{table}
	
	\begin{table}
		\begin{center}
			\caption{Statistically significant population variability of second-order quantities, by $m$ value. The ratios of estimated population variance to reference values are indicated. The adjusted $p$-value is computed using the $\chi^2$ distribution. Parentheses indicate lack of statistical significance when all $m$ values are considered together using the $\chi^2$ distribution. Daggers indicate estimated statistical significance when accounting for non-normality with all $m$ values considered together.}
			\label{tab:pop-var-2}
			\begin{tblr}{|Q[c,m]|Q[c,m]|Q[c,m]||Q[c,m]|Q[c,m]|Q[c,m]|}
				Quantity & Ratio & Adj.\ $p$-value & Quantity & Ratio & Adj.\ $p$-value \\ \hline
				$\langle (\Delta s^0_2)^2 \rangle$ & 0.31 & $4.1 \times 10^{-3}$ & $\Re \langle \dot s^1_3 {v^1}^* \rangle$ & 0.10 & ${7.1 \times 10^{-3}}^\dagger$ \\ \hline
				$\langle \Delta s^0_2 \Delta s^0_3 \rangle$ & 0.25 & $(0.030)$ & $\langle \dot \theta^2 \rangle$ & 0.62 & ${{} < 10^{-8}}^\dagger$ \\ \hline
				$\langle (\Delta s^0_3)^2 \rangle$ & 0.27 & $(0.016)$ & $\Re \langle \dot \theta e^{i \phi} {v^1}^* \rangle$ & 0.30 & ${{} < 10^{-8}}^\dagger$ \\ \hline
				$L(\Delta s^0_2, \Delta s^0_3)$ & 0.24 & ${1.6 \times 10^{-3}}^\dagger$ & $\langle v^1 {v^1}^* \rangle$ & 0.34 & ${{} < 10^{-8}}^\dagger$ \\ \hline
				$\langle \Delta s^0_2 \Delta v^0 \rangle$ & 1.43 & ${1.5 \times 10^{-8}}^\dagger$ & $\Re L(\dot s^1_2, {{}\dot s^1_3}^*)$ & 0.19 & ${6.0 \times 10^{-3}}^\dagger$ \\ \hline
				$\langle \Delta s^0_3 \Delta v^0 \rangle$ & 0.59 & $(0.023)$ & $\Re L(\dot s^1_2, {v^1}^*)$ & 0.10 & $0.015^\dagger$ \\ \hline
				$\langle (\dot s^0_2)^2 \rangle$ & 0.40 & ${{} < 10^{-8}}^\dagger$ & $D^1_{22}$ & 0.36 & ${1.4 \times 10^{-8}}^\dagger$ \\ \hline
				$\langle \dot s^0_2 \dot s^0_3 \rangle$ & 0.09 & ${4.7 \times 10^{-3}}^\dagger$ & $\Re D^1_{23}$ & 0.19 & ${6.7 \times 10^{-3}}^\dagger$ \\ \hline
				$\langle (\dot s^0_3)^2 \rangle$ & 0.42 & ${{} < 10^{-8}}^\dagger$ & $\Re D^1_{2 \theta}$ & 0.23 & ${2.8 \times 10^{-3}}^\dagger$ \\ \hline
				$\langle \dot s^0_3 \Delta v^0 \rangle$ & 0.22 & ${{} < 10^{-8}}^\dagger$ & $D^1_{33}$ & 0.43 & ${1.5 \times 10^{-5}}^\dagger$ \\ \hline
				$\langle (\Delta v^0)^2 \rangle$ & 0.52 & ${{} < 10^{-8}}^\dagger$ & $D^1_{\theta\theta}$ & 0.59 & ${{} < 10^{-8}}^\dagger$ \\ \hline
				$L(\dot s^0_2, \dot s^0_3)$ & 0.26 & ${{} < 10^{-8}}^\dagger$ & $\Re D^1_{\theta v}$ & 0.13 & ${3.0 \times 10^{-3}}^\dagger$ \\ \hline
				$L(\dot s^0_2, \Delta v^0)$ & 0.16 & ${1.7 \times 10^{-6}}^\dagger$ & $D^1_{vv}$ & 0.25 & ${{} < 10^{-8}}^\dagger$ \\ \hline
				$D^0_{22}$ & 0.50 & ${{} < 10^{-8}}^\dagger$ & $\langle s^2_2 {s^2_2}^* \rangle$ & 0.33 & $(0.041)$ \\ \hline
				$D^0_{33}$ & 0.44 & ${{} < 10^{-8}}^\dagger$ & $\Re \langle s^2_2 {s^2_3}^* \rangle$ & 0.30 & $(0.036)$ \\ \hline
				$D^0_{3v}$ & 0.06 & ${3.1 \times 10^{-3}}^\dagger$ & $\langle s^2_3 {s^2_3}^* \rangle$ & 0.86 & ${{} < 10^{-8}}^\dagger$ \\ \hline
				$D^0_{vv}$ & 0.22 & ${2.5 \times 10^{-5}}^\dagger$ & $\Re L(s^2_2, {s^2_3}^*)$ & 0.19 & $(0.037)$ \\ \hline
				$\langle s^1_3 {s^1_3}^* \rangle$ & 0.75 & ${{} < 10^{-8}}^\dagger$ & $\langle \dot s^2_2 {{}\dot s^2_2}^* \rangle$ & 0.53 & ${{} < 10^{-8}}^\dagger$ \\ \hline
				$\Re L(s^1_2, {s^1_3}^*)$ & 1.49 & ${{} < 10^{-8}}^\dagger$ & $\Re \langle \dot s^2_2 {{}\dot s^2_3}^* \rangle$ & 0.20 & ${2.6 \times 10^{-6}}^\dagger$ \\ \hline
				$\Re \langle s^1_2 {v^1}^* \rangle$ & 3.24 & $(0.040)$ & $\langle \dot s^2_3 {{}\dot s^2_3}^* \rangle$ & 0.54 & ${{} < 10^{-8}}^\dagger$ \\ \hline
				$\Re \langle s^1_3 {v^1}^* \rangle$ & 0.60 & $0.027$ & $\Re L(\dot s^2_2, {{}\dot s^2_3}^*)$ & 0.20 & ${6.2 \times 10^{-5}}^\dagger$ \\ \hline
				$\langle \dot s^1_2 {{}\dot s^1_2}^* \rangle$ & 0.34 & ${{} < 10^{-8}}^\dagger$ & $D^2_{22}$ & 0.95 & ${{} < 10^{-8}}^\dagger$ \\ \hline
				$\Re \langle \dot s^1_2 {{}\dot s^1_3}^* \rangle$ & 0.11 & ${7.3 \times 10^{-3}}^\dagger$ & $\Re D^2_{23}$ & 0.31 & ${2.7 \times 10^{-7}}^\dagger$ \\ \hline
				$\Re \langle \dot s^1_2 {v^1}^* \rangle$ & 0.05 & $(0.038)$ & $D^2_{33}$ & 0.77 & ${{} < 10^{-8}}^\dagger$ \\ \hline
				$\langle \dot s^1_3 {{}\dot s^1_3}^* \rangle$ & 0.45 & ${{} < 10^{-8}}^\dagger$ & & & \\ \hline
			\end{tblr}
		\end{center}
	\end{table}
	
	Next, we investigated population covariance of quantities whose population variance was statistically significant. Attaining statistical significance to demonstrate non-zero correlations was infeasible because of the small population size, so we instead considered the converse problem where the null hypothesis tested is that of perfect linear correlation between two quantities. For every pair of quantities, we simulated $5 \times 10^6$ trials of joint Gaussian variables having the estimated covariance due to within-trajectory stochasticity alone (Eq.\ \eqref{eq:cov-pop}) with no population variability, and evaluated the differences between the experimentally measured population covariance matrix and the simulated ones. A result that was not positive definite was considered null. We later estimated adjustments of statistical significance due to non-Gaussianity (see Appendix D). For evaluating quantitative significance, we used for every pair of quantities a ``reference'' ($2 \times 2$) covariance matrix $\mathbf M$ and an inner product defined by:
	\begin{equation}
		\langle \mathbf x, \mathbf y \rangle_{\mathbf M} := \mathbf x^{\mathsf T} \mathbf M^{-1} \mathbf y = (\mathbf W \mathbf x)^{\mathsf T} (\mathbf W \mathbf y), \quad \mathbf W^{\mathsf T} \mathbf W = \mathbf M^{-1}.
	\end{equation}
	(If $\mathbf M$ were a covariance matrix, $\mathbf W$ would be a whitening matrix.) To do calculations with respect to this inner product, we work with transformed variables $\mathbf W \mathbf x$ of the original variables $\mathbf x$. A matrix $\mathbf K$ representing a covariance matrix or a linear combination of covariance matrices in the original coordinates becomes $\mathbf W \mathbf K \mathbf W^{\mathsf T}$ in transformed coordinates. Its eigenvalues are equal to those of $\mathbf K \mathbf W^{\mathsf T} \mathbf W = \mathbf K \mathbf M^{-1}$ \cite{MatrixAnalysis}. The signs of its eigenvalues are independent of $\mathbf M$ as these correspond to the possible signs of $(\mathbf W^{\mathsf T} \mathbf x)^{\mathsf T} \mathbf K (\mathbf W^{\mathsf T} \mathbf x)$ for $\mathbf x \ne \mathbf 0$ ($\mathbf W$ is square and invertible). For the unbiased estimate of population covariance $\mathbf K$, we computed the lesser and greater eigenvalues, $\kappa_<$ and $\kappa_>$ respectively, of $\mathbf K \mathbf M^{-1}$. A result of $\kappa_</\max(\kappa_>,1) > 0.09$ was considered quantitatively significant. Results involving first-order quantities (Tables \ref{tab:pairs0}\textendash \ref{tab:pairs1-4}) were in line with what might be expected from the results for population variability of single quantities. Pairs of second-order quantities had similar fractions of quantitatively significant results regardless of the $m$ values (statistically significant results reported in Tables \ref{tab:pairs2-1}\textendash \ref{tab:pairs2-3}).
	
	\section{Conclusions}
	We have elucidated the ``laws of motion'' obeyed by T cell morphodynamics and migration in 3D collagen matrices. We have corrected previous understanding which incorrectly posited discrete structure of the probability distribution of dynamics and periodic oscillations of shape. We have introduced a new method of 3D shape description relative to the polarization axis, preserving different possible modes of shape variation, and have described dynamics using an underdamped Langevin equation. This approach reveals different correlation times for different modes: the $m=0$ modes have correlation times approximately 60 s, the $(l,m)=(2,1)$ and $(l,m)=(2,2)$ modes approximately 90 s, and the $(l,m)=(3,1)$ and $(l,m)=(3,2)$ modes approximately 30 s. In addition, we have quantified time-irreversibility, which has been qualitatively observed in previous studies. Furthermore, we have extracted novel coefficients describing non-Gaussianity and have found patterns in the signs of the coefficients. Also, we did not find statistically significant third-order time-antisymmetric quantities. However, our analysis is limited by small sample size. Still, we have been able to determine the presence of some non-Gaussian effects by normalizing values when calculating statistics.
	
	We also addressed the possibility of temporal and population heterogeneity. Such effects have been found in previous cell migration studies \cite{MetznerSuperstatisticalRandomWalks2015,3DCellMigrationNotRandomWalk2014,StochasticHeterogeneousCellMigration2019,MitterwallnerNonMarkovianMotility2020}. In our study, we found substantial population heterogeneity but no evidence of temporal heterogeneity. Furthermore, we did not find any chiral effects, either on average or varying from cell to cell.
	
	While mechanistic insights likely cannot be gleaned at this stage, we have characterized biology at the level of emergent behavior, which is an important step in understanding a complex system such as cell migration. Future work could involve analyzing larger datasets and comparing different biological conditions such as addition of chemokine \cite{AmselemDictyosteliumChemotaxis2012}. In particular, multiple biological replicates would be desired to confirm the findings.
		
	\section{Appendix A}
	In this appendix, we describe the procedure for converting discrete-time to continuous-time, and give the values of the parameters of the linear Gaussian model. We write the linear Gaussian underdamped Langevin equation as:
	\begin{equation} \label{eq:linear-gaussian-langevin}
		\begin{pmatrix}
			\dot{\mathbf x} \\
			\ddot{\mathbf x}
		\end{pmatrix} = \boldsymbol{\Gamma} \begin{pmatrix}
		\mathbf x \\
		\dot{\mathbf x}
		\end{pmatrix} + \begin{pmatrix}
		\mathbf 0 \\
		\boldsymbol{\xi}
		\end{pmatrix}, \quad \langle \boldsymbol{\xi}(t) \boldsymbol{\xi}(t')^{\mathsf T} \rangle = 2 \mathbf D \delta(t - t')
	\end{equation}
	(using Hermitian conjugate as appropriate for complex variables), where only variables possessing a stationary distribution are included in the state vector $(\mathbf x^{\mathsf T}, \dot{\mathbf x}^{\mathsf T})^{\mathsf T}$, so that $\boldsymbol{\Gamma}$ only has eigenvalues with strictly negative real part. We introduce the covariance matrix:
	\begin{equation}
		\mathbf C := \begin{pmatrix}
			\langle \mathbf x \mathbf x^{\mathsf T} \rangle & \langle \mathbf x {{}\dot{\mathbf x}}^{\mathsf T} \rangle \\
			\langle \dot{\mathbf x} \mathbf x^{\mathsf T} \rangle & \langle \dot{\mathbf x} {{}\dot{\mathbf x}}^{\mathsf T} \rangle
		\end{pmatrix},
	\end{equation}
	which can be determined from $\boldsymbol{\Gamma}$ and $\mathbf D$ using the Lyapunov equation \cite{low2023second}. We have:
	\begin{align}
		\int_0^{\tau} \mathrm d t \, \langle \dot x^i(t) x^j(0) \rangle &= [\boldsymbol{\Gamma}^{-1} (e^{\boldsymbol{\Gamma}\tau} - \mathds 1) \mathbf C]^{\dot x^i x^j}, \\
		\int_0^{\tau} \mathrm d t \, \langle x^i(t) \dot x^j(0) \rangle &= [\boldsymbol{\Gamma}^{-1} (e^{\boldsymbol{\Gamma}\tau} - \mathds 1) \mathbf C]^{x^i \dot x^j}, \\
		\int_0^{\tau} \mathrm d t \int_0^{\tau} \mathrm d t' \, \langle \dot x^i(t) \dot x^j(t') \rangle &= [\boldsymbol{\Gamma}^{-2} (e^{\boldsymbol{\Gamma}\tau} - \mathds 1 - \boldsymbol{\Gamma} \tau) \mathbf C]^{\dot x^i \dot x^j} + [\boldsymbol{\Gamma}^{-2} (e^{\boldsymbol{\Gamma}\tau} - \mathds 1 - \boldsymbol{\Gamma} \tau) \mathbf C]^{\dot x^j \dot x^i}, \\
		\int_{n\tau}^{(n+1)\tau} \mathrm d t \int_0^{\tau} \mathrm d t' \, \langle \dot x^i(t) \dot x^j(t') \rangle &= [\boldsymbol{\Gamma}^{-2} (e^{\boldsymbol{\Gamma}\tau} - \mathds 1)^2 e^{(n-1) \boldsymbol{\Gamma}\tau} \mathbf C]^{\dot x^i \dot x^j}, \quad n \ge 1,
	\end{align}
	where $\mathds 1$ is the identity matrix.
	
	For $m=0$, we have for the linear Gaussian model, after taking into account finite trajectory length:
	\begin{align}
		\mathbf C^0 &= \begin{pmatrix}
			4.6 \times 10^{-3} & -8.3 \times 10^{-4} & 0 & -1.4 \times 10^{-4} & 3.4 \times 10^{-3} \\
			-8.3 \times 10^{-4} & 2.6 \times 10^{-3} & 1.4 \times 10^{-4} & 0 & -1.4 \times 10^{-3} \\
			0 & 1.4 \times 10^{-4} & 1.8 \times 10^{-4} & -1.2 \times 10^{-5} & -6.4 \times 10^{-4} \\
			-1.4 \times 10^{-4} & 0 & -1.2 \times 10^{-5} & 1.2 \times 10^{-4} & -3.4 \times 10^{-5} \\
			3.4 \times 10^{-3} & -1.4 \times 10^{-3} & -6.4 \times 10^{-4} & -3.4 \times 10^{-5} & 4.9 \times 10^{-2}
		\end{pmatrix} \\
		\mathbf A_{\mathbf x}^0 &= \begin{pmatrix}
			-0.021 & 0.020 \\
			-0.021 & -0.054 \\
			0.36 & -0.20
		\end{pmatrix} \\
		\mathbf A_{\mathbf v}^0 &= \begin{pmatrix}
			-0.51 & 0.27 & -9.5 \times 10^{-3} \\
			-0.05 & -0.63 & -6.1 \times 10^{-3} \\
			0.77 & -4.80 & -0.54
		\end{pmatrix} \\
		\mathbf D^0 &= \begin{pmatrix}
			8.7 \times 10^{-5} & -1.8 \times 10^{-5} & -1.3 \times 10^{-4} \\
			-1.8 \times 10^{-5} & 7.2 \times 10^{-5} & 4.3 \times 10^{-4} \\
			-1.3 \times 10^{-4} & 4.3 \times 10^{-4} & 2.5 \times 10^{-2}
		\end{pmatrix}
	\end{align}
	where time is measured in units of frames and velocities are measured in \textmu m/fr. We have the eigendecomposition:
	\begin{align}
		\boldsymbol{\Gamma}^0 &= \mathbf S^0 \boldsymbol{\Lambda}^0 (\mathbf S^0)^{-1} \\
		\boldsymbol{\Lambda}^0 &= \operatorname{diag}(-0.063, -0.22+0.06i, c.c., -0.45, -0.71) \\
		\mathbf S^0 &= \begin{pmatrix}
			0.64 & 0.40+0.04i & c.c. & -0.87 & -0.00 \\
			-0.72 & 0.20+0.09i & c.c. & -0.01 & -0.05 \\
			-0.040 & -0.091+0.013i & c.c. & 0.389 & +0.000 \\
			0.045 & -0.051-0.009i & c.c. & 0.004 & 0.039 \\
			0.26 & 0.88 & c.c. & -0.31 & 1.00
		\end{pmatrix}
	\end{align}
	where c.c.\ denotes complex conjugate of the previous column. For $m=1$, we have:
	\begin{align}
		\mathbf C^1 &= \begin{pmatrix}
			8.3 \times 10^{-3} & 9.9 \times 10^{-4} & 0 & -1.8 \times 10^{-4} & 1.2 \times 10^{-3} & 9.3 \times 10^{-3} \\
			9.9 \times 10^{-4} & 3.0 \times 10^{-3} & 1.8 \times 10^{-4} & 0 & -1.8 \times 10^{-4} & 9.8 \times 10^{-4} \\
			0 & 1.8 \times 10^{-4} & 2.7 \times 10^{-4} & -3.5 \times 10^{-5} & -2.2 \times 10^{-4} & -4.9 \times 10^{-4} \\
			-1.8 \times 10^{-4} & 0 & -3.5 \times 10^{-5} & 2.8 \times 10^{-4} & 2.3 \times 10^{-4} & 7.0 \times 10^{-4} \\
			1.2 \times 10^{-3} & -1.8 \times 10^{-4} & -2.2 \times 10^{-4} & 2.3 \times 10^{-4} & 1.6 \times 10^{-3} & 6.8 \times 10^{-3} \\
			9.3 \times 10^{-3} & 9.8 \times 10^{-4} & -4.9 \times 10^{-4} & 7.0 \times 10^{-4} & 6.8 \times 10^{-3} & 0.087
		\end{pmatrix} \\
		\mathbf A_{\mathbf x}^1 &= \begin{pmatrix}
			-0.070 & 0.105 \\
			0.039 & -0.143 \\
			0.18 & -0.28 \\
			1.13 & -0.97
		\end{pmatrix} \\
		\mathbf A_{\mathbf v}^1 &= \begin{pmatrix}
			-1.33 & 0.85 & -1.9 \times 10^{-3} & 0.040 \\
			0.71 & -1.19 & -0.048 & -0.033 \\
			1.7 & -1.6 & -0.81 & -0.034 \\
			14 & -11 & 1.4 & -1.3
		\end{pmatrix} \\
		\mathbf D^1 &= \begin{pmatrix}
			3.9 \times 10^{-4} & -2.8 \times 10^{-4} & -6.6 \times 10^{-4} & -4.2 \times 10^{-3} \\
			-2.8 \times 10^{-4} & 3.9 \times 10^{-4} & 7.0 \times 10^{-4} & 4.2 \times 10^{-3} \\
			-6.6 \times 10^{-4} & 7.0 \times 10^{-4} & 2.0 \times 10^{-3} & 9.7 \times 10^{-3} \\
			-4.2 \times 10^{-3} & 4.2 \times 10^{-3} & 9.7 \times 10^{-3} & 0.104
		\end{pmatrix} \\
		\boldsymbol{\Gamma}^1 &= \mathbf S^1 \boldsymbol{\Lambda}^1 (\mathbf S^1)^{-1} \\
		\boldsymbol{\Lambda}^1 &= \operatorname{diag}(-0.049, -0.15, -0.35, -0.44, -0.99, -2.60) \\
		\mathbf S^1 &= \begin{pmatrix}
			0.78 & -0.042 & -0.37 & -0.21 & -0.026 & 0.021 \\
			-0.25 & -0.87 & -0.65 & -0.10 & 0.022 & -0.044 \\
			-0.039 & 0.007 & 0.13 & 0.091 & 0.067 & -0.020 \\
			0.012 & 0.13 & 0.22 & 0.046 & -0.058 & 0.044 \\
			0.14 & 0.080 & -0.001 & 0.12 & -0.13 & 0.32 \\
			0.55 & -0.46 & -0.61 & 0.96 & -0.99 & -0.95
		\end{pmatrix}.
	\end{align}
	For $m=2$, we have:
	\begin{align}
		\mathbf C^2 &= \begin{pmatrix}
			5.3 \times 10^{-3} & 2.1 \times 10^{-3} & 0 & -8.4 \times 10^{-5} \\
			2.1 \times 10^{-3} & 2.0 \times 10^{-3} & 8.4 \times 10^{-5} & 0 \\
			0 & 8.4 \times 10^{-5} & 1.1 \times 10^{-4} & 5.7 \times 10^{-5} \\
			-8.4 \times 10^{-5} & 0 & 5.7 \times 10^{-5} & 1.1 \times 10^{-4}
		\end{pmatrix} \\
		\mathbf A_{\mathbf x}^2 &= \begin{pmatrix}
			-0.029 & 0.030 \\
			0.006 & -0.062
		\end{pmatrix} \\
		\mathbf A_{\mathbf v}^2 &= \begin{pmatrix}
			-0.67 & 0.24 \\
			-0.01 & -0.50
		\end{pmatrix} \\
		\mathbf D^2 &= \begin{pmatrix}
			5.8 \times 10^{-5} & 2.2 \times 10^{-5} \\
			2.2 \times 10^{-5} & 5.5 \times 10^{-5}
		\end{pmatrix} \\
		\boldsymbol{\Gamma}^2 &= \mathbf S^2 \boldsymbol{\Lambda}^2 (\mathbf S^2)^{-1} \\
		\boldsymbol{\Lambda}^2 &= \operatorname{diag}(-0.042, -0.20, -0.31, -0.61) \\
		\mathbf S^2 &= \begin{pmatrix}
			0.99 & -0.27 & 0.46 & 0.85 \\
			0.14 & -0.94 & 0.84 & 0.06 \\
			-0.042 & 0.055 & -0.14 & -0.52 \\
			-0.006 & 0.19 & -0.26 & -0.036
		\end{pmatrix}.
	\end{align}

	\section{Appendix B}
	In this appendix, we calculate the contribution of measurement error to the estimation of $L(\dot x^i \dot x^j, \dot x^k)$. Let $\mathbf y := \mathbf x + \boldsymbol{\eta}$ denote the measured values, with $\boldsymbol{\eta}$ the measurement error. We only consider second moments of $\boldsymbol{\eta}$, denoted by $\boldsymbol{\Lambda}$, and may be state-dependent. Following \cite{Bruckner2020}, we make the assumption $\boldsymbol{\Lambda} \ll \langle \dot{\mathbf x} {{}\dot{\mathbf x}}^{\mathsf T} \rangle (\Delta t)^2$, where $\Delta t$ is the time-step, necessary for validity of the estimation procedure. We have verified that this is the case for our dataset. We define $\widetilde{\boldsymbol{\Lambda}} := \boldsymbol{\Lambda} (\Delta t)^{-2}$ so that $\widetilde{\boldsymbol{\Lambda}} = \mathcal O((\Delta t)^0)$. The estimate is given by:
	\begin{equation}
		\begin{aligned}
			L(\dot x^i \dot x^j, \dot x^k) &\approx (\Delta t)^{-4} \langle (y^i(t+\Delta t)-y^i(t))(y^j(t+\Delta t)-y^j(t))(y^k(t+2\Delta t)-y^k(t+\Delta t)) \\
			&\qquad{}-(y^i(t+2\Delta t)-y^i(t+\Delta t))(y^j(t+2\Delta t)-y^j(t+\Delta t))(y^k(t+\Delta t)-y^k(t)) \rangle.
		\end{aligned}
	\end{equation}
	The contribution of measurement error is seen to be:
	\begin{equation} \label{eq:Lvvv}
		\begin{aligned}
			&(\Delta t)^{-2} \langle (\widetilde{\Lambda}^{ij} (t+\Delta t) + \widetilde{\Lambda}^{ij}(t))(x^k(t+2\Delta t)-x^k(t+\Delta t)) \\
			&\qquad{} - (\widetilde{\Lambda}^{ij} (t+2\Delta t) + \widetilde{\Lambda}^{ij}(t+\Delta t))(x^k(t+\Delta t)-x^k(t)) \\
			&\qquad{} - \widetilde{\Lambda}^{ik} (t+\Delta t) (x^j(t+\Delta t)-x^j(t)) - \widetilde{\Lambda}^{jk}(t+\Delta t) (x^i(t+\Delta t)-x^i(t)) \\
			&\qquad{} + \widetilde{\Lambda}^{ik}(t+\Delta t) (x^j(t+2\Delta t)-x^j(t+\Delta t)) + \widetilde{\Lambda}^{jk}(t+\Delta t) (x^i(t+2\Delta t)-x^i(t+\Delta t)) \rangle.
		\end{aligned}
	\end{equation}
	Now, we have:
	\begin{equation}
		(\Delta t)^{-2} \langle \widetilde{\Lambda}^{ij} (t+\Delta t) (x^k(t+2\Delta t)-2 x^k(t+\Delta t) + x^k(t)) \rangle = \langle \widetilde{\Lambda}^{ij} \circ \ddot x^k \rangle + \mathcal O(\Delta t),
	\end{equation}
	where $\circ$ denotes Stratonovich convention. The last two lines of Eq.\ \eqref{eq:Lvvv} therefore contribute:
	\begin{equation}
		\langle \widetilde{\Lambda}^{ik} \circ \ddot x^j \rangle + \langle \widetilde{\Lambda}^{jk} \circ \ddot x^i \rangle + \mathcal O(\Delta t).
	\end{equation}
	The remaining terms can be written:
	\begin{equation} \label{eq:Lvvv2}
		\begin{aligned}
			&(\Delta t)^{-2} \langle (\widetilde{\Lambda}^{ij}(t) + \widetilde{\Lambda}^{ij}(t+2\Delta t)) (x^k(t+2\Delta t)-2x^k(t+\Delta t)+x^k(t)) \\
			&\qquad{} + \widetilde{\Lambda}^{ij}(t) (x^k(t+\Delta t)-x^k(t)) - \widetilde{\Lambda}^{ij}(t+2 \Delta t) (x^k(t+2\Delta t)-x^k(t+\Delta t)) \rangle.
		\end{aligned}
	\end{equation}
	The second line of Eq.\ \eqref{eq:Lvvv2} evaluates to $\langle \widetilde{\Lambda}^{ij} \circ \ddot x^k \rangle + \mathcal O(\Delta t)$. The first line of Eq.\ \eqref{eq:Lvvv2} can be evaluated by using an It\^o\textendash Taylor expansion \cite{Bruckner2020}, resulting in:
	\begin{equation}
		(\Delta t)^{-2} \langle (\widetilde{\Lambda}^{ij}(t) + \widetilde{\Lambda}^{ij}(t+2\Delta t)) (x^k(t+2\Delta t)-2x^k(t+\Delta t)+x^k(t)) \rangle = 2 \langle \widetilde{\Lambda}^{ij} \circ \ddot x^k \rangle + \mathcal O(\Delta t).
	\end{equation}
	The final result is therefore:
	\begin{equation}
		\langle 4 \widetilde{\Lambda}^{ij} \circ \ddot x^k + \widetilde{\Lambda}^{ik} \circ \ddot x^j + \widetilde{\Lambda}^{jk} \circ \ddot x^i \rangle + \mathcal O(\Delta t).
	\end{equation}
	Finally, we need to estimate $\langle \widetilde{\Lambda}^{ij} \circ \ddot x^k \rangle$ from time-lapse data. However, we cannot do this by using the estimator for $\Lambda$ described in \cite{Bruckner2020,low2023second} together with the direct estimator for $\ddot x^k$, because this will introduce terms involving $\Lambda^{ik}$ and $\Lambda^{jk}$. Instead, we need to infer $\langle {} \circ \ddot {\mathbf x} \mid \mathbf x, \dot{\mathbf x} \rangle$ and plug in the obtained function of $\mathbf x$ and $\dot{\mathbf x}$.
	
	\section{Appendix C}
\begin{figure}
		\begin{center}
			\includegraphics[scale=0.7]{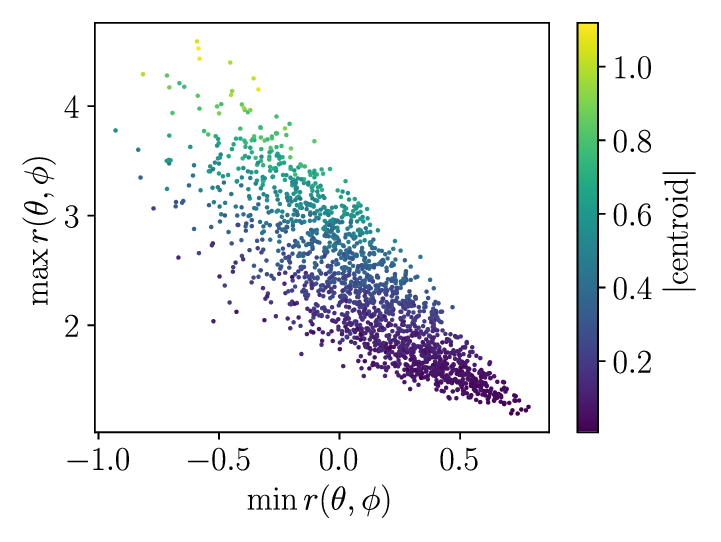}
		\end{center}
		\caption{Minimum and maximum $r(\theta,\phi)$, along with distance from origin to centroid, for simulated data.}
		\label{fig:minmax}
	\end{figure}

	In this appendix, we describe the estimation of contributions of the shape mapping to non-Gaussian effects. We consider a distance function:
	\begin{equation}
		r(\theta, \phi) = 1 + \sum_{l=2}^3 \sum_{m=-l}^l r_l^m Y_l^m(\theta, \phi)
	\end{equation}
	(in actuality, we use real spherical harmonics) and investigate the map $r_l^m \mapsto s_l^m$. First, we attempted to solve for each $s_l^m$ the corresponding $r_l^m$. However, in many cases the obtained solutions had large negative values of $r(\theta,\phi)$ and were thus unreliable. Therefore, we took another approach. We then tried to estimate the distribution of $r_l^m$ by generating for each cell a Gaussian distribution with the same mean and variance as $s_l^m$, and setting the $r_l^m$ values to these multiplied by some factor $\gamma_l^m$ and for $m=0$ adding an offset $\delta_l$, to be determined by equating the resulting mean and variance of the corresponding $s_l^m$ for $m=0$ and the resulting variance for $m>0$. The calculated factors $\gamma_l^m$ and offsets $\delta_l$ are tabulated in Table \ref{tab:gamma}. The resulting $r(\theta, \phi)$ often had negative minimum values, but these were generally small compared to the maximum values (Fig.\ \ref{fig:minmax}), especially when raised to a power (after taking centroid offset into account) and when combined with the fact that larger $r$ values take up more surface area, so we deemed this to be acceptable. The statistics for the calculated $s_l^m$ for the entire population are $\operatorname{Var}(s^0_2) = 0.014$, $\operatorname{Var}(s^0_3) = 4.6 \times 10^{-3}$ (fitted), and $\operatorname{Cov}(s^0_2, s^0_3) = -2.7 \times 10^{-3}$, compared to a true value of $-4.0 \times 10^{-3}$. The cell-wise statistics are $\operatorname{Var}(s^0_2) = 4.5 \times 10^{-3}$, $\operatorname{Var}(s^0_3) = 3.3 \times 10^{-3}$, $\operatorname{Cov}(s^0_2, s^0_3) = -0.5 \times 10^{-3}$, $\Re \langle s^1_2 {s^1_3}^* \rangle = 0.6 \times 10^{-3}$, and $\Re \langle s^2_2 {s^2_3}^* \rangle = 2.7 \times 10^{-3}$ (compare with Table \ref{tab:corning-vars}).
	
	\begin{table}
		\begin{center}
			\caption{Correction factors and offsets associated with nonlinear shape map.}
			\label{tab:gamma}
			\begin{tblr}{|Q[c,m]|Q[c,m]|Q[c,m]|Q[c,m]|Q[c,m]|Q[c,m]|Q[c,m]|Q[c,m]|Q[c,m]|}
				$\gamma_2^0$ & $\delta_2$ & $\gamma_3^0$ & $\delta_3$ & $\gamma_2^1$ & $\gamma_3^1$ & $\gamma_2^2$ & $\gamma_3^2$ & $\gamma_3^3$ \\ \hline
				1.33 & $-0.023$ & 3.05 & 0.083 & 1.55 & 1.95 & 2.04 & 0.54 & 2.53 \\ \hline
			\end{tblr}
		\end{center}
	\end{table}
	
	Next, we fit the obtained $s_l^m$ to quadratic polynomials in $r_l^m$. Along with azimuthal and chiral symmetry, the shape map obeys the symmetry of flipping the $z$-axis. In addition, a spherical shape must have $s_l^m = 0$. This leads to:
	\begin{align}
		s^0_2 &\approx \alpha^0_2 r^0_2 + \beta^{00}_{22} (r^0_2)^2 + \beta^{00}_{33} (r^0_3)^2 + \beta^{11^*}_{22} r^1_2 {r^1_2}^* + \beta^{11^*}_{33} r^1_3 {r^1_3}^* + \beta^{22^*}_{22} r^2_2 {r^2_2}^* + \beta^{22^*}_{33} r^2_3 {r^2_3}^* + \beta^{33^*}_{33} r^3_3 {r^3_3}^*, \label{eq:quad-first} \\
		s^0_3 &\approx \alpha^0_3 r^0_3 + \beta^{00}_{23} r^0_2 r^0_3 + \beta^{11^*}_{23} \Re [r^1_2 {r^1_3}^*] + \beta^{22^*}_{23} \Re [r^2_2 {r^2_3}^*], \\
		s^1_2 &\approx \alpha^1_2 r^1_2 + \beta^{01}_{22} r^0_2 r^1_2 + \beta^{01}_{33} r^0_3 r^1_3 + \beta^{1^*2}_{22} {r^1_2}^* r^2_2 + \beta^{1^*2}_{33} {r^1_3}^* r^2_3 + \beta^{2^*3}_{33} {r^2_3}^* r^3_3, \\
		s^1_3 &\approx \alpha^1_3 r^1_3 + \beta^{01}_{23} r^0_2 r^1_3 + \beta^{01}_{32} r^0_3 r^1_2 + \beta^{1^*2}_{23} {r^1_2}^* r^2_3 + \beta^{1^*2}_{32} {r^1_3}^* r^2_2 + \beta^{2^*3}_{23} {r^2_2}^* r^3_3, \\
		s^2_2 &\approx \alpha^2_2 r^2_2 + \beta^{02}_{22} r^0_2 r^2_2 + \beta^{02}_{33} r^0_3 r^2_3 + \beta^{11}_{22} (r^1_2)^2 + \beta^{11}_{33} (r^1_3)^2 + \beta^{1^*3}_{33} {r^1_3}^* r^3_3, \\
		s^2_3 &\approx \alpha^2_3 r^2_3 + \beta^{02}_{23} r^0_2 r^2_3 + \beta^{02}_{32} r^0_3 r^2_2 + \beta^{11}_{23} r^1_2 r^1_3 + \beta^{1^*3}_{23} {r^1_2}^* r^3_3, \\
		s^3_3 &\approx \alpha^3_3 r^3_3 + \beta^{03}_{23} r^0_2 r^3_3 + \beta^{12}_{23} r^1_2 r^2_3 + \beta^{12}_{32} r^1_3 r^2_2. \label{eq:quad-last}
	\end{align}
	The coefficients of the fit, along with the root mean squared magnitude of the regressors, are given in Table \ref{tab:quadshape}, in the order written in Eqs.\ \eqref{eq:quad-first}\textendash \eqref{eq:quad-last}. Despite high $R^2$ values ($\gtrsim 0.9$), the fitted coefficients changed when the distribution of $r_l^m$ was modified (not shown), indicating a lack of predictive power and nonlinearity beyond quadratic.
	
	\begin{table}
		\begin{center}
			\caption{Coefficients of a quadratic fit for the shape mapping.}
			\label{tab:quadshape}
			\begin{tblr}{Q[c,m]|Q[c,m]|Q[c,m]|}
				\SetCell[r=2]{c} & $\alpha$ & $\beta$ \\ \hline
				& \SetCell[c=2]{c} RMS of regressors \\ \hline
				\SetCell[r=2]{c} $s^0_2$ & $1.02$ & $-0.73, 0.08, 0.18, 0.31, -0.34, -0.82, -0.34$ \\ \hline
				& 0.247 & 0.092, 0.178, 0.031, 0.020, 0.043, 0.001, 0.009 \\ \hline
				\SetCell[r=2]{c} $s^0_3$ & 0.56 & $-0.74, -0.19, -1.82$ \\ \hline
				& 0.342 & 0.106, 0.013, 0.005 \\ \hline
				\SetCell[r=2]{c} $s^1_2$ & 0.66 & $-0.05, 0.24, 0.76, 0.52, 2.54$ \\ \hline
				& 0.140 & 0.034, 0.042, 0.024, 0.003, 0.002 \\ \hline
				\SetCell[r=2]{c} $s^1_3$ & 0.56 & $-0.43, 0.05, 0.77, 0.65, -0.47$ \\ \hline
				& 0.111 & 0.029, 0.049, 0.004, 0.019, 0.012 \\ \hline
				\SetCell[r=2]{c} $s^2_2$ & 0.63 & $-1.09, 0.07, 0.27, 0.20, -0.25$ \\ \hline
				& 0.167 & 0.037, 0.010, 0.031, 0.020, 0.009 \\ \hline
				\SetCell[r=2]{c} $s^2_3$ & 0.88 & $-1.23, -0.60, 0.32, 0.68$ \\ \hline
				& 0.025 & 0.007, 0.055, 0.017, 0.011 \\ \hline
				\SetCell[r=2]{c} $s^3_3$ & 0.51 & $-1.02, 0.56, -0.42$ \\ \hline
				& 0.071 & 0.016, 0.004, 0.019 \\ \hline
			\end{tblr}
		\end{center}
	\end{table}
	
\begin{figure}
		\begin{center}
			\includegraphics[scale=0.7]{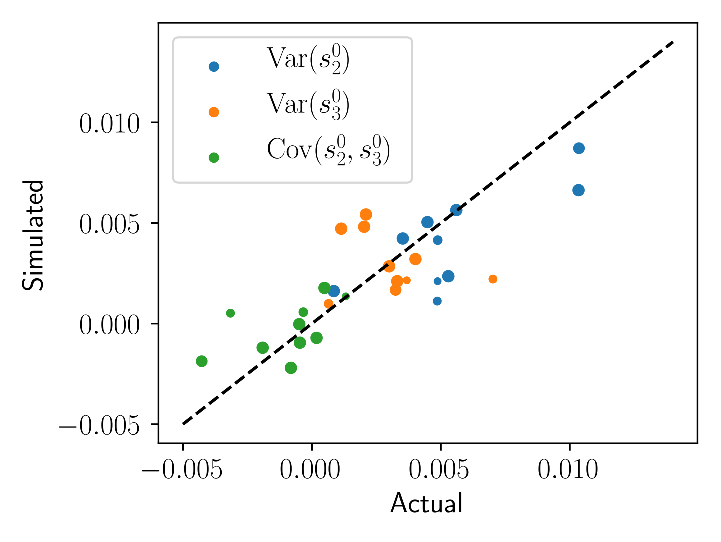}
		\end{center}
		\caption{Variance and covariance for individual cells, actual and simulated. The size of the marker is proportional to the length of the time-series.}
		\label{fig:varcovs02s03}
	\end{figure}

	To properly estimate dynamics of the $r_l^m$, we would have to fit not only moments of the distribution of $s_l^m$, as we have done, but also all the dynamical statistics. However, this was infeasible and thus we are not able to make a serious estimate. Furthermore,  statistics for individual cells were not well reproduced (Fig.\ \ref{fig:varcovs02s03}). Thus, as an approximation, we resorted to using the dynamics of $s_l^m$ multiplied by $\gamma_l^m$ and for $m=0$ adding $\delta_l$. We ran simulations of this dynamics for $r_l^m$ and calculated the corresponding $s_l^m$. The simulated covariance matrices, angular momenta $L(v^i,v^j)$ and diffusivities were mostly in agreement with the true values (not shown; also note the large uncertainties in Table \ref{tab:corning-vars}), which indicates that we have a reasonable approximation. We then used the quadratic fit (Eqs.\ \eqref{eq:quad-first}\textendash \eqref{eq:quad-last}) to calculate theoretically the terms in the third-order quantities up to quadratic order in $\beta$ (i.e., combining $\alpha$ and $\beta$ into a single linear coefficient when writing in terms of demeaned variables) under assumption of a linear Gaussian process for $r_l^m$, orientational changes, and velocity. We ignore population heterogeneity in this calculation. The advantage of the theoretical approach is that it decomposes the value into a sum of terms, thus allowing sensitivity to parameters to be more easily examined. (This becomes forbiddingly complicated for fourth-order quantities, so we did not examine those.) In no case did a small quantity result from the subtraction of large quantities. On the other hand, simulation allows the full nonlinearity to be accounted for. Results from simulations were in line with theoretical estimates. This procedure gives a rough order-of-magnitude estimate for the contribution of nonlinear shape mapping to measured quantities.
	
	\section{Appendix D}
	In this appendix, we address issues related to statistical testing. Consider a quantity evolving in time with an eigenvalue $\lambda$ for its dynamics (i.e., the conditional expectation $\langle \dot x \mid x \rangle = \lambda x$, or the covariance function $\langle x(\tau) x(0) \rangle \propto e^{\lambda|\tau|}$ \cite{low2023second}; $|\lambda|$ is a decay rate). From Eq.\ \eqref{eq:adjust}, the time-average of such a quantity may be approximated by the average of $N$ i.i.d.\ variables with $N \approx |\lambda|T/2$, where $T$ is the trajectory length. Thus, we consider $X_1,X_2,\ldots,X_N$ i.i.d.\ with $\langle X_1 \rangle = 0$, $\langle X_1^2 \rangle = 1$, and let:
	\begin{align}
		Y &:= \frac{1}{\sqrt{N}} \sum_{i=1}^N X_i, \\
		\Psi &:= \left({\frac 1 N \sum_{i=1}^N X_i^2}\right)^{1/2}.
	\end{align}
	We assume that moments of $X_1$ of all orders exist. To examine moments of $Y/\Psi$, we expand:
	\begin{equation}
		\left({\frac{Y}{\Psi}}\right)^n = Y^n \left[{1 - \frac{n}{2} (\Psi^2-1) + \frac{n(n+2)}{8}(\Psi^2-1)^2 - \cdots}\right],
	\end{equation}
	where by the central limit theorem:
	\begin{equation}
		\Psi^2 - 1 = \mathcal O(N^{-1/2}).
	\end{equation}
	For $n \ge 2$ even, we have:
	\begin{equation}
		\langle Y^n (\Psi^2 - 1) \rangle = \mathcal O(N^{-1}),
	\end{equation}
	and we know that $\langle Y^n \rangle$ differs from the Gaussian value with $\mathcal O(N^{-1})$ correction. For $n \ge 1$ odd, we have in general:
	\begin{equation}
		\langle Y^{2n} \rangle = \mathcal O(N^0), \quad \langle Y^n \rangle = \mathcal O(N^{-1/2}), \quad \langle Y^n (\Psi^2 - 1) \rangle = \mathcal O(N^{-1/2}), \quad \langle Y^n (\Psi^2 - 1)^2 \rangle = \mathcal O(N^{-3/2}).
	\end{equation}
	Thus we see that the approach of $Y/\Psi$ to normality scales as $\mathcal O(N^{-1/2})$. However, if $\langle X_1^3 \rangle = 0$, then we have instead for $n \ge 1$ odd:
	\begin{equation}
		\langle Y^n \rangle = \mathcal O(N^{-3/2}), \quad \langle Y^n (\Psi^2 - 1) \rangle = \mathcal O(N^{-3/2}),
	\end{equation}
	and the approach to normality is in this case $\mathcal O(N^{-1})$. We expect similar results to hold for similar quantities.
	
	For zero-mean Gaussian processes, the decay rate associated with a product of quantities (actually, the multidimensional probabilist's Hermite polynomial \cite{low2023second}) is the sum of their decay rates. To evaluate the appropriateness of such an approximation, we can investigate fourth-order quantities or quantities of the form of Eq.\ \eqref{eq:exp-int}, but as mentioned in the main text, the latter did not have statistically significant deviations from a Gaussian process. Fourth-order quantities satisfy the relations:
	\begin{align}
		\langle x^i x^j x^k \dot x^l \rangle + \langle x^i x^j \dot x^k x^l \rangle + \langle x^i \dot x^j x^k x^l \rangle + \langle \dot x^i x^j x^k x^l \rangle &= 0, \\
		\widetilde{L}(x^i, \dot x^j \dot x^k, \dot x^l) + \widetilde{L}(x^i, \dot x^k \dot x^l, \dot x^j) + \widetilde{L}(x^i, \dot x^l \dot x^j, \dot x^k) &= \langle \dot x^i \dot x^j \dot x^k \dot x^l \rangle.
	\end{align}
	Similarly to third-order quantities, we can partition fourth-order quantities based on whether velocities or accelerations occur at most twice, three times, or four times. Subject to the above relations, these groups comprise, respectively, 1621 real and 1438 imaginary (independent) quantities, 2684 real and 2540 imaginary quantities, and 1922 real and 1594 imaginary quantities. Imaginary parts were not statistically significant, and no statistically significant trends were found by using linear regression with time. The results for real parts are given in Table \ref{tab:4th-order}, sorted by unadjusted $p$-values. Based on the sign of the $t$-value, we inferred the $p$-values to be either approximate, underestimates, or overestimates (see later). After the first double horizontal line, results are restricted to quantities involving at most one velocity. After the second double horizontal line, results are restricted to quantities involving no velocities.
	
	\begin{table}
		\begin{center}
			\caption{Fourth-order quantities.}
			\label{tab:4th-order}
			\begin{tblr}{|Q[c,m]|Q[c,m]|Q[c,m]||Q[c,m]|Q[c,m]|Q[c,m]|}
				Quantity & $t$-value & $p$-value & Quantity & $t$-value & $p$-value \\ \hline
				$\Re \langle \dot s^0_2 (\dot \theta e^{i \phi})^2 {{}\dot s^2_3}^* \rangle$ & $-14.65$ & $\approx 1.39 \times 10^{-7}$ & $\Re \langle \dot s^0_2 (\dot \theta e^{i \phi})^2 {{}\dot s^2_2}^* \rangle$ & $-6.43$ & $\approx 1.21 \times 10^{-4}$ \\ \hline
				$\Re \langle \dot s^0_2 \dot \theta e^{i \phi} v^1 {{}\dot s^2_2}^* \rangle$ & $-10.85$ & $\approx 1.81 \times 10^{-6}$ & $\Re \langle \Delta s^0_3 s^1_2 \, \mathrm d[\dot s^0_2, {v^1}^*]/\mathrm d t \rangle$ & 6.43 & $\approx 1.21 \times 10^{-4}$ \\ \hline
				$\Re \langle \dot s^0_2 (v^1)^2 {{}\dot s^2_2}^* \rangle$ & $-9.48$ & $\approx 5.57 \times 10^{-6}$ & $\Re \langle s^1_2 {s^2_3}^* \dot s^0_3 \dot \theta e^{i \phi} \rangle$ & 6.40 & $\approx 1.25 \times 10^{-4}$ \\ \hline
				$\Re \langle s^1_2 {s^2_3}^* \dot \theta e^{-i \phi} \dot s^2_3 \rangle$ & $-9.29$ & $\approx 6.55 \times 10^{-6}$ & $\Re \langle s^1_2 \dot s^0_2 \dot \theta e^{i \phi} {{}\dot s^2_3}^* \rangle$ & $-6.32$ & $\approx 1.37 \times 10^{-4}$ \\ \hline
				$\Re \langle s^1_2 {s^2_3}^* {v^1}^* \dot s^2_2 \rangle$ & $-8.68$ & $> 1.15 \times 10^{-5}$ & $\Re L(\dot s^0_2 \dot \theta e^{i \phi} {{}\dot s^2_3}^*, \dot s^1_2)$ & 6.16 & $\approx 1.66 \times 10^{-4}$ \\ \hline
				$\Re \widetilde{L}(s^2_3, \dot s^0_3 {{}\dot s^1_2}^*, {{}\dot s^1_3}^*)$ & $-7.92$ & $\approx 2.39 \times 10^{-5}$ & $\langle v^1 {v^1}^* \dot s^2_3 {{}\dot s^2_3}^* \rangle$ & 6.15 & $< 1.68 \times 10^{-4}$ \\ \hline
				$\Re \langle s^1_2 {s^1_2}^* s^2_3 {{}\dot s^2_2}^* \rangle$ & $-7.89$ & $> 2.46 \times 10^{-5}$ & $\Re \widetilde{L}(s^2_3, \dot \theta e^{i \phi} {{}\dot s^2_2}^*, {{}\dot s^1_2}^*)$ & 6.12 & $\approx 1.75 \times 10^{-4}$ \\ \hline 
				$\Re \widetilde{L}(s^2_3, {{}\dot s^1_2}^* {{}\dot s^2_2}^*, \dot s^1_2)$ & $7.76$ & $\approx 2.82 \times 10^{-5}$ & $\Re \langle \dot s^0_2 v^1 \, \mathrm d[\dot s^1_2, {{}\dot s^2_2}^*]/\mathrm d t \rangle$ & 6.03 & $\approx 1.94 \times 10^{-4}$ \\ \hline
				$\langle v^1 {v^1}^* \dot s^2_2 {{}\dot s^2_2}^* \rangle$ & $7.59$ & $< 3.38 \times 10^{-5}$ & $\Re \langle s^2_2 {s^2_3}^* v^1 {{}\dot s^1_3}^* \rangle$ & 5.97 & $> 2.09 \times 10^{-4}$ \\ \hline
				$\Re \langle \dot \theta e^{i \phi} {v^1}^* \dot s^2_2 {{}\dot s^2_2}^* \rangle$ & 7.54 & $< 3.55 \times 10^{-5}$ & $\Re \langle s^2_3 (\dot s^0_3)^2 {{}\dot s^2_3}^* \rangle$ & $-5.96$ & $\approx 2.14 \times 10^{-4}$ \\ \hline
				$\Re L(\dot s^0_2 \dot s^0_3 \dot \theta e^{i \phi}, {{}\dot s^1_3}^*)$ & $-7.15$ & $\approx 5.39 \times 10^{-5}$ & $\Re \langle v^1 {v^1}^* \dot s^2_2 {{}\dot s^2_3}^* \rangle$ & 5.92 & $< 2.23 \times 10^{-4}$ \\ \hline 
				$\Re \langle s^1_3 \dot s^0_2 \dot s^0_3 \dot \theta e^{-i \phi} \rangle$ & 6.93 & $\approx 6.84 \times 10^{-5}$ & $\Re \langle s^1_2 {s^2_3}^* \dot s^0_3 v^1 \rangle$ & 5.85 & $\approx 2.42 \times 10^{-4}$ \\ \hline
				$\Re \langle (v^1)^2 \, \mathrm d[{{}\dot s^1_3}^*, {{}\dot s^1_3}^*]\mathrm d t \rangle$ & 6.76 & $\approx 8.30 \times 10^{-5}$ & $\Re \langle \dot s^0_2 \dot \theta e^{i \phi} v^1 {{}\dot s^2_3}^* \rangle$ & $-5.81$ & $\approx 2.57 \times 10^{-4}$ \\ \hline
				$\Re \langle s^1_3 {s^2_3}^* \dot \theta e^{-i \phi} \dot s^2_3 \rangle$ & $-6.66$ & $\approx 9.23 \times 10^{-5}$ & $\Re \widetilde{L}({s^2_3}^*, \dot s^0_2 v^1, v^1)$ & $-5.78$ & $\approx 2.65 \times 10^{-4}$ \\ \hline
				$\langle (\Delta s^0_3)^2 \dot s^0_3 \Delta v^0 \rangle$ & 6.45 & $> 1.18 \times 10^{-4}$ & $\Re \langle s^2_3 {s^2_3}^* s^1_2 {{}\dot s^1_3}^* \rangle$ & 5.78 & $> 2.65 \times 10^{-4}$ \\ \hline \hline
				$\Re \langle s^2_2 {s^2_3}^* s^1_2 {{}\dot s^1_3}^* \rangle$ & 4.72 & $> 1.09 \times 10^{-3}$ & $\Re \langle s^1_2 {s^1_3}^* s^1_3 {v^1}^* \rangle$ & $-3.86$ & $> 3.83 \times 10^{-3}$ \\ \hline \hline
				$\Re \langle s^1_2 {s^1_3}^* s^1_2 {s^1_3}^* \rangle$ & 3.42 & $> 7.63 \times 10^{-3}$ & $\Re \langle s^1_2 {s^1_3}^* s^2_3 {s^2_2}^* \rangle$ & 2.76 & $< 2.22 \times 10^{-2}$ \\ \hline
			\end{tblr}
		\end{center}
	\end{table}
	
\begin{figure}
		\begin{center}
			\includegraphics[scale=0.7]{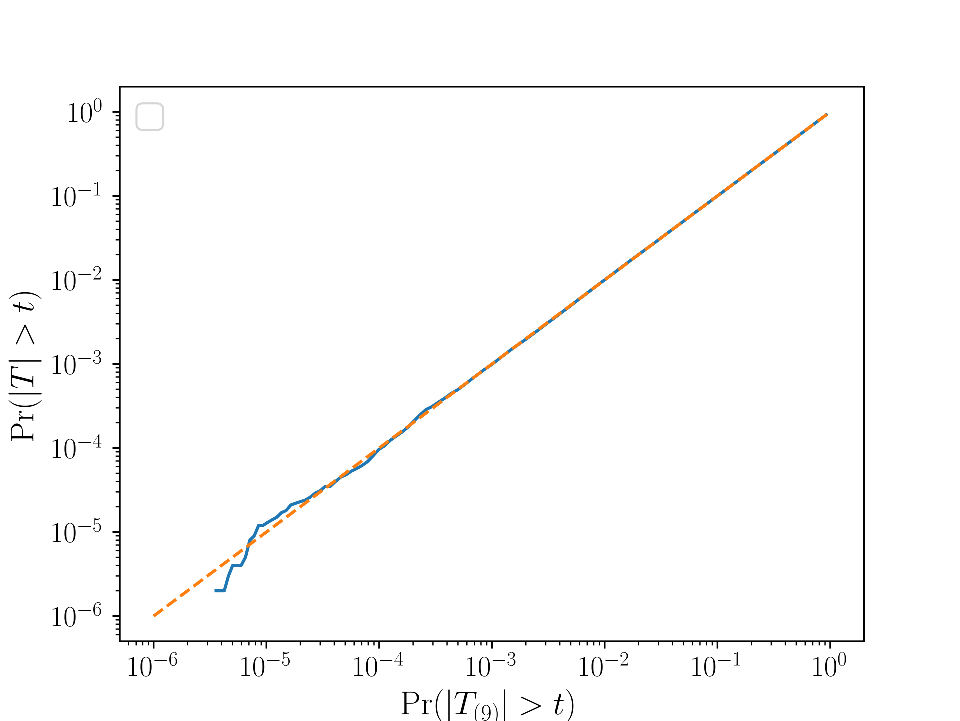}
		\end{center}
		\caption{Results for weighted $t$-test for Gaussian variables with variances inversely proportional to the weights, for $10^6$ samples. Dashed line is the identity function.}
		\label{fig:t}
	\end{figure}

	We used the \texttt{ttest\_ind} function from Python's \texttt{statsmodels.stats.weightstats} module. We tested this on a case of Gaussian variables with variances inversely proportional to the weights and obtained the expected results (Fig.\ \ref{fig:t}). The number of samples exceeding a given threshold is approximately Poisson distributed.
	
\begin{figure}
		\begin{center}
			\includegraphics[scale=0.7]{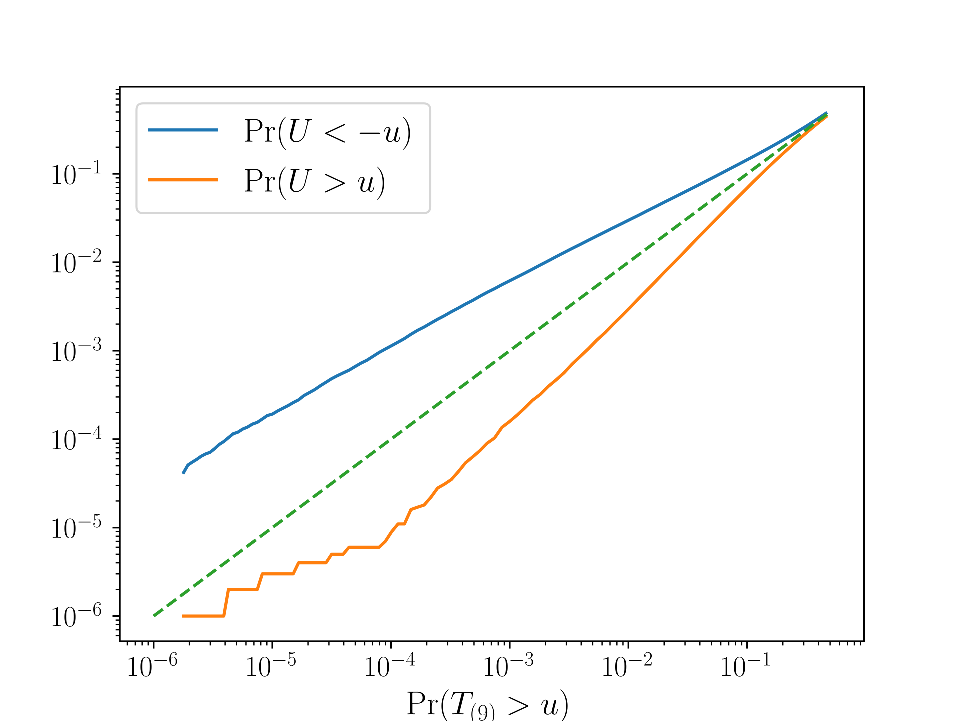}
		\end{center}
		\caption{Results for $\overline{X^2}$ for a zero-mean Gaussian variable $X$ with decay rate $|\lambda| = 0.084/\Delta t$, for $10^6$ samples. Dashed line is the identity function.}
		\label{fig:t2}
	\end{figure}

\begin{figure}
		\begin{center}
			\includegraphics[scale=0.7]{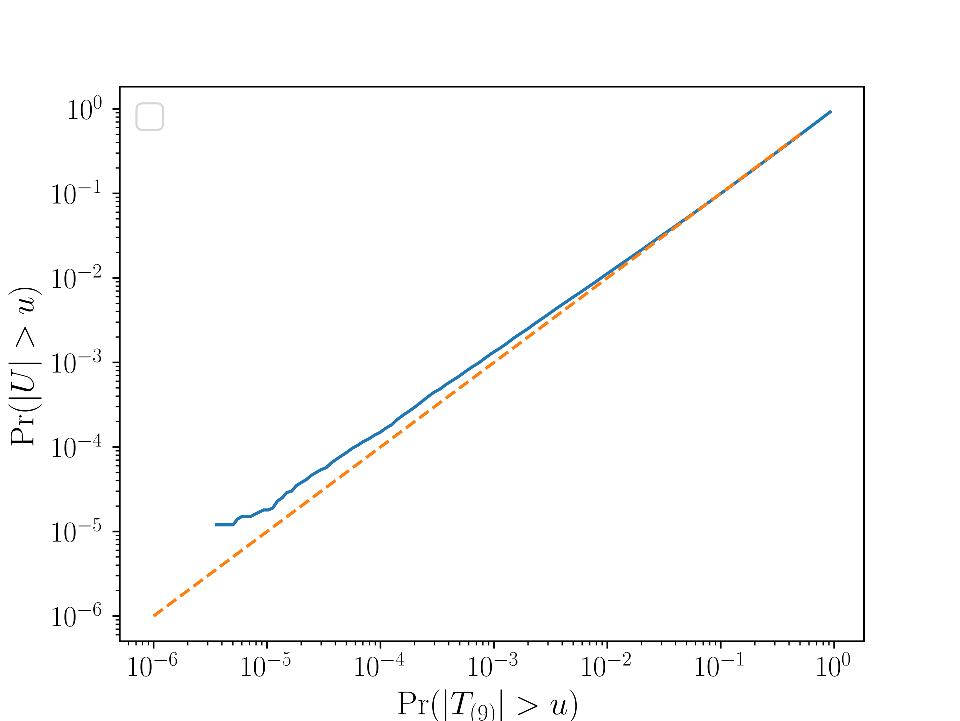}
		\end{center}
		\caption{Results for $\overline{XY}/\sqrt{\overline{X^2} \cdot \overline{Y^2}}$ for independent zero-mean Gaussians $X$ and $Y$ with decay rate $|\lambda| = 0.084/\Delta t$, for $10^6$ samples. Dashed line is the identity function.}
		\label{fig:t2sym}
	\end{figure}

	From the results of \cite{low2023second}, we estimate the decay rates for velocities by higher-magnitude eigenvalues of $\boldsymbol{\Gamma}$. We give the decay rates in units of the frame interval $\Delta t$. For quantities involving accelerations, we take the number of independent samples to be equal to the trajectory length (denoted in the figures by ``$N = T/\Delta t$''). We show results for $\overline{X^2}$ for a zero-mean Gaussian variable $X$ (Fig.\ \ref{fig:t2}) and $\overline{XY}/\sqrt{\overline{X^2} \cdot \overline{Y^2}}$ for independent zero-mean Gaussians $X$ and $Y$ (Fig.\ \ref{fig:t2sym}). We denote the $t$-statistic obtained from such quantities by the variable $U$. We see that the latter case is very close to a $t$-distribution.
	
\begin{figure}
		\begin{center}
			\includegraphics[scale=0.7]{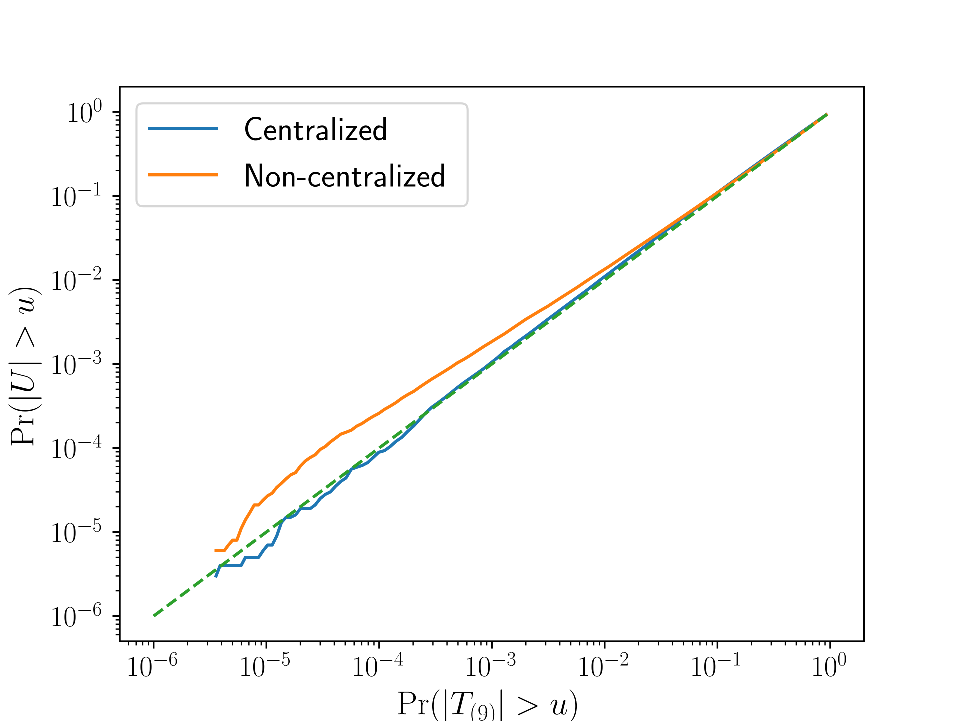}
		\end{center}
		\caption{Results for cubic functions of a zero-mean Gaussian variable (details in text), with decay rate $|\lambda|=0.141/\Delta t$, for $10^6$ samples. Dashed line is the identity function.}
		\label{fig:t3rev}
	\end{figure}

	Next, we show in Fig.\ \ref{fig:t3rev}, for a zero-mean Gaussian variable $X$, the quantities $\overline{(X-\overline{X})^3}/\overline{(X-\overline{X})^2}^{3/2}$ (centralized) and $\overline{X^3}/(\overline{X^2})^{3/2}$ (non-centralized). The centralized case is similar to a $t$-distribution; the non-centralized case would be further from a $t$-distribution than if using complex variables or decay rates appropriate to velocities.
	
\begin{figure}
		\begin{center}
			\includegraphics[scale=0.7]{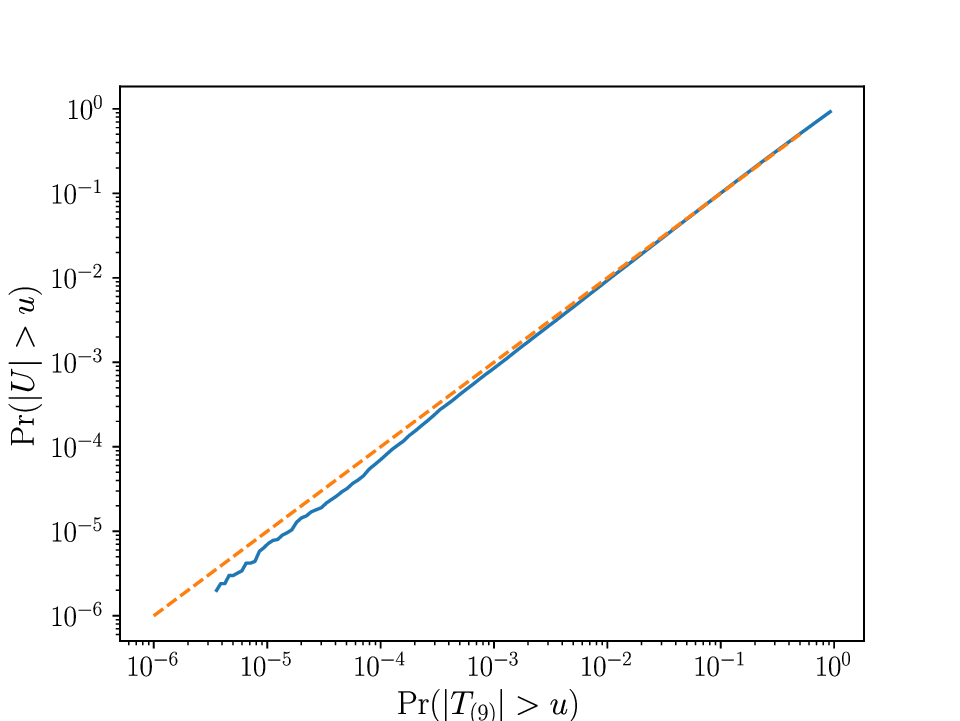}
		\end{center}
		\caption{Results for a fourth-order quantity of three zero-mean Gaussian variables (details in text), with decay rate $|\lambda| = 0.452/\Delta t$, for $5 \times 10^6$ samples. Dashed line is the identity function.}
		\label{fig:t4-3r}
	\end{figure}

\begin{figure}
		\begin{center}
			\includegraphics[scale=0.7]{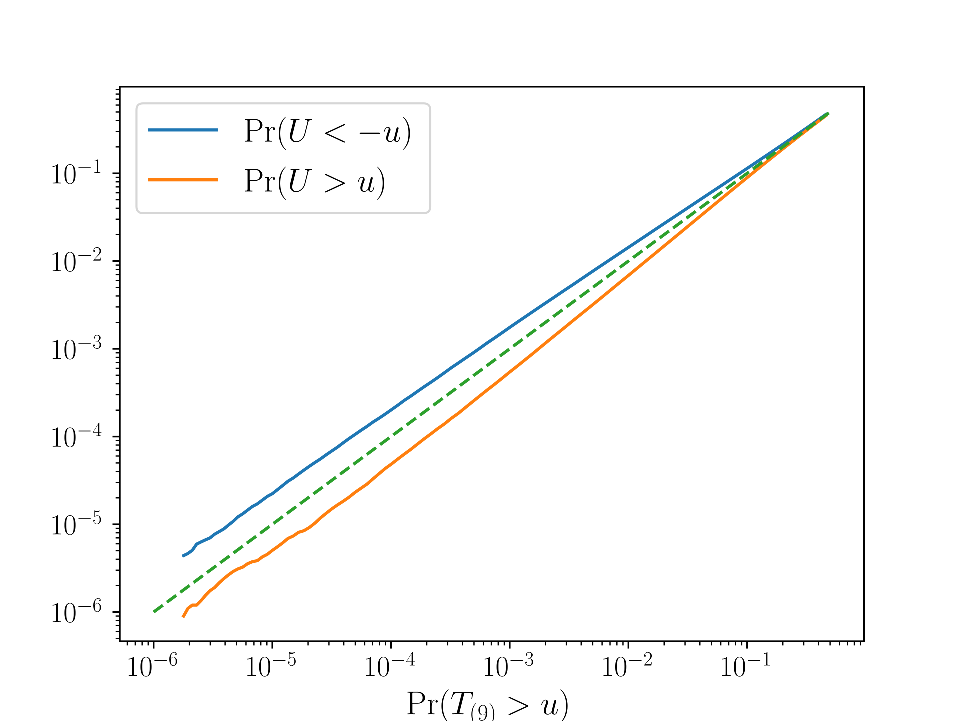}
		\end{center}
		\caption{Results for fourth-order quantities of two complex Gaussian variables (details in text), with decay rate $|\lambda| = 1.314/\Delta t$, for $2 \times 10^7$ samples. Dashed line is the identity function.}
		\label{fig:t4-2cq}
	\end{figure}

	For fourth-order quantities, we first consider a quantity with vanishing third moment. For this we take independent zero-mean Gaussians (real unless otherwise specified) $X$, $Y$, and $Z$ and consider $\overline{X^2 Y Z}/(\overline{X^2} \sqrt{\overline{Y^2} \cdot \overline{Z^2}})$ (Fig.\ \ref{fig:t4-3r}). We use the decay rate corresponding to $s^1_2 {s^1_2}^* s^2_3 {{}\dot s^2_2}^*$, as this is the slowest decay rate in Table \ref{tab:4th-order}, except the last row. This case is close to a $t$-distribution. For $\langle v^1 {v^1}^* \dot s^2_2 {{}\dot s^2_2}^* \rangle$, we then consider the case of independent complex Gaussian variables $W$ and $Z$ isotropic in the complex plane, and consider the quantity $\overline{|W|^2 |Z|^2}/(\overline{|W|^2} \cdot \overline{|Z|^2})$ (Fig.\ \ref{fig:t4-2cq}). We can use these results to estimate a lower bound for an adjustment to the $p$-values of quantities. Quantities in Table \ref{tab:4th-order} involving accelerations can be considered to be small compared to their respective normalization factors.
	
\begin{figure}
		\begin{center}
			\includegraphics[scale=0.7]{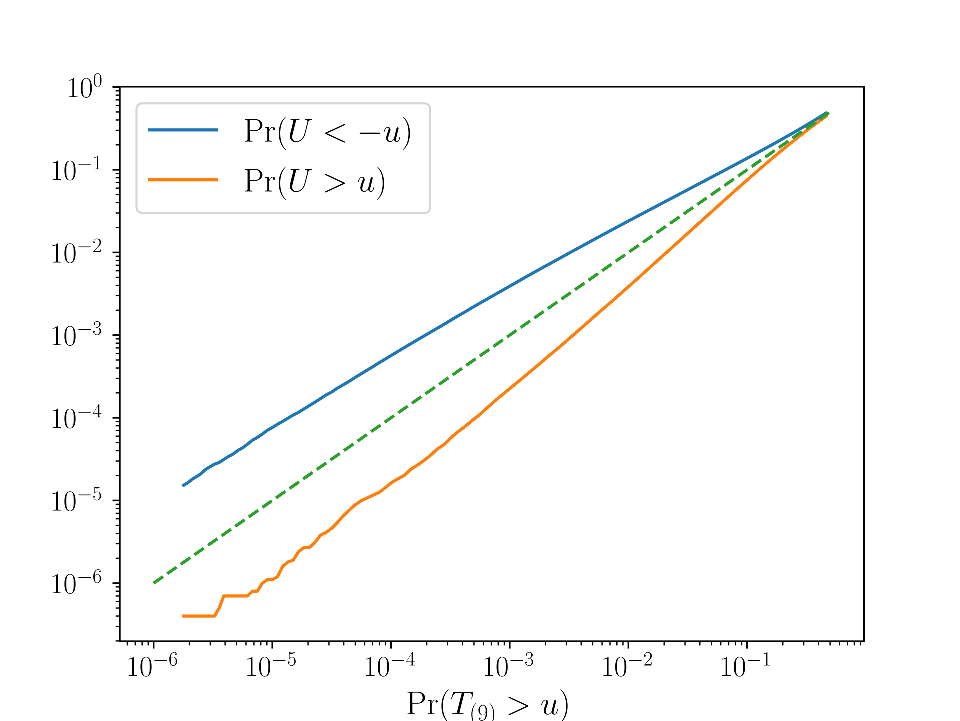}
		\end{center}
		\caption{Results for $\overline{X^2 Y^2}/(\overline{X^2} \cdot \overline{Y^2})$ for independent zero-mean Gaussians $X$ and $Y$ for $|\lambda| = 0.532/\Delta t$ with $10^7$ trials. Dashed line is the identity function.}
		\label{fig:t4-2r}
	\end{figure}

\begin{figure}
		\begin{center}
			\includegraphics[scale=0.7]{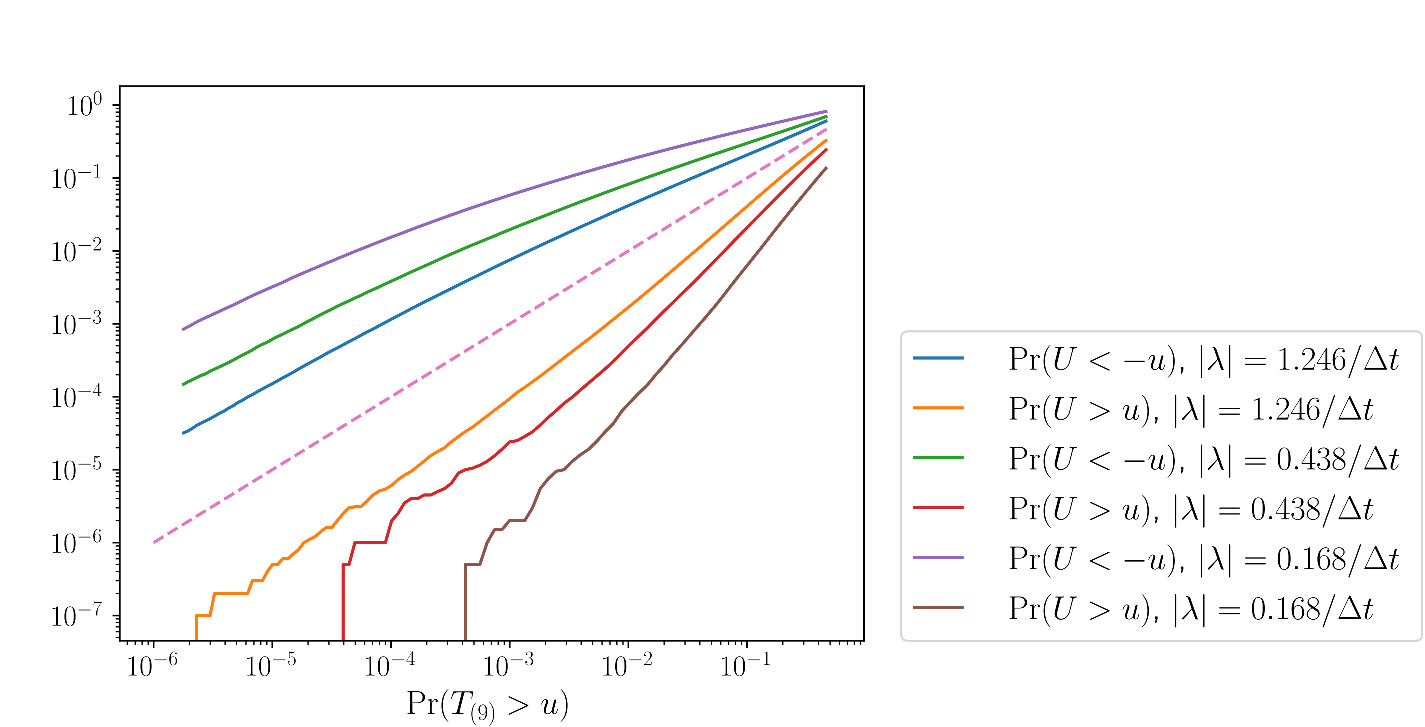}
		\end{center}
		\caption{Results for $\overline{|Z|^4}/\overline{|Z|^2}^2$ for a complex Gaussian $Z$ isotropic in the complex plane with $10^7$ trials for $|\lambda| = 1.246/\Delta t$ and $2 \times 10^6$ trials for the other cases. Dashed line is the identity function.}
		\label{fig:t4-2}
	\end{figure}

\begin{figure}
		\begin{center}
			\includegraphics[scale=0.7]{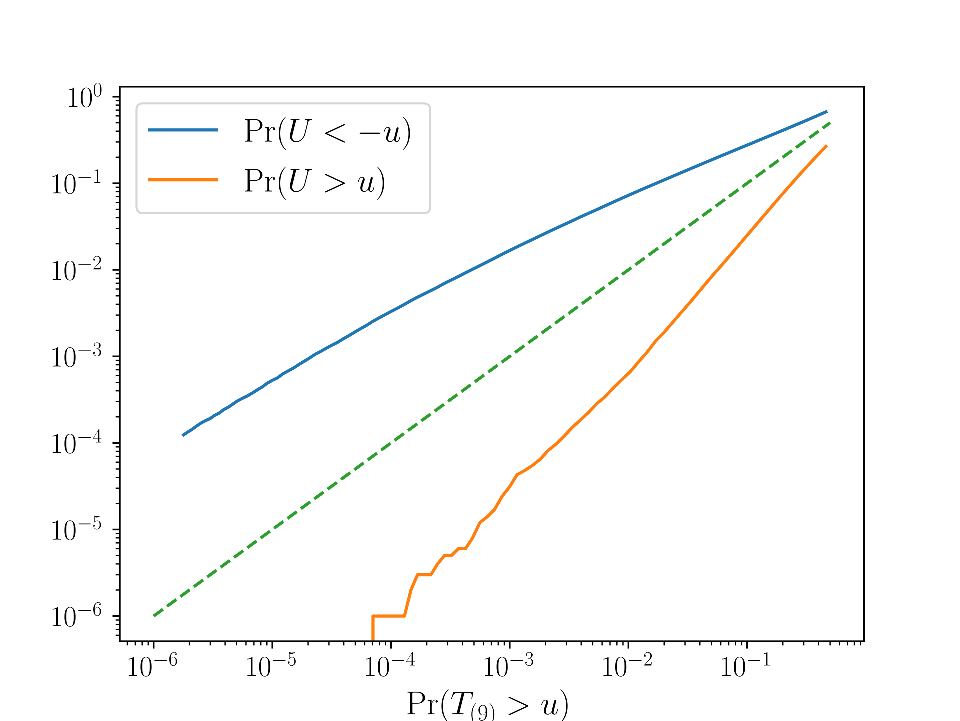}
		\end{center}
		\caption{Results for $\overline{X^4}/\overline{X^2}^2$ for a zero-mean Gaussian $X$ for $|\lambda| = 0.894/\Delta t$ with $10^6$ trials. Dashed line is the identity function.}
		\label{fig:t4-0v}
	\end{figure}

	To address the remaining quantities involving at least two velocities, we consider two cases: (1) quantities involving two state positions and two velocities, and (2) quantities involving four velocities with the same value of $|m|$. We use the approximation that state positions are weakly correlated with velocities \cite{low2023second}. Results for case (1) are depicted in Fig.\ \ref{fig:t4-2r}. The most statistically significant result for case (2) with $m \ne 0$ has a $t$-value of 5.42 and an (unadjusted) $p$-value of $4.23 \times 10^{-4}$, whereas the most statistically significant result with $m = 0$ has a $t$-value of 3.56 and a $p$-value of $6.10 \times 10^{-3}$. These cases are depicted in Figs.\ \ref{fig:t4-2} and \ref{fig:t4-0v}.
	
\begin{figure}
		\begin{center}
			\includegraphics[scale=0.7]{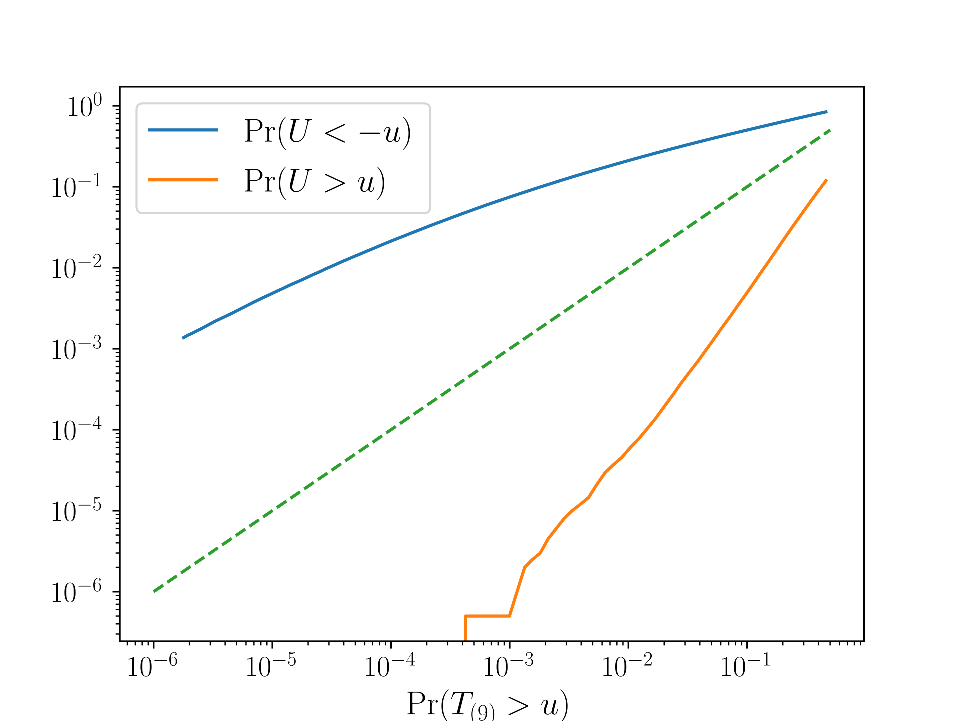}
		\end{center}
		\caption{Results for $\overline{(X-\overline{X})^4}/\overline{(X-\overline{X})^2}^2$ for a zero-mean Gaussian $X$ with $|\lambda| = 0.250/\Delta t$ with $2 \times 10^6$ trials. Dashed line is the identity function.}
		\label{fig:t4-0}
	\end{figure}

	The case for quantities involving a single velocity without an $m=0$ variable was given in Fig.\ \ref{fig:t4-2}. No quantities involving a single velocity and an $m=0$ variable had $p$-values below $2.22 \times 10^{-2}$. Simulations indicate that the $p$-value for $\Re \langle s^1_2 {s^1_3}^* s^1_2 {s^1_3}^* \rangle$ based on the $t$-distribution is a slight underestimate, contrary to what would be expected based on the sign of the $t$-value. We think this may be due to the smallness of $\langle s^1_2 {s^1_3}^* \rangle$; we believe that our procedure should give reasonable approximations. For the remaining cases, see Figs.\ \ref{fig:t4-2} and \ref{fig:t4-0}. These results indicate that, for the most part, a Gaussian approximation for first-order quantities is reasonable for describing the dynamics of second-order quantities, within statistical uncertainty. We thus use this approximation to estimate statistical significance for population variability.
	
	Now we address population variability. Let $Y_i \sim \mathcal N(\mu, \sigma_i^2)$, $1 \le i \le N_{\text{cells}}$, be independent with common mean $\mu$ and individual variances $\sigma_i^2$. The minimum-variance unbiased estimator for $\mu$ is given by:
	\begin{equation}
		\overline{Y} := \frac{\sum_{i=1}^{N_{\text{cells}}} \sigma_i^{-2} Y_i}{\sum_{i=1}^{N_{\text{cells}}} \sigma_i^{-2}}.
	\end{equation}
	We investigate the quantity:
	\begin{equation} \label{eq:chi-sq}
		V := \sum_{i=1}^{N_{\text{cells}}} \frac{(Y_i - \overline{Y})^2}{\sigma_i^2}.
	\end{equation}
	By Cochran's theorem, $V \sim \chi^2 (N_{\text{cells}} - 1)$. Explicitly, we have the relation:
	\begin{equation}
		\sum_{i=1}^{N_{\text{cells}}} \left({\frac{Y_i - \mu}{\sigma_i}}\right)^2 = \left({\sum_{i=1}^{N_{\text{cells}}} \frac{1}{\sigma_i^2}}\right) (\overline{Y} - \mu)^2 + V
	\end{equation}
	which establishes the independence of $\overline{Y}$ and $V$ as well as the claimed distribution of $V$.
	
	To consider deviations from normality, let $X_{ij}$, $1 \le i \le N_{\text{cells}}$, $1 \le j \le N_i$ be i.i.d., with $\langle X_{1,1} \rangle = 0$, $\langle X_{1,1}^2 \rangle = 1$. We again assume that moments of $X_{1,1}$ of all orders exist. We also assume that $N_{\text{cells}}$ is fixed and that all the $N_i$ are proportional, i.e., $N_i \propto N$. We define:
	\begin{equation}
		Y_i := \frac{\sqrt{N}}{N_i} \sum_{j=1}^{N_i} X_{ij}, \quad 1 \le i \le N_{\text{cells}},
	\end{equation}
	so that $\langle Y_i \rangle = 0$, $\langle Y_i^2 \rangle = N/N_i$. We investigate, as before:
	\begin{equation}
		V := \sum_{i=1}^{N_{\text{cells}}} \frac{N_i}{N} (Y_i - \overline{Y})^2, \quad \overline{Y} := \frac{\sum_{i=1}^{N_{\text{cells}}} N_i Y_i}{\sum_{i=1}^{N_{\text{cells}}} N_i}.
	\end{equation}
	We first have:
	\begin{equation}
		\langle V \rangle = N_{\text{cells}} - 1.
	\end{equation}
	For higher moments, the second moment $\langle V^2 \rangle$ depends on $\langle Y_i^4 \rangle$, the third moment $\langle V^3 \rangle$ additionally depends on $\langle Y_i^6 \rangle$ and $\langle Y_i^3 \rangle \langle Y_j^3 \rangle$, the fourth moment $\langle V^4 \rangle$ additionally depends on $\langle Y_i^8 \rangle$ and $\langle Y_i^5 \rangle \langle Y_j^3 \rangle$, etc. Thus from the previous discussion, we conclude that the approach of $V$ to $\chi^2 (N_{\text{cells}} - 1)$ is $\mathcal O(N^{-1})$. Although this seems improved from the previous case of $\mathcal O(N^{-1/2})$, in practice the deviation from a Gaussian case is quite large, as we will see.
	
	We consider the case where the $X_{ij}$ are products of two zero-mean jointly Gaussian-distributed variables. To construct a pair of zero-mean jointly Gaussian-distributed variables with arbitrary correlation coefficient, we take independent $X, Y \sim \mathcal N(0, 1)$ and a number $\nu$, and consider the pair $(X, Y + \nu X)$. We have:
	\begin{equation}
		\langle (Y + \nu X)^2 \rangle = 1 + \nu^2,
	\end{equation}
	and therefore the correlation coefficient $\rho$ between $X$ and $Y + \nu X$ is related to $\nu$ via:
	\begin{equation} \label{eq:rho-nu}
		\rho = \frac{\nu}{\sqrt{1 + \nu^2}}, \quad \nu = \frac{\rho}{\sqrt{1 - \rho^2}}.
	\end{equation}
	We calculate the centralized second and fourth moments:
	\begin{align}
		\langle [X(Y + \nu X) - \nu]^2 \rangle &= 1 + 2 \nu^2, \\
		\langle [X(Y + \nu X) - \nu]^4 \rangle &= 9 + 60 \nu^2 + 60 \nu^4.
	\end{align}
	Note that the excess kurtosis for $|\rho| = 1$ is twice that for $\rho = 0$.
	
\begin{figure}
		\begin{center}
			\includegraphics[scale=0.7]{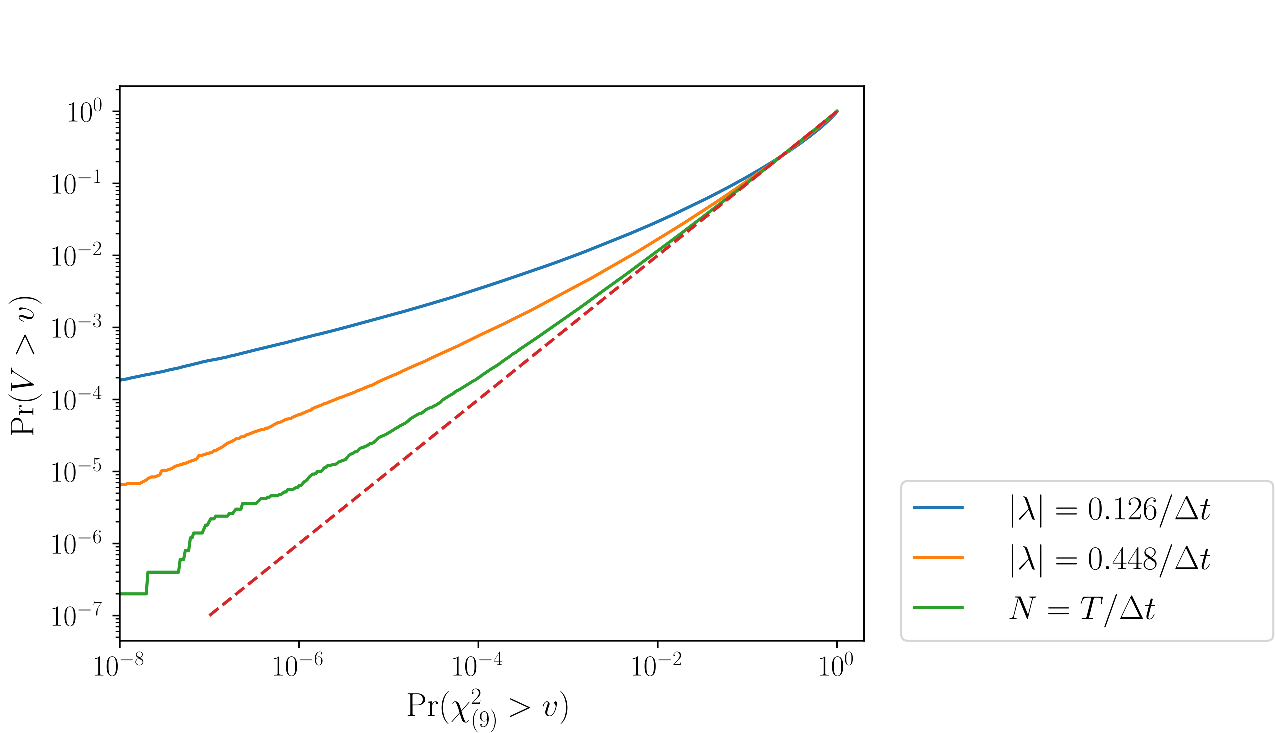}
		\end{center}
		\caption{Results for the square of a zero-mean Gaussian, for $5 \times 10^6$ trials. Dashed line is the identity function.}
		\label{fig:chi2-1}
	\end{figure}

\begin{figure}
		\begin{center}
			\includegraphics[scale=0.7]{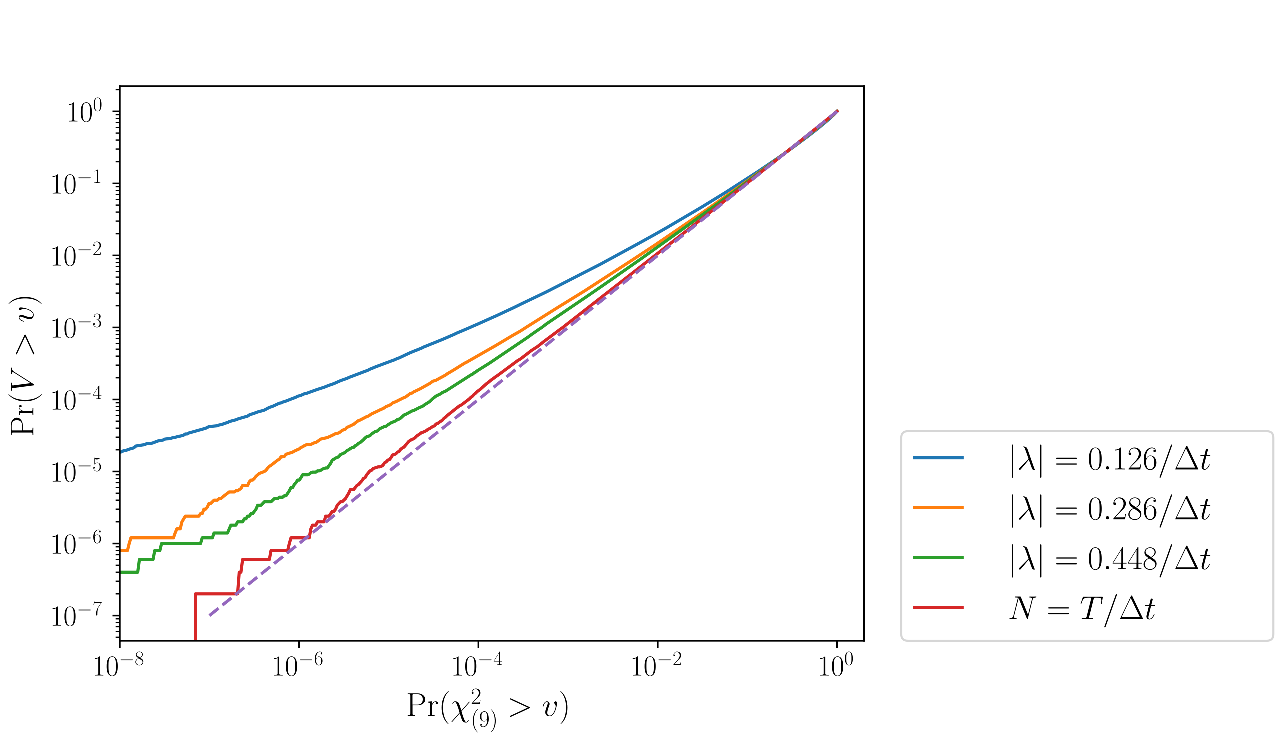}
		\end{center}
		\caption{Results for the product of two uncorrelated zero-mean Gaussians, for $5 \times 10^6$ trials. Dashed line is the identity function.}
		\label{fig:chi2-0}
	\end{figure}

\begin{figure}
		\begin{center}
			\includegraphics[scale=0.7]{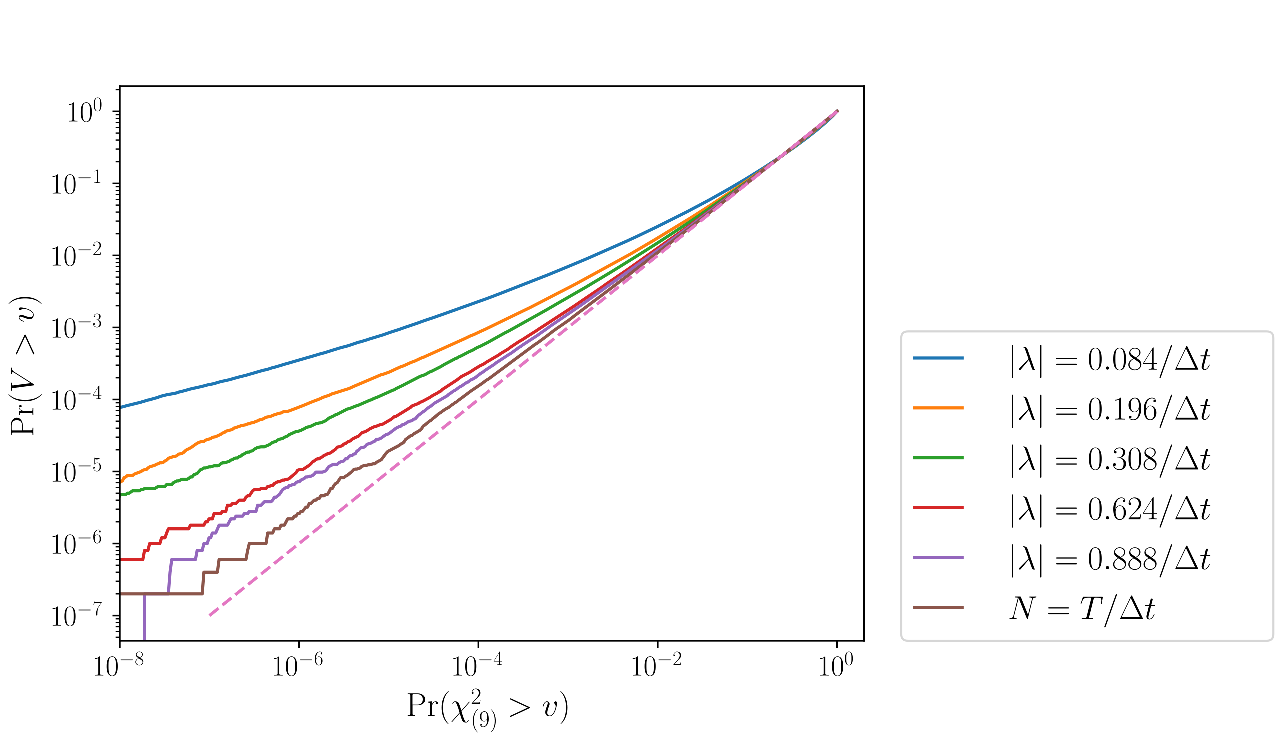}
		\end{center}
		\caption{Results for the squared magnitude of a complex Gaussian, for $5 \times 10^6$ trials. Dashed line is the identity function.}
		\label{fig:chi2-c1}
	\end{figure}

\begin{figure}
		\begin{center}
			\includegraphics[scale=0.7]{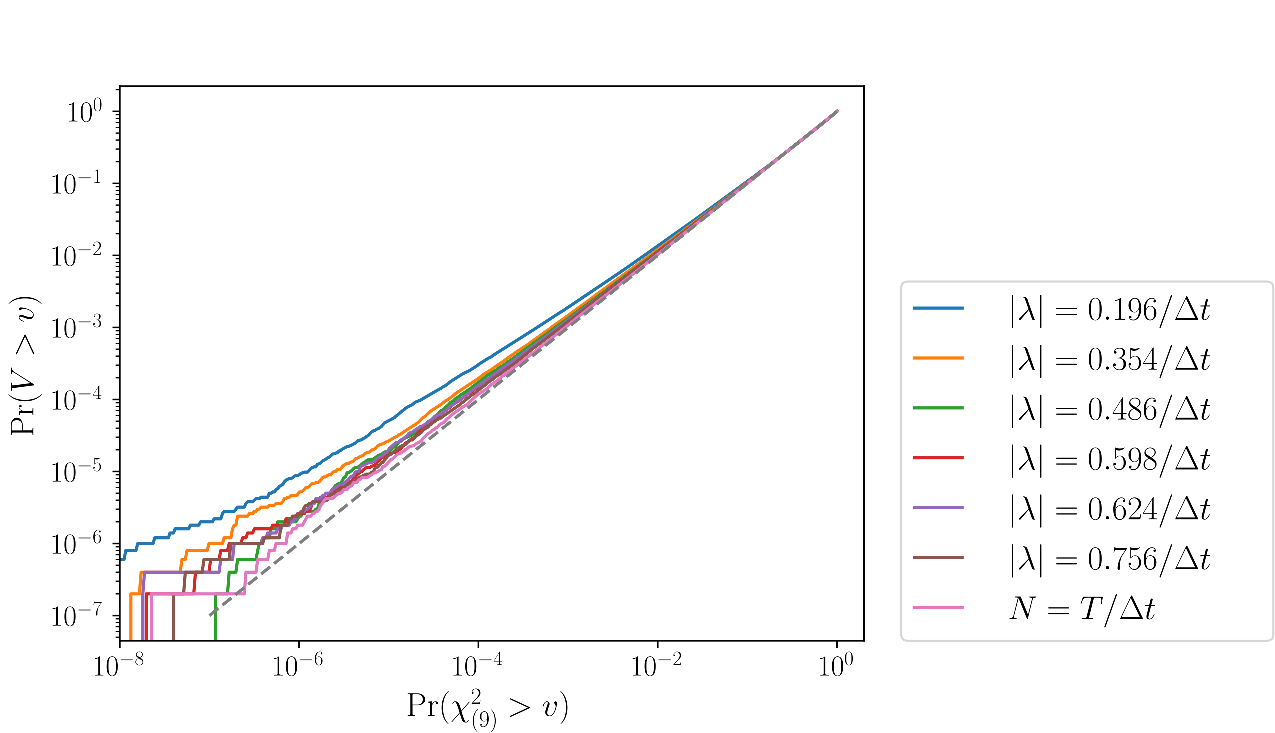}
		\end{center}
		\caption{Results for the real part of the product of two uncorrelated complex Gaussians, for $5 \times 10^6$ trials.}
		\label{fig:chi2-c0}
	\end{figure}

	For the complex case, we consider independent complex Gaussian random variables $W$ and $Z$, isotropic in the complex plane, and normalized such that $\langle |W|^2 \rangle = \langle |Z|^2 \rangle = 1$. Similarly, for a number $\nu$, we consider the pair $(W, Z + \nu W)$. We have:
	\begin{equation}
		\langle |Z + \nu W|^2 \rangle = 1 + |\nu|^2,
	\end{equation}
	so that for a real correlation coefficient $\rho$ between $W$ and $Z + \nu W$, Eq.\ \eqref{eq:rho-nu} again relates $\rho$ and $\nu$. We now have for the centralized second and fourth moments:
	\begin{align}
		\langle \{\Re [W^*(Z + \nu W)] - \nu\}^2 \rangle &= \frac 1 2 + \nu^2, \\
		\langle \{\Re [W^*(Z + \nu W)] - \nu\}^4 \rangle &= \frac 3 2 + 9 \nu^2 + 9 \nu^4.
	\end{align}
	Note that the excess kurtosis for the complex case is one-half that for the real case. The case $|\rho| \le 0.3$ can be approximated by $\rho = 0$, and the case $|\rho| \ge 0.5$ can be approximated by $|\rho| = 1$. Results for the real case are shown in Figs.\ \ref{fig:chi2-1} and \ref{fig:chi2-0}, while results for the complex case are shown in Figs.\ \ref{fig:chi2-c1} and \ref{fig:chi2-c0}.

	For the tests of the covariance of two quantities, both of which are non-Gaussian, we take the higher corresponding $p$-value on the vertical axis corresponding to the Gaussian $p$-value on the horizontal axis. The results are shown in Tables \ref{tab:pairs0}\textendash \ref{tab:pairs2-3}.
	
	\begin{table}
		\begin{center}
			\caption{Results for pairs of quantities (1). The quantities $\kappa_{<}/\max(\kappa_{>},1)$ (details in text) (top), the number of null results out of $5 \times 10^6$ trials (middle), and the scaled covariances (bottom).}
			\label{tab:pairs0}
			\begin{tblr}{Q[c,m]|Q[c,m]|Q[c,m]|Q[c,m]|}
				& $\langle s^0_2 \rangle$ & $\langle s^0_3 \rangle$ & $\langle v^0 \rangle$ \\ \hline
				$\langle s^0_0 \rangle$ & {0.51 \\ 0 \\ 0.38} & {0.15 \\ 50 \\ 0.23} & {0.07 \\ 0 \\ 0.24} \\ \hline
				$\langle s^0_2 \rangle$ & & {0.08 \\ 74172 \\ $-0.91$} & {0.07 \\ 0 \\ 0.45} \\ \hline
				$\langle s^0_3 \rangle$ & & & {0.02 \\ $> 10^6$ \\ $-0.38$} \\ \hline
			\end{tblr}
		\end{center}
	\end{table}
	
	\begin{table}
		\begin{center}
			\caption{Results for pairs of quantities (2). The quantities $\kappa_{<}/\max(\kappa_{>},1)$ (details in text) (top), the number of null results out of $5 \times 10^6$ trials (middle), and the scaled covariances (bottom). Asterisks denote statistical significance when estimated using a Gaussian distribution, while daggers denote statistical significance estimated to account for non-normality.}
			\label{tab:pairs1-1}
			\begin{tblr}{colspec = {Q[c,m]|Q[c,m]|Q[c,m]|Q[c,m]|Q[c,m]|Q[c,m]|Q[c,m]|Q[c,m]|Q[c,m]|Q[c,m]|Q[c,m]|},
					cells = {font=\fontsize{8pt}{9pt}\selectfont},
				}
				& $\langle (\Delta s^0_2)^2 \rangle$ & $L(\Delta s^0_2, \Delta s^0_3)$ & $\langle \Delta s^0_2 \Delta v^0 \rangle$ & $\langle (\dot s^0_2)^2 \rangle$ & $\langle \dot s^0_2 \dot s^0_3 \rangle$ & $\langle (\dot s^0_3)^2 \rangle$ & $\langle \dot s^0_3 \Delta v^0 \rangle$ & $\langle (\Delta v^0)^2 \rangle$ & $L(\dot s^0_2, \dot s^0_3)$ & $L(\dot s^0_2, \Delta v^0)$ \\ \hline
				$\langle s^0_0 \rangle$ & {0.09 \\ 2249 \\ $-0.26$} & {0.05 \\ 7974 \\ 0.35} & {0.39 \\ $0^{*\dagger}$ \\ $-0.04$} & {0.06 \\ $83^{*\dagger}$ \\ $-0.78$} & {0.02 \\ 15292 \\ 0.24} & {0.10 \\ $0^{*\dagger}$ \\ $-0.40$} & {0.05 \\ $0^{*\dagger}$ \\ $-0.31$} & {0.12 \\ $0^{*\dagger}$ \\ 0.46} & {0.05 \\ $69^{*\dagger}$ \\ 0.53} & {0.04 \\ $44^{*\dagger}$ \\ $-0.31$} \\ \hline
				$\langle s^0_2 \rangle$ & {0.11 \\ 27370 \\ 0.44} & {0.07 \\ 41964 \\ $-0.39$} & {0.53 \\ $0^{*\dagger}$ \\ 0.44} & {0.21 \\ $0^{*\dagger}$ \\ 0.04} & {0.04 \\ 7142 \\ 0.15} & {0.18 \\ $0^{*\dagger}$ \\ 0.31} & {0.03 \\ 2800 \\ $-0.53$} & {0.28 \\ $0^{*\dagger}$ \\ $-0.08$} & {0.13 \\ $0^{*\dagger}$ \\ $-0.19$} & {0.09 \\ $0^{*\dagger}$ \\ 0.07} \\ \hline
				$\langle s^0_3 \rangle$ & {0.00 \\ $> 10^6$ \\ $-0.45$} & {0.08 \\ 420071 \\ 0.28} & {0.16 \\ 57234 \\ $-0.61$} & {0.35 \\ $31^{*\dagger}$ \\ $-0.11$} & {0.07 \\ 47534 \\ $-0.09$} & {0.26 \\ $799^*$ \\ $-0.23$} & {0.07 \\ 180842 \\ 0.29} & {0.51 \\ $13^{*\dagger}$ \\ 0.04} & {0.21 \\ $314^{*\dagger}$ \\ 0.14} & {0.14 \\ $1340^*$ \\ $-0.11$} \\ \hline
				$\langle v^0 \rangle$ & {0.17 \\ 10241 \\ 0.12} & {0.12 \\ 12221 \\ $-0.12$} & {0.16 \\ $1^{*\dagger}$ \\ 0.18} & {0.16 \\ $0^{*\dagger}$ \\ 0.15} & {0.09 \\ 4162 \\ $-0.01$} & {0.10 \\ $5^{*\dagger}$ \\ 0.22} & {0.13 \\ $0^{*\dagger}$ \\ $-0.11$} & {0.14 \\ $0^{*\dagger}$ \\ 0.20} & {0.13 \\ $0^{*\dagger}$ \\ $-0.13$} & {0.07 \\ $393^{*\dagger}$ \\ 0.13} \\ \hline
			\end{tblr}
		\end{center}
	\end{table}
	
	\begin{table}
		\begin{center}
			\caption{Results for pairs of quantities (3). The quantities $\kappa_{<}/\max(\kappa_{>},1)$ (details in text) (top), the number of null results out of $5 \times 10^6$ trials (middle), and the scaled covariances (bottom). Asterisks denote statistical significance when estimated using a Gaussian distribution, while daggers denote statistical significance estimated to account for non-normality.}
			\label{tab:pairs1-2}
			\begin{tblr}{colspec = {Q[c,m]|Q[c,m]|Q[c,m]|Q[c,m]|Q[c,m]|Q[c,m]|Q[c,m]|Q[c,m]|Q[c,m]|Q[c,m]|Q[c,m]|Q[c,m]|},
					cells = {font=\fontsize{8pt}{9pt}\selectfont},
				}
				& $D^0_{22}$ & $D^0_{33}$ & $D^0_{3v}$ & $D^0_{vv}$ & $\langle s^1_3 {s^1_3}^* \rangle$ & $\Re L(s^1_2, {s^1_3}^*)$ & $\Re \langle s^1_3 {v^1}^* \rangle$ & $\langle \dot s^1_2 {{}\dot s^1_2}^* \rangle$ & $\Re \langle \dot s^1_2 {{}\dot s^1_3}^* \rangle$ & $\langle \dot s^1_3 {{}\dot s^1_3}^* \rangle$ & $\Re \langle \dot s^1_3 {v^1}^* \rangle$ \\ \hline
				$\langle s^0_0 \rangle$ & {0.06 \\ $188^{*\dagger}$ \\ $-0.95$} & {0.12 \\ $0^{*\dagger}$ \\ $-0.21$} & {0.02 \\ 2146 \\ $-0.07$} & {0.06 \\ $23^{*\dagger}$ \\ 0.12} & {0.08 \\ 3253 \\ $-1.20$} & {0.39 \\ $0^{*\dagger}$ \\ 0.84} & {0.15 \\ 22302 \\ $-0.43$} & {0.08 \\ $0^{*\dagger}$ \\ $-0.37$} & {0.03 \\ 5299 \\ $-0.16$} & {0.09 \\ $2^{*\dagger}$ \\ $-0.59$} & {0.01 \\ 143759 \\ 0.41} \\ \hline
				$\langle s^0_2 \rangle$ & {0.27 \\ $0^{*\dagger}$ \\ $-0.00$} & {0.23 \\ $0^{*\dagger}$ \\ 0.13} & {0.01 \\ 204739 \\ $-0.24$} & {0.08 \\ 3190 \\ $-0.31$} & {0.33 \\ $0^{*\dagger}$ \\ 0.36} & {0.37 \\ $0^{*\dagger}$ \\ $-0.76$} & {0.03 \\ $> 10^6$ \\ 0.97} & {0.14 \\ $0^{*\dagger}$ \\ 0.33} & {0.05 \\ 3657 \\ $-0.16$} & {0.16 \\ $2^{*\dagger}$ \\ 0.45} & {0.05 \\ 2279 \\ 0.04} \\ \hline
				$\langle s^0_3 \rangle$ & {0.42 \\ $58^{*\dagger}$ \\ $-0.13$} & {0.33 \\ $172^{*\dagger}$ \\ $-0.17$} & {0.04 \\ 151786 \\ 0.11} & {0.17 \\ 4589 \\ 0.14} & {0.19 \\ 16254 \\ $-0.46$} & {0.12 \\ 30134 \\ 0.69} & {0.03 \\ $> 10^6$ \\ $-0.57$} & {0.16 \\ 10780 \\ $-0.29$} & {0.11 \\ 5597 \\ 0.01} & {0.07 \\ 290344 \\ $-0.45$} & {0.08 \\ 20643 \\ 0.09} \\ \hline
				$\langle v^0 \rangle$ & {0.20 \\ $0^{*\dagger}$ \\ 0.13} & {0.12 \\ $0^{*\dagger}$ \\ 0.21} & {0.06 \\ 2147 \\ $-0.01$} & {0.17 \\ $17^{*\dagger}$ \\ 0.06} & {0.24 \\ $0^{*\dagger}$ \\ 0.09} & {0.14 \\ $2^{*\dagger}$ \\ $-0.38$} & {0.14 \\ 69909 \\ 0.23} & {0.04 \\ 37998 \\ 0.25} & {0.11 \\ 4180 \\ 0.03} & {0.09 \\ $11^{*\dagger}$ \\ 0.24} & {0.08 \\ 2891 \\ $-0.04$} \\ \hline
			\end{tblr}
		\end{center}
	\end{table}
	
	\begin{table}
		\begin{center}
			\caption{Results for pairs of quantities (4). The quantities $\kappa_{<}/\max(\kappa_{>},1)$ (details in text) (top), the number of null results out of $5 \times 10^6$ trials (middle), and the scaled covariances (bottom). Asterisks denote statistical significance when estimated using a Gaussian distribution, while daggers denote statistical significance estimated to account for non-normality.}
			\label{tab:pairs1-3}
			\begin{tblr}{colspec = {Q[c,m]|Q[c,m]|Q[c,m]|Q[c,m]|Q[c,m]|Q[c,m]|Q[c,m]|Q[c,m]|Q[c,m]|Q[c,m]|Q[c,m]|},
					cells = {font=\fontsize{8pt}{9pt}\selectfont},
				}
				& $\langle \dot \theta^2 \rangle$ & $\Re \langle \dot \theta e^{i \theta} {v^1}^* \rangle$ & $\langle v^1 {v^1}^* \rangle$ & $\Re L(\dot s^1_2, {{}\dot s^1_3}^*)$ & $\Re L(\dot s^1_2, {v^1}^*)$ & $D^1_{22}$ & $\Re D^1_{23}$ & $\Re D^1_{2 \theta}$ & $D^1_{33}$ & $D^1_{\theta \theta}$ \\ \hline
				$\langle s^0_0 \rangle$ & {0.16 \\ $0^{*\dagger}$ \\ $-0.28$} & {0.08 \\ $0^{*\dagger}$ \\ $-0.27$} & {0.08 \\ $0^{*\dagger}$ \\ $-0.33$} & {0.03 \\ 46032 \\ 0.48} & {0.03 \\ 5755 \\ $-0.09$} & {0.09 \\ $0^{*\dagger}$ \\ $-0.27$} & {0.05 \\ 5985 \\ $-0.23$} & {0.06 \\ 3455 \\ $-0.28$} & {0.11 \\ $25^{*\dagger}$ \\ $-0.36$} & {0.14 \\ $0^{*\dagger}$ \\ $-0.48$} \\ \hline
				$\langle s^0_2 \rangle$ & {0.19 \\ $0^{*\dagger}$ \\ $-0.57$} & {0.11 \\ $0^{*\dagger}$ \\ $-0.38$} & {0.17 \\ $0^{*\dagger}$ \\ $-0.18$} & {0.10 \\ 1667 \\ 0.03} & {0.05 \\ 8578 \\ $-0.10$} & {0.14 \\ $0^{*\dagger}$ \\ 0.38} & {0.05 \\ 32383 \\ $-0.40$} & {0.01 \\ $> 10^6$ \\ $-0.63$} & {0.18 \\ $61^{*\dagger}$ \\ 0.35} & {0.19 \\ $0^{*\dagger}$ \\ $-0.53$} \\ \hline
				$\langle s^0_3 \rangle$ & {0.29 \\ $292^{*\dagger}$ \\ 0.32} & {0.24 \\ $80^{*\dagger}$ \\ 0.15} & {0.34 \\ $18^{*\dagger}$ \\ $-0.01$} & {0.18 \\ 2590 \\ 0.07} & {0.10 \\ 7124 \\ $-0.01$} & {0.18 \\ 8525 \\ $-0.28$} & {0.18 \\ 3258 \\ 0.09} & {0.05 \\ 582617 \\ 0.31} & {0.12 \\ 107798 \\ $-0.39$} & {0.37 \\ $60^{*\dagger}$ \\ 0.23} \\ \hline
				$\langle v^0 \rangle$ & {0.25 \\ $0^{*\dagger}$ \\ 0.03} & {0.20 \\ $0^{*\dagger}$ \\ 0.07} & {0.14 \\ $43^{*\dagger}$ \\ 0.16} & {0.12 \\ 2120 \\ $-0.10$} & {0.05 \\ 71604 \\ 0.10} & {0.05 \\ 39786 \\ 0.26} & {0.19 \\ 3053 \\ 0.02} & {0.11 \\ 4471 \\ $-0.13$} & {0.07 \\ 3600 \\ 0.26} & {0.25 \\ $0^{*\dagger}$ \\ 0.01} \\ \hline
			\end{tblr}
		\end{center}
	\end{table}
	
	\begin{table}
		\begin{center}
			\caption{Results for pairs of quantities (5). The quantities $\kappa_{<}/\max(\kappa_{>},1)$ (details in text) (top), the number of null results out of $5 \times 10^6$ trials (middle), and the scaled covariances (bottom). Asterisks denote statistical significance when estimated using a Gaussian distribution, while daggers denote statistical significance estimated to account for non-normality.}
			\label{tab:pairs1-4}
			\begin{tblr}{colspec = {Q[c,m]|Q[c,m]|Q[c,m]|Q[c,m]|Q[c,m]|Q[c,m]|Q[c,m]|Q[c,m]|Q[c,m]|Q[c,m]|Q[c,m]|},
					cells = {font=\fontsize{8pt}{9pt}\selectfont},
				}
				& $\Re D^1_{\theta v}$ & $D^1_{vv}$ & $\langle s^2_3 {s^2_3}^* \rangle$ & $\langle \dot s^2_2 {{}\dot s^2_2}^* \rangle$ & $\Re \langle \dot s^2_2 {{}\dot s^2_3}^* \rangle$ & $\langle \dot s^2_3 {{}\dot s^2_3}^* \rangle$ & $\Re L(\dot s^2_2, {{}\dot s^2_3}^*)$ & $D^2_{22}$ & $\Re D^2_{23}$ & $D^2_{33}$ \\ \hline
				$\langle s^0_0 \rangle$ & {0.02 \\ 43714 \\ $-0.42$} & {0.04 \\ $113^{*\dagger}$ \\ $-0.55$} & {0.09 \\ 4569 \\ $-1.24$} & {0.06 \\ 1687 \\ $-0.99$} & {0.03 \\ 7286 \\ $-0.54$} & {0.10 \\ $0^{*\dagger}$ \\ $-0.71$} & {0.03 \\ 57284 \\ 0.60} & {0.13 \\ $12^{*\dagger}$ \\ $-1.19$} & {0.06 \\ $17^{*\dagger}$ \\ $-0.61$} & {0.17 \\ $0^{*\dagger}$ \\ $-0.68$} \\ \hline
				$\langle s^0_2 \rangle$ & {0.06 \\ $774^{*\dagger}$ \\ $-0.11$} & {0.14 \\ $0^{*\dagger}$ \\ 0.04} & {0.39 \\ $1^{*\dagger}$ \\ 0.33} & {0.24 \\ $0^{*\dagger}$ \\ $-0.33$} & {0.09 \\ $48^{*\dagger}$ \\ 0.21} & {0.26 \\ $0^{*\dagger}$ \\ 0.25} & {0.11 \\ $97^{*\dagger}$ \\ $-0.03$} & {0.40 \\ $0^{*\dagger}$ \\ $-0.41$} & {0.14 \\ $0^{*\dagger}$ \\ 0.32} & {0.39 \\ $0^{*\dagger}$ \\ 0.24} \\ \hline
				$\langle s^0_3 \rangle$ & {0.13 \\ 2150 \\ 0.01} & {0.23 \\ $152^{*\dagger}$ \\ $-0.09$} & {0.20 \\ 16708 \\ $-0.47$} & {0.50 \\ $19^{*\dagger}$ \\ $-0.06$} & {0.09 \\ 19264 \\ $-0.24$} & {0.22 \\ 6102 \\ $-0.36$} & {0.20 \\ $217^{*\dagger}$ \\ 0.06} & {0.60 \\ $13^{*\dagger}$ \\ $-0.07$} & {0.13 \\ 2042 \\ $-0.32$} & {0.24 \\ 2264 \\ $-0.42$} \\ \hline
				$\langle v^0 \rangle$ & {0.11 \\ 12159 \\ 0.05} & {0.13 \\ $438^{*\dagger}$ \\ 0.12} & {0.25 \\ $1^{*\dagger}$ \\ 0.07} & {0.23 \\ $0^{*\dagger}$ \\ 0.08} & {0.10 \\ 1752 \\ 0.13} & {0.12 \\ $332^{*\dagger}$ \\ 0.24} & {0.14 \\ $107^{*\dagger}$ \\ $-0.09$} & {0.24 \\ $0^{*\dagger}$ \\ 0.12} & {0.11 \\ $6^{*\dagger}$ \\ 0.18} & {0.12 \\ $35^{*\dagger}$ \\ 0.31} \\ \hline
			\end{tblr}
		\end{center}
	\end{table}
	
	\begin{table}
		\begin{center}
			\caption{Results for pairs of quantities (6). The ``Ratio'' column contains $\kappa_{<}/\max(\kappa_{>},1)$ (details in text) and the ``No.''\ column contains the number of null results out of $5 \times 10^6$ trials. Only statistically significant results estimated using a Gaussian distribution are shown, while daggers denote statistical significance estimated to account for non-normality.}
			\label{tab:pairs2-1}
			\begin{tblr}{colspec = {|Q[c,m]|Q[c,m]|Q[c,m]|Q[c,m]||Q[c,m]|Q[c,m]|Q[c,m]|Q[c,m]|},
					cells = {font=\fontsize{8pt}{9pt}\selectfont},
				}
				\SetCell[c=2]{c} Quantities & & Ratio & No. & \SetCell[c=2]{c} Quantities & & Ratio & No. \\ \hline
				$\langle \Delta s^0_2 \Delta v^0 \rangle$ & $\langle (\dot s^0_2)^2 \rangle$ & 0.27 & $0^\dagger$ & $\langle s^1_3 {s^1_3}^* \rangle$ & $\Re \langle \dot \theta e^{i \phi} {v^1}^* \rangle$ & 0.30 & $0^\dagger$ \\ \hline
				$\langle \Delta s^0_2 \Delta v^0 \rangle$ & $\langle (\dot s^0_3)^2 \rangle$ & 0.26 & $0^\dagger$ & $\langle s^1_3 {s^1_3}^* \rangle$ & $\langle v^1 {v^1}^* \rangle$ & 0.32 & $0^\dagger$ \\ \hline
				$\langle \Delta s^0_2 \Delta v^0 \rangle$ & $\langle \dot s^0_3 \Delta v^0 \rangle$ & 0.10 & 23 & $\langle s^1_3 {s^1_3}^* \rangle$ & $D^1_{22}$ & 0.26 & 8 \\ \hline
				$\langle \Delta s^0_2 \Delta v^0 \rangle$ & $\langle (\Delta v^0)^2 \rangle$ & 0.27 & $0^\dagger$ & $\langle s^1_3 {s^1_3}^* \rangle$ & $D^1_{\theta \theta}$ & 0.55 & $0^\dagger$ \\ \hline
				$\langle \Delta s^0_2 \Delta v^0 \rangle$ & $L(\dot s^0_2, \dot s^0_3)$ & 0.16 & $0^\dagger$ & $\langle s^1_3 {s^1_3}^* \rangle$ & $D^1_{vv}$ & 0.21 & $0^\dagger$ \\ \hline
				$\langle \Delta s^0_2 \Delta v^0 \rangle$ & $L(\dot s^0_2, \Delta v^0)$ & 0.03 & $6^\dagger$ & $\Re L(s^1_2, {s^1_3}^*)$ & $\langle \dot s^1_3 {{}\dot s^1_3}^* \rangle$ & 0.38 & $0^\dagger$ \\ \hline
				$\langle \Delta s^0_2 \Delta v^0 \rangle$ & $D^0_{22}$ & 0.36 & $0^\dagger$ & $\Re L(s^1_2, {s^1_3}^*)$ & $\Re \langle \dot s^1_3 {v^1}^* \rangle$ & 0.20 & $0^\dagger$ \\ \hline
				$\langle \Delta s^0_2 \Delta v^0 \rangle$ & $D^0_{33}$ & 0.27 & $2^\dagger$ & $\Re L(s^1_2, {s^1_3}^*)$ & $\langle \dot \theta^2 \rangle$ & 0.19 & $0^\dagger$ \\ \hline
				$\langle \Delta s^0_2 \Delta v^0 \rangle$ & $D^0_{vv}$ & 0.15 & 35 & $\Re L(s^1_2, {s^1_3}^*)$ & $\Re L(\dot s^1_2, {{}\dot s^1_3}^*)$ & 0.09 & $8^\dagger$ \\ \hline
				$\langle (\dot s^0_2)^2 \rangle$ & $\langle (\dot s^0_3)^2 \rangle$ & 0.10 & 31 & $\Re L(s^1_2, {s^1_3}^*)$ & $D^1_{\theta \theta}$ & 0.34 & $0^\dagger$ \\ \hline
				$\langle (\dot s^0_2)^2 \rangle$ & $\langle \dot s^0_3 \Delta v^0 \rangle$ & 0.19 & $0^\dagger$ & $\Re L(s^1_2, {s^1_3}^*)$ & $D^1_{vv}$ & 0.16 & $0^\dagger$ \\ \hline
				$\langle (\dot s^0_2)^2 \rangle$ & $\langle (\Delta v^0)^2 \rangle$ & 0.24 & $0^\dagger$ & $\langle \dot s^1_2 {{} \dot s^1_2}^* \rangle$ & $\langle \dot \theta^2 \rangle$ & 0.27 & $0^\dagger$ \\ \hline
				$\langle (\dot s^0_2)^2 \rangle$ & $D^0_{33}$ & 0.17 & $0^\dagger$ & $\langle \dot s^1_2 {{} \dot s^1_2}^* \rangle$ & $D^1_{22}$ & 0.06 & $15^\dagger$ \\ \hline
				$\langle (\dot s^0_3)^2 \rangle$ & $\langle \dot s^0_3 \Delta v^0 \rangle$ & 0.20 & $0^\dagger$ & $\langle \dot s^1_2 {{} \dot s^1_2}^* \rangle$ & $D^1_{\theta \theta}$ & 0.31 & $0^\dagger$ \\ \hline
				$\langle (\dot s^0_3)^2 \rangle$ & $\langle (\Delta v^0)^2 \rangle$ & 0.35 & $0^\dagger$ & $\langle \dot s^1_3 {{} \dot s^1_3}^* \rangle$ & $\langle \dot \theta^2 \rangle$ & 0.45 & $0^\dagger$ \\ \hline
				$\langle (\dot s^0_3)^2 \rangle$ & $L(\dot s^0_2, \Delta v^0)$ & 0.09 & 16 & $\langle \dot s^1_3 {{} \dot s^1_3}^* \rangle$ & $\Re \langle \dot \theta e^{i \phi} {v^1}^* \rangle$ & 0.24 & $0^\dagger$ \\ \hline
				$\langle (\dot s^0_3)^2 \rangle$ & $D^0_{22}$ & 0.11 & $0^\dagger$ & $\langle \dot s^1_3 {{} \dot s^1_3}^* \rangle$ & $\langle v^1 {v^1}^* \rangle$ & 0.16 & 74 \\ \hline
				$\langle (\dot s^0_3)^2 \rangle$ & $D^0_{vv}$ & 0.20 & 12 & $\langle \dot s^1_3 {{} \dot s^1_3}^* \rangle$ & $D^1_{\theta \theta}$ & 0.44 & $0^\dagger$ \\ \hline
				$\langle \dot s^0_3 \Delta v^0 \rangle$ & $\langle (\Delta v^0)^2 \rangle$ & 0.21 & $0^\dagger$ & $\langle \dot \theta^2 \rangle$ & $D^1_{22}$ & 0.27 & $0^\dagger$ \\ \hline
				$\langle \dot s^0_3 \Delta v^0 \rangle$ & $L(\dot s^0_2, \dot s^0_3)$ & 0.21 & $0^\dagger$ & $\langle \dot \theta^2 \rangle$ & $D^1_{33}$ & 0.38 & $5^\dagger$ \\ \hline
				$\langle \dot s^0_3 \Delta v^0 \rangle$ & $L(\dot s^0_2, \Delta v^0)$ & 0.16 & $0^\dagger$ & $\langle \dot \theta^2 \rangle$ & $D^1_{vv}$ & 0.13 & $0^\dagger$ \\ \hline
				$\langle \dot s^0_3 \Delta v^0 \rangle$ & $D^0_{22}$ & 0.18 & $0^\dagger$ & $\Re \langle \dot \theta e^{i \phi} {v^1}^* \rangle$ & $D^1_{22}$ & 0.17 & 75 \\ \hline
				$\langle \dot s^0_3 \Delta v^0 \rangle$ & $D^0_{33}$ & 0.20 & $0^\dagger$ & $\Re \langle \dot \theta e^{i \phi} {v^1}^* \rangle$ & $D^1_{vv}$ & 0.07 & $17^\dagger$ \\ \hline
				$\langle (\Delta v^0)^2 \rangle$ & $L(\dot s^0_2, \dot s^0_3)$ & 0.23 & $0^\dagger$ & $D^1_{22}$ & $D^1_{\theta \theta}$ & 0.23 & $7^\dagger$ \\ \hline
				$\langle (\Delta v^0)^2 \rangle$ & $D^0_{22}$ & 0.33 & $0^\dagger$ & $D^1_{33}$ & $D^1_{\theta \theta}$ & 0.39 & $2^\dagger$ \\ \hline
				$\langle (\Delta v^0)^2 \rangle$ & $D^0_{33}$ & 0.32 & $0^\dagger$ & $D^1_{\theta \theta}$ & $D^1_{vv}$ & 0.15 & $31^\dagger$ \\ \hline
				$L(\dot s^0_2, \dot s^0_3)$ & $D^0_{33}$ & 0.12 & $0^\dagger$ & $\langle s^2_3 {s^2_3}^* \rangle$ & $\langle \dot s^2_2 {{}\dot s^2_2}^* \rangle$ & 0.24 & 16 \\ \hline
				$L(\dot s^0_2, \dot s^0_3)$ & $D^0_{vv}$ & 0.18 & $17^\dagger$ & $\langle s^2_3 {s^2_3}^* \rangle$ & $D^2_{22}$ & 0.32 & 8 \\ \hline
				$L(\dot s^0_2, \Delta v^0)$ & $D^0_{33}$ & 0.12 & $1^\dagger$ & $\langle \dot s^2_2 {{}\dot s^2_2}^* \rangle$ & $\Re \langle \dot s^2_2 {{}\dot s^2_3}^* \rangle$ & 0.18 & 29 \\ \hline
				$D^0_{22}$ & $D^0_{33}$ & 0.19 & $0^\dagger$ & $\langle \dot s^2_2 {{}\dot s^2_2}^* \rangle$ & $\Re D^2_{23}$ & 0.16 & $1^\dagger$ \\ \hline
				$D^0_{22}$ & $D^0_{vv}$ & 0.15 & $23^\dagger$ & $\Re \langle \dot s^2_2 {{}\dot s^2_3}^* \rangle$ & $D^2_{22}$ & 0.12 & 93 \\ \hline
				$D^0_{33}$ & $D^0_{vv}$ & 0.17 & $20^\dagger$ & $\langle \dot s^2_3 {{}\dot s^2_3}^* \rangle$ & $D^2_{22}$ & 0.15 & 67 \\ \hline
				$\langle s^1_3 {s^1_3}^* \rangle$ & $\Re L(s^1_2, {s^1_3}^*)$ & 0.14 & 75 & $D^2_{22}$ & $\Re D^2_{23}$ & 0.27 & $0^\dagger$ \\ \hline
				$\langle s^1_3 {s^1_3}^* \rangle$ & $\langle \dot s^1_2 {{} \dot s^1_2}^* \rangle$ & 0.17 & 57 & $D^2_{22}$ & $D^2_{33}$ & 0.08 & $3^\dagger$ \\ \hline
				$\langle s^1_3 {s^1_3}^* \rangle$ & $\langle \dot \theta^2 \rangle$ & 0.54 & $0^\dagger$ & & & & \\ \hline
			\end{tblr}
		\end{center}
	\end{table}
	
	\begin{table}
		\begin{center}
			\caption{Results for pairs of quantities (7). The ``Ratio'' column contains $\kappa_{<}/\max(\kappa_{>},1)$ (details in text) and the ``No.''\ column contains the number of null results out of $5 \times 10^6$ trials. Only statistically significant results estimated using a Gaussian distribution are shown, while daggers denote statistical significance estimated to account for non-normality.}
			\label{tab:pairs2-2}
			\begin{tblr}{colspec = {|Q[c,m]|Q[c,m]|Q[c,m]|Q[c,m]||Q[c,m]|Q[c,m]|Q[c,m]|Q[c,m]|},
					cells = {font=\fontsize{8pt}{9pt}\selectfont},
				}
				\SetCell[c=2]{c} Quantities & & Ratio & No. & \SetCell[c=2]{c} Quantities & & Ratio & No. \\ \hline
				$\langle \Delta s^0_2 \Delta v^0 \rangle$ & $\langle \dot s^1_2 {{}\dot s^1_2}^* \rangle$ & 0.20 & $0^\dagger$ & $\langle (\Delta v^0)^2 \rangle$ & $\langle \dot \theta^2 \rangle$ & 0.20 & 3 \\ \hline
				$\langle \Delta s^0_2 \Delta v^0 \rangle$ & $\langle \dot \theta^2 \rangle$ & 0.39 & $0^\dagger$ & $\langle (\Delta v^0)^2 \rangle$ & $D^1_{22}$ & 0.26 & $0^\dagger$ \\ \hline
				$\langle \Delta s^0_2 \Delta v^0 \rangle$ & $\Re \langle \dot \theta e^{i \phi} {v^1}^* \rangle$ & 0.21 & $0^\dagger$ & $\langle (\Delta v^0)^2 \rangle$ & $D^1_{33}$ & 0.32 & 5 \\ \hline
				$\langle \Delta s^0_2 \Delta v^0 \rangle$ & $\langle v^1 {v^1}^* \rangle$ & 0.23 & $2^\dagger$ & $\langle (\Delta v^0)^2 \rangle$ & $D^1_{\theta \theta}$ & 0.33 & $0^\dagger$ \\ \hline
				$\langle \Delta s^0_2 \Delta v^0 \rangle$ & $D^1_{22}$ & 0.26 & $2^\dagger$ & $L(\dot s^0_2, \dot s^0_3)$ & $\langle s^1_3 {s^1_3}^* \rangle$ & 0.21 & $0^\dagger$ \\ \hline
				$\langle \Delta s^0_2 \Delta v^0 \rangle$ & $D^1_{\theta \theta}$ & 0.30 & 71 & $L(\dot s^0_2, \dot s^0_3)$ & $\Re L(s^1_2, {s^1_3}^*)$ & 0.11 & $10^\dagger$ \\ \hline
				$\langle \Delta s^0_2 \Delta v^0 \rangle$ & $D^1_{vv}$ & 0.18 & $0^\dagger$ & $L(\dot s^0_2, \dot s^0_3)$ & $\langle \dot s^1_3 {{}\dot s^1_3}^* \rangle$ & 0.13 & 71 \\ \hline
				$\langle (\dot s^0_2)^2 \rangle$ & $\langle s^1_3 {s^1_3}^* \rangle$ & 0.24 & 3 & $L(\dot s^0_2, \dot s^0_3)$ & $\langle \dot \theta^2 \rangle$ & 0.26 & $0^\dagger$ \\ \hline
				$\langle (\dot s^0_2)^2 \rangle$ & $\Re L(s^1_2, {s^1_3}^*)$ & 0.14 & 49 & $L(\dot s^0_2, \dot s^0_3)$ & $\Re \langle \dot \theta e^{i \phi} {v^1}^* \rangle$ & 0.16 & $18^\dagger$ \\ \hline
				$\langle (\dot s^0_2)^2 \rangle$ & $\langle \dot s^1_3 {{}\dot s^1_3}^* \rangle$ & 0.13 & 86 & $L(\dot s^0_2, \dot s^0_3)$ & $D^1_{33}$ & 0.16 & 90 \\ \hline
				$\langle (\dot s^0_2)^2 \rangle$ & $\langle \dot \theta^2 \rangle$ & 0.21 & $0^\dagger$ & $L(\dot s^0_2, \dot s^0_3)$ & $D^1_{\theta \theta}$ & 0.24 & $0^\dagger$ \\ \hline
				$\langle (\dot s^0_2)^2 \rangle$ & $D^1_{22}$ & 0.11 & 12 & $L(\dot s^0_2, \Delta v^0)$ & $\Re L(s^1_2, {s^1_3}^*)$ & 0.11 & 124 \\ \hline
				$\langle (\dot s^0_2)^2 \rangle$ & $D^1_{\theta \theta}$ & 0.20 & $0^\dagger$ & $L(\dot s^0_2, \Delta v^0)$ & $\langle \dot \theta^2 \rangle$ & 0.13 & $16^\dagger$ \\ \hline
				$\langle (\dot s^0_3)^2 \rangle$ & $\langle s^1_3 {s^1_3}^* \rangle$ & 0.35 & $0^\dagger$ & $L(\dot s^0_2, \Delta v^0)$ & $D^1_{\theta \theta}$ & 0.15 & $9^\dagger$ \\ \hline
				$\langle (\dot s^0_3)^2 \rangle$ & $\Re L(s^1_2, {s^1_3}^*)$ & 0.20 & $0^\dagger$ & $D^0_{22}$ & $\langle s^1_3 {s^1_3}^* \rangle$ & 0.28 & $0^\dagger$ \\ \hline
				$\langle (\dot s^0_3)^2 \rangle$ & $\langle \dot s^1_3 {{}\dot s^1_3}^* \rangle$ & 0.15 & $0^\dagger$ & $D^0_{22}$ & $\Re L(s^1_2, {s^1_3}^*)$ & 0.20 & $2^\dagger$ \\ \hline
				$\langle (\dot s^0_3)^2 \rangle$ & $\langle \dot \theta^2 \rangle$ & 0.35 & $0^\dagger$ & $D^0_{22}$ & $\langle \dot s^1_2 {{}\dot s^1_2}^* \rangle$ & 0.12 & $3^\dagger$ \\ \hline
				$\langle (\dot s^0_3)^2 \rangle$ & $\Re \langle \dot \theta e^{i \phi} {v^1}^* \rangle$ & 0.19 & $0^\dagger$ & $D^0_{22}$ & $\langle \dot s^1_3 {{}\dot s^1_3}^* \rangle$ & 0.16 & $1^\dagger$ \\ \hline
				$\langle (\dot s^0_3)^2 \rangle$ & $\langle v^1 {v^1}^* \rangle$ & 0.12 & 89 & $D^0_{22}$ & $\langle \dot \theta^2 \rangle$ & 0.27 & $0^\dagger$ \\ \hline
				$\langle (\dot s^0_3)^2 \rangle$ & $D^1_{\theta \theta}$ & 0.30 & $0^\dagger$ & $D^0_{22}$ & $D^1_{\theta \theta}$ & 0.20 & $10^\dagger$ \\ \hline
				$\langle \dot s^0_3 \Delta v^0 \rangle$ & $\langle s^1_3 {s^1_3}^* \rangle$ & 0.18 & $0^\dagger$ & $D^0_{33}$ & $\langle s^1_3 {s^1_3}^* \rangle$ & 0.41 & $0^\dagger$ \\ \hline
				$\langle \dot s^0_3 \Delta v^0 \rangle$ & $\Re L(s^1_2, {s^1_3}^*)$ & 0.10 & 41 & $D^0_{33}$ & $\Re L(s^1_2, {s^1_3}^*)$ & 0.27 & $0^\dagger$ \\ \hline
				$\langle \dot s^0_3 \Delta v^0 \rangle$ & $\langle \dot s^1_2 {{}\dot s^1_2}^* \rangle$ & 0.20 & $0^\dagger$ & $D^0_{33}$ & $\langle \dot s^1_3 {{}\dot s^1_3}^* \rangle$ & 0.18 & $0^\dagger$ \\ \hline
				$\langle \dot s^0_3 \Delta v^0 \rangle$ & $\langle \dot s^1_3 {{}\dot s^1_3}^* \rangle$ & 0.17 & $0^\dagger$ & $D^0_{33}$ & $\langle \dot \theta^2 \rangle$ & 0.28 & $0^\dagger$ \\ \hline
				$\langle \dot s^0_3 \Delta v^0 \rangle$ & $\Re \langle \dot \theta e^{i \phi} {v^1}^* \rangle$ & 0.09 & 46 & $D^0_{33}$ & $D^1_{\theta \theta}$ & 0.24 & $3^\dagger$ \\ \hline
				$\langle \dot s^0_3 \Delta v^0 \rangle$ & $\langle v^1 {v^1}^* \rangle$ & 0.15 & $0^\dagger$ & $D^0_{vv}$ & $\Re L(s^1_2, {s^1_3}^*)$ & 0.14 & 100 \\ \hline
				$\langle \dot s^0_3 \Delta v^0 \rangle$ & $D^1_{22}$ & 0.22 & $0^\dagger$ & $D^0_{vv}$ & $\langle \dot s^1_2 {{}\dot s^1_2}^* \rangle$ & 0.17 & 21 \\ \hline
				$\langle \dot s^0_3 \Delta v^0 \rangle$ & $D^1_{33}$ & 0.20 & $18^\dagger$ & $D^0_{vv}$ & $D^1_{22}$ & 0.17 & 35 \\ \hline
				$\langle \dot s^0_3 \Delta v^0 \rangle$ & $D^1_{vv}$ & 0.15 & $0^\dagger$ & $\langle \Delta s^0_2 \Delta v^0 \rangle$ & $\langle s^2_3 {s^2_3}^* \rangle$ & 0.51 & $0^\dagger$ \\ \hline
				$\langle (\Delta v^0)^2 \rangle$ & $\langle s^1_3 {s^1_3}^* \rangle$ & 0.49 & $0^\dagger$ & $\langle \Delta s^0_2 \Delta v^0 \rangle$ & $\langle \dot s^2_2 {{}\dot s^2_2}^* \rangle$ & 0.38 & $0^\dagger$  \\ \hline
				$\langle (\Delta v^0)^2 \rangle$ & $\Re L(s^1_2, {s^1_3}^*)$ & 0.33 & $0^\dagger$ & $\langle \Delta s^0_2 \Delta v^0 \rangle$ & $\langle \dot s^2_3 {{}\dot s^2_3}^* \rangle$ & 0.35 & $1^\dagger$ \\ \hline
				$\langle (\Delta v^0)^2 \rangle$ & $\langle \dot s^1_2 {{}\dot s^1_2}^* \rangle$ & 0.19 & $0^\dagger$ & $\langle \Delta s^0_2 \Delta v^0 \rangle$ & $D^2_{22}$ & 0.68 & $0^\dagger$ \\ \hline
				$\langle (\Delta v^0)^2 \rangle$ & $\langle \dot s^1_3 {{}\dot s^1_3}^* \rangle$ & 0.39 & $0^\dagger$ & $\langle \Delta s^0_2 \Delta v^0 \rangle$ & $\Re D^2_{23}$ & 0.16 & 32 \\ \hline
			\end{tblr}
		\end{center}
	\end{table}
	
	\begin{table}
		\begin{center}
			\caption{Results for pairs of quantities (8). The ``Ratio'' column contains $\kappa_{<}/\max(\kappa_{>},1)$ (details in text) and the ``No.''\ column contains the number of null results out of $5 \times 10^6$ trials. Only statistically significant results estimated using a Gaussian distribution are shown, while daggers denote statistical significance estimated to account for non-normality.}
			\label{tab:pairs2-3}
			\begin{tblr}{colspec = {|Q[c,m]|Q[c,m]|Q[c,m]|Q[c,m]||Q[c,m]|Q[c,m]|Q[c,m]|Q[c,m]|},
					cells = {font=\fontsize{8pt}{9pt}\selectfont},
				}
				\SetCell[c=2]{c} Quantities & & Ratio & No. & \SetCell[c=2]{c} Quantities & & Ratio & No. \\ \hline
				$\langle \Delta s^0_2 \Delta v^0 \rangle$ & $D^2_{33}$ & 0.46 & $0^\dagger$ & $\langle s^1_3 {s^1_3}^* \rangle$ & $D^2_{33}$ & 0.17 & $0^\dagger$ \\ \hline
				$\langle (\dot s^0_2)^2 \rangle$ & $\langle s^2_3 {s^2_3}^* \rangle$ & 0.31 & $0^\dagger$ & $\Re L(s^1_2, {s^1_3}^*)$ & $\langle s^2_3 {s^2_3}^* \rangle$ & 0.28 & 111 \\ \hline
				$\langle (\dot s^0_3)^2 \rangle$ & $\langle s^2_3 {s^2_3}^* \rangle$ & 0.27 & 12 & $\Re L(s^1_2, {s^1_3}^*)$ & $\langle \dot s^2_2 {{}\dot s^2_2}^* \rangle$ & 0.28 & $0^\dagger$ \\ \hline
				$\langle (\dot s^0_3)^2 \rangle$ & $\langle \dot s^2_2 {{}\dot s^2_2}^* \rangle$ & 0.17 & 96 & $\Re L(s^1_2, {s^1_3}^*)$ & $D^2_{22}$ & 0.50 & $0^\dagger$ \\ \hline
				$\langle (\dot s^0_3)^2 \rangle$ & $D^2_{22}$ & 0.17 & $0^\dagger$ & $\langle \dot s^1_2 {{}\dot s^1_2}^* \rangle$ & $\langle s^2_3 {s^2_3}^* \rangle$ & 0.25 & $1^\dagger$ \\ \hline
				$\langle (\dot s^0_3)^2 \rangle$ & $\Re D^2_{23}$ & 0.13 & 23 & $\langle \dot s^1_2 {{}\dot s^1_2}^* \rangle$ & $D^2_{22}$ & 0.16 & 20 \\ \hline
				$\langle (\dot s^0_3)^2 \rangle$ & $D^2_{33}$ & 0.12 & $0^\dagger$ & $\langle \dot s^1_3 {{}\dot s^1_3}^* \rangle$ & $D^2_{22}$ & 0.20 & $7^\dagger$ \\ \hline
				$\langle \dot s^0_3 \Delta v^0 \rangle$ & $\langle s^2_3 {s^2_3}^* \rangle$ & 0.21 & $0^\dagger$ & $\langle \dot \theta^2 \rangle$ & $\langle s^2_3 {s^2_3}^* \rangle$ & 0.51 & $0^\dagger$ \\ \hline
				$\langle \dot s^0_3 \Delta v^0 \rangle$ & $\langle \dot s^2_2 {{}\dot s^2_2}^* \rangle$ & 0.12 & $11^\dagger$ & $\langle \dot \theta^2 \rangle$ & $\langle \dot s^2_2 {{}\dot s^2_2}^* \rangle$ & 0.20 & $0^\dagger$ \\ \hline
				$\langle \dot s^0_3 \Delta v^0 \rangle$ & $\langle \dot s^2_3 {{}\dot s^2_3}^* \rangle$ & 0.21 & $0^\dagger$ & $\langle \dot \theta^2 \rangle$ & $\Re \langle \dot s^2_2 {{}\dot s^2_3}^* \rangle$ & 0.20 & $5^\dagger$ \\ \hline
				$\langle \dot s^0_3 \Delta v^0 \rangle$ & $D^2_{22}$ & 0.12 & $0^\dagger$ & $\langle \dot \theta^2 \rangle$ & $\langle \dot s^2_3 {{}\dot s^2_3}^* \rangle$ & 0.47 & $0^\dagger$ \\ \hline
				$\langle \dot s^0_3 \Delta v^0 \rangle$ & $\Re D^2_{23}$ & 0.16 & $0^\dagger$ & $\langle \dot \theta^2 \rangle$ & $D^2_{22}$ & 0.19 & $0^\dagger$ \\ \hline
				$\langle \dot s^0_3 \Delta v^0 \rangle$ & $D^2_{33}$ & 0.22 & $0^\dagger$ & $\langle \dot \theta^2 \rangle$ & $\Re D^2_{23}$ & 0.34 & $0^\dagger$ \\ \hline
				$\langle (\Delta v^0)^2 \rangle$ & $\langle s^2_3 {s^2_3}^* \rangle$ & 0.39 & 15 & $\langle \dot \theta^2 \rangle$ & $D^2_{33}$ & 0.52 & $0^\dagger$ \\ \hline
				$\langle (\Delta v^0)^2 \rangle$ & $\langle \dot s^2_2 {{}\dot s^2_2}^* \rangle$ & 0.35 & $0^\dagger$ & $\Re \langle \dot \theta e^{i \phi} {v^1}^* \rangle$ & $\langle s^2_3 {s^2_3}^* \rangle$ & 0.30 & $0^\dagger$ \\ \hline
				$\langle (\Delta v^0)^2 \rangle$ & $\Re \langle \dot s^2_2 {{}\dot s^2_3}^* \rangle$ & 0.19 & 3 & $\Re \langle \dot \theta e^{i \phi} {v^1}^* \rangle$ & $\Re \langle \dot s^2_2 {{}\dot s^2_3}^* \rangle$ & 0.17 & $3^\dagger$ \\ \hline
				$\langle (\Delta v^0)^2 \rangle$ & $\langle \dot s^2_3 {{}\dot s^2_3}^* \rangle$ & 0.43 & $0^\dagger$ & $\Re \langle \dot \theta e^{i \phi} {v^1}^* \rangle$ & $\langle \dot s^2_3 {{}\dot s^2_3}^* \rangle$ & 0.20 & $0^\dagger$ \\ \hline
				$\langle (\Delta v^0)^2 \rangle$ & $D^2_{22}$ & 0.34 & $0^\dagger$ & $\Re \langle \dot \theta e^{i \phi} {v^1}^* \rangle$ & $\Re D^2_{23}$ & 0.24 & $0^\dagger$ \\ \hline
				$\langle (\Delta v^0)^2 \rangle$ & $\Re D^2_{23}$ & 0.30 & $0^\dagger$ & $\Re \langle \dot \theta e^{i \phi} {v^1}^* \rangle$ & $D^2_{33}$ & 0.20 & $0^\dagger$ \\ \hline
				$\langle (\Delta v^0)^2 \rangle$ & $D^2_{33}$ & 0.42 & $0^\dagger$ & $\langle v^1 {v^1}^* \rangle$ & $\langle s^2_3 {s^2_3}^* \rangle$ & 0.33 & $0^\dagger$ \\ \hline
				$L(\dot s^0_2, \dot s^0_3)$ & $D^2_{33}$ & 0.13 & $52^\dagger$ & $\langle v^1 {v^1}^* \rangle$ & $\Re \langle \dot s^2_2 {{}\dot s^2_3}^* \rangle$ & 0.12 & 53 \\ \hline
				$L(\dot s^0_2, \Delta v^0)$ & $\langle s^2_3 {s^2_3}^* \rangle$ & 0.15 & 3 & $\langle v^1 {v^1}^* \rangle$ & $\langle \dot s^2_3 {{}\dot s^2_3}^* \rangle$ & 0.12 & 105 \\ \hline
				$D^0_{22}$ & $\langle s^2_3 {s^2_3}^* \rangle$ & 0.27 & 10 & $\langle v^1 {v^1}^* \rangle$ & $\Re D^2_{23}$ & 0.17 & $0^\dagger$ \\ \hline
				$D^0_{22}$ & $D^2_{33}$ & 0.14 & $2^\dagger$ & $D^1_{22}$ & $\langle s^2_3 {s^2_3}^* \rangle$ & 0.24 & 85 \\ \hline
				$D^0_{33}$ & $\langle s^2_3 {s^2_3}^* \rangle$ & 0.31 & 13 & $D^1_{\theta \theta}$ & $\langle s^2_3 {s^2_3}^* \rangle$ & 0.59 & $0^\dagger$ \\ \hline
				$D^0_{33}$ & $D^2_{22}$ & 0.15 & 92 & $D^1_{\theta \theta}$ & $\langle \dot s^2_2 {{}\dot s^2_2}^* \rangle$ & 0.14 & $7^\dagger$ \\ \hline
				$D^0_{33}$ & $\Re D^2_{23}$ & 0.19 & $2^\dagger$ & $D^1_{\theta \theta}$ & $\Re \langle \dot s^2_2 {{}\dot s^2_3}^* \rangle$ & 0.20 & $6^\dagger$ \\ \hline
				$D^0_{33}$ & $D^2_{33}$ & 0.12 & $3^\dagger$ & $D^1_{\theta \theta}$ & $\langle \dot s^2_3 {{}\dot s^2_3}^* \rangle$ & 0.43 & $0^\dagger$ \\ \hline
				$D^0_{vv}$ & $\Re \langle \dot s^2_2 {{}\dot s^2_3}^* \rangle$ & 0.19 & 108 & $D^1_{\theta \theta}$ & $D^2_{22}$ & 0.10 & $19^\dagger$ \\ \hline
				$D^0_{vv}$ & $\langle \dot s^2_3 {{}\dot s^2_3}^* \rangle$ & 0.21 & $14^\dagger$ & $D^1_{\theta \theta}$ & $\Re D^2_{23}$ & 0.34 & $0^\dagger$ \\ \hline
				$D^0_{vv}$ & $\Re D^2_{23}$ & 0.22 & $18^\dagger$ & $D^1_{\theta \theta}$ & $D^2_{33}$ & 0.48 & $0^\dagger$ \\ \hline
				$D^0_{vv}$ & $D^2_{33}$ & 0.20 & $15^\dagger$ & $D^1_{vv}$ & $\langle s^2_3 {s^2_3}^* \rangle$ & 0.19 & 32 \\ \hline
				$\langle s^1_3 {s^1_3}^* \rangle$ & $\langle \dot s^2_2 {{}\dot s^2_2}^* \rangle$ & 0.28 & 66 & $D^1_{vv}$ & $\Re D^2_{23}$ & 0.13 & $32^\dagger$ \\ \hline
				$\langle s^1_3 {s^1_3}^* \rangle$ & $D^2_{22}$ & 0.39 & $0^\dagger$ & $D^1_{vv}$ & $D^2_{33}$ & 0.12 & 81 \\ \hline
			\end{tblr}
		\end{center}
	\end{table}

	\section{Code availability}
	Code and data files are available at \texttt{https://github.com/yeerenlow/tcells\_paper\_code}.
	
	\section{Acknowledgments}
	Y.I.L.\ acknowledges support from the McGill University Faculty of Science and supervision from Paul Fran\c{c}ois (Universit\'{e} de Montr\'{e}al) and Paul Wiseman (McGill University) for the analysis of a similar dataset.
	
	\bibliographystyle{unsrt}
	\bibliography{document4}
\end{document}